\documentclass[12pt, a4paper]{article}
\usepackage{graphicx}
\usepackage{mathrsfs}
\usepackage{amssymb}
\usepackage{color}
\usepackage{amsmath}
\usepackage{ascmac}
\usepackage{amsthm}
\usepackage{booktabs}
\usepackage{lscape}
\usepackage{dcolumn}
\usepackage{longtable}
\usepackage{dcolumn}
\usepackage{arydshln}
\usepackage{bm}
\usepackage{url}
\usepackage{caption}
\usepackage{subfig}
\usepackage{setspace}
\usepackage{cases}
\usepackage{comment}
\usepackage{multirow}
\usepackage{bigfoot}

\newcommand{\vx}{\mathbf{x}}

\newcommand{\eps}{\epsilon}
\newcommand{\vbeta}{\bm{\beta}}

\newcommand{\vgamma}{\bm{\gamma}}

\newcommand{\vphi}{\bm{\phi}}

\newcommand{\vtheta}{\bm{\theta}}

\begin{document}

\title {
\LARGE{Persistence of Natural Disasters on Children's Health:}\\
\LARGE{Evidence from the Great Kant\=o Earthquake of 1923}}

\author{Kota Ogasawara\thanks{
Department of Industrial Engineering, School of Engineering, Tokyo Institute of Technology, 2-12-1, Ookayama, Meguro-ku, Tokyo 152-8552, Japan (E-mail: ogasawara.k.ab@m.titech.ac.jp).\newline
I would like to thank Dan Bogart, Neil Cummins, Fabian Drixler, Giovanni Federico, Bernard Harris, Janet Hunter, Volha Lazuka, Kazushige Matsuda, Yukitoshi Matsushita, Stephen Morgan, Eric Schneider, Anthony Wray, and the participants at the CSG seminar (Chiba), EHES conference (T\"ubingen), Stunting conference (LSE), AED seminar (Kyoto), KIER workshop (Kyoto), WEHC (MIT), EHS conference (Belfast), and Tokyo Tech seminar for their helpful comments on the paper.
The work was supported by the Tokyo Institute of Technology (grant in 2015; 2019).
I wish to thank Motoya Taura and Tatsuki Inoue for their outstanding research assistance.
There are no conflicts of interest to declare.
All errors are my own.
}
\\
}
\date{\today}
\maketitle

\begin{abstract}
\begin{spacing}{0.85}
This study uses a catastrophic earthquake in 1923 to analyse the long-term effects of a natural disaster on children's health.
I find that fetal exposure to Japan's Great Kant\=o Earthquake had stunting effects on girls in the devastated area.
Disaster relief spending helped remediate stunting among boys by late primary school age, whereas it did not ameliorate girls' stunting, suggesting a prenatal selection mechanism and compensating investment after birth.
While the maternal mental stress via the fear of vibrations and anticipation of future aftershocks played a role in the adverse health effects, the maternal nutritional stress via physical disruption also enhanced those effects.
\end{spacing}
\bigskip

\bigskip


\end{abstract}
\newpage
\section{Introduction}

Natural disasters have significant and lasting impacts on economies.
In Japan, the Great Kant\=o Earthquake of 1923, an extremely huge quake with a moment magnitude scale of 7.9, was an unprecedented crisis, leaving 156,000 people killed, injured, or missing.
The massive earthquake of 1923 had positive long-term effects on industries in the vicinity of Tokyo through the so-called creative destruction mechanism, which includes the technological upgradation of machinery and the selection into efficient firms.\footnote{\setstretch{0.84}Okazaki et al., `Creative'. See also Pereira, `The opportunity' for the positive long-term economic impacts of the Lisbon Earthquake of 1755.}
However, a growing body of the literature indicates that fetal exposure to disasters can impede normal human development and thus lead to negative consequences for later-life health and socioeconomic outcomes.\footnote{\setstretch{0.84} See Almond and Currie, `Killing'; Currie and Vogl, `Early-life'; Prinz et al.~`Health' for comprehensive reviews in the related literature.}
The weight of evidence from this literature implies that little is known about the adverse long-term developmental effects of Japan's 1923 earthquake.

To bridge this gap in the body of knowledge, the present study investigates the impacts of fetal exposure to the Great Kant\=o Earthquake on children's health.
For this purpose, I select Chiba prefecture, where the physical disruption was primarily caused by strong vibrations, and establish a series of datasets on children's health using a set of physical examination records.\footnote{\setstretch{0.84} While Tokyo and Kanagawa prefectures were also greatly impacted by the earthquake hit, Chiba prefecture is considered to be the most appropriate research area. Tokyo experienced vibrations and an enormous fire at the same time, whereas Kanagawa suffered both vibrations and an enormous tsunami; Hunter, ```Extreme'. These incidents complicate the identification because separating their effects is difficult. Moreover, physical examination records for both affected prefectures are unavailable. However, fortunately, a set of records for Chiba prefecture are scattered but still remaining.}
Specifically, I construct school-year-age-level panel datasets on the height and weight of primary schoolchildren aged 6--11 years born between 1914 and 1929.
To better identify the impacts of the earthquake, I exploit the geospatial variation in the physical devastation and then interact the variation with the children who potentially experienced the earthquake \textit{in utero}.

I find that fetal exposure to the Great Kant\=o Earthquake negatively affected the growth of children.
Primary school girls aged 9--11 exposed \textit{in utero} in the area extremely affected by the earthquake with the maximum seismic intensity scale were roughly one cm shorter than those in surrounding cohorts.
Given this result, I attempt to distinguish different pathways of adverse health effects: mental and nutritional stress.
First, I test whether disaster relief supplies can mitigate the adverse health effects on children by absorbing nutritional stress in the physically devastated area in which the strongest vibration brought about unavoidable mental stress.
Second, I employ the regional heterogeneity in seismic activity and market disruption to test which stress is more likely to be plausible in the area with little physical damage.
From these exercises, I find suggestive evidence that while mental stress plays an important role in the adverse health effects, nutritional stress could be another pathway that enhances the adverse health effects on the exposed children in the physically devastated area.

Utilizing an earthquake to investigate the long-term impacts of fetal damage makes this study different from existing studies using other disasters such as pandemics, famines, and wars.
One feature is that forecasting the timing, location, and magnitude of the earthquakes is undoubtedly difficult.\footnote{\setstretch{0.84} For instance, while historical patterns of weather shocks such as floods are more predictive, earthquake prediction in this year was certainly limited; Takemura, `Kant\=o'.}
Pandemics, famines, and wars spread gradually in space and time, leading to endogenous responses by people to some degree and encouraging governments to prepare for forthcoming shocks by promoting hygiene behaviors, instigating preventive food rations, and planning for evacuation.\footnote{\setstretch{0.84} Almond and Currie, `Killing', p. 7.}
This implies that earthquakes can provide an ideal quasi-experimental environment in which the identification can be improved.
Another feature is that the strong vibrations and physical disruptions due to an earthquake influence not only physical but also mental health.
Unlike earthquakes, famines and exposure to infectious diseases are mainly associated with nutritional losses.
Although wars can also influence mental health, systematic preventive responses such as mass evacuation may make it difficult to provide clean experimental conditions.\footnote{\setstretch{0.84} Currie and Vogl, `Early-life', p. 27.}
In this light, earthquakes offer a favorable condition for disentangling mental and nutritional stresses despite not thus far being implemented in the literature.

Given these advantages, this study's contributions to the literature can be highlighted in the following three ways.
First, it adds evidence on the long-term health impacts of fetal earthquake exposure.
Previous studies have documented the adverse effects of fetal earthquake exposure on pregnancy outcomes such as low birth weight.\footnote{\setstretch{0.84} Glynn et al., `When'; Torche, `The effect'; Kim, `Maternal'.}
However, limited evidence is available on the lasting effects of fetal earthquake exposure on later-life health.\footnote{\setstretch{0.84}One exception is Caruso and Miller, `Long', who test the link between fetal exposure to the Ancash Earthquake of 1970 and self-reported disability status in adulthood, showing little evidence on such a link in modern Peru.}
This study bridges this gap by examining the adverse effects of fetal earthquake exposure on the development and diseases of the juvenile population.
It is the first to show the long-term human costs of the earthquake of 1923 that resulted in an unprecedented shock to the Japanese economy, given that previous studies have predominantly analyzed its impacts on industries and market functions.\footnote{A couple of studies investigate the effects on the spatio-temporal distribution of the industries and the creative destruction mechanism in the vicinity of Tokyo (Imaizumi et al., `Impact'; Okazaki et al., `Creative'), whereas another study investigates the convergence of prices in the cities after the earthquake (Hunter and Ogasawara, `Price').}

Second, by providing evidence on the ameliorating effects of disaster relief on child stunting, this study contributes to the literature on the optimal timing of child investment.\footnote{Heckman, `The developmental'.}
Empirical evidence on how the complementarities in parental investments respond to early-life shocks is at best mixed.\footnote{Almond and Mazumder, `Fetal'.}
Within the limited volume of evidence, the recent study finds that New Deal-related spending ameliorated the adverse long-term effects of Dust Bowl exposure on human capital formation.\footnote{Vellore, `The dust'.}
The consistent evidence found in the present study implies that prenatal adverse effects can be mitigated by the end of primary school.
However, I find a clear gender imbalance in the compensating effects of the disaster relief, which may be driven by the prenatal selection mechanism and postnatal-biased resource allocation regulated by the institutional context.

Third, this study enriches our understanding of the mechanisms behind the adverse health effects of fetal health shocks.
While a growing body of evidence documents the long-term effects of early-life health shocks on human capital, identifying the pathways of those effects has remained a challenge.\footnote{Currie and Vogl, `Early-life'.}
I provide suggestive evidence not only on the role of the maternal mental stress underlying the adverse health effects of fetal earthquake exposure, but also on the enhancing role of maternal nutritional stress in the physically disrupted area.
This is in line with the call for future research by Prinz et al., who suggest that mental health issues are becoming increasingly important for understanding labor market outcomes in adulthood.\footnote{Prinz et al., `Health'.}

\section{Background} \label{sec:sec2}

\subsection{Great Kant\=o Earthquake} \label{sec:sec21}

\begin{figure}[]
\centering
\includegraphics[width=6cm]{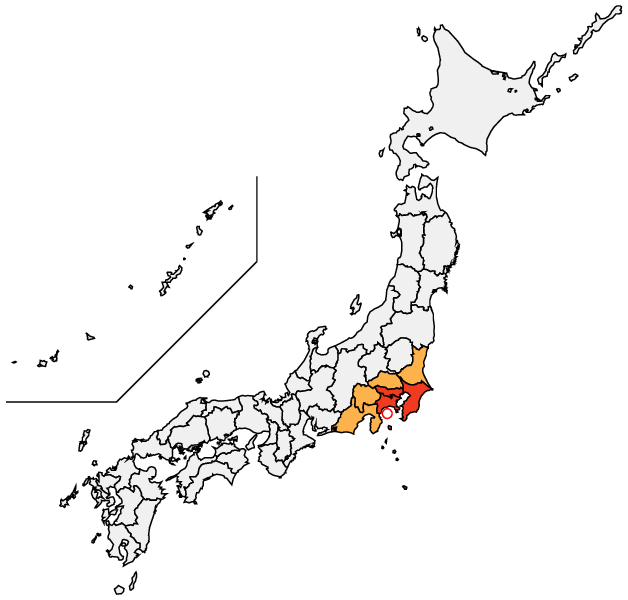}
\includegraphics[width=5.5cm]{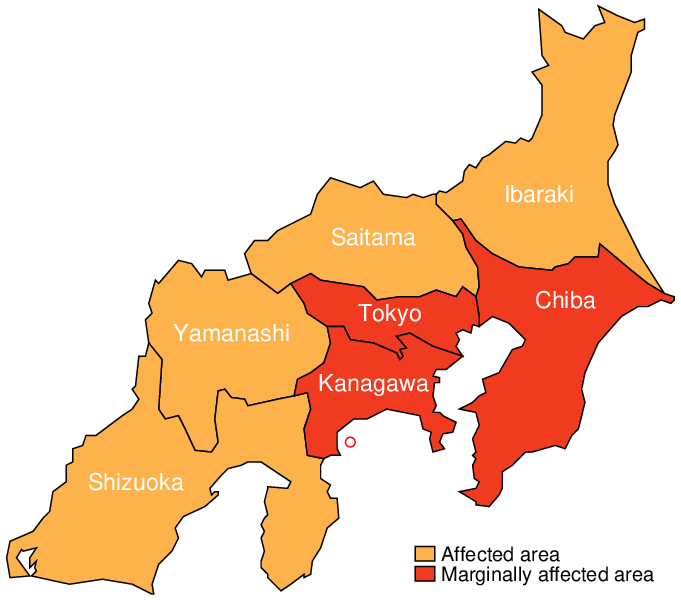}
\caption{Affected area and hypocenter}
\label{fig:prefecture}
\scriptsize{\begin{minipage}{350pt}
\setstretch{0.85}Notes: 
The red circle shows the hypocenter of the earthquake. 
The affected area includes Tokyo, Kanagawa, and Chiba prefectures. 
The marginally affected area includes Saitama, Shizuoka, Yamanashi, and Ibaraki prefectures.\\
Sources: Created by the author from Tokyo City Office, \textit{Kant\=ochih\=oshinsai}, p.~161.
The location of the hypocenter was based on the official database of the Japan Meteorological Agency, \textit{Shindo}.
\end{minipage}}
\end{figure}

The Great Kant\=o Earthquake hit the southern area of the Kant\=o district including the seven prefectures shown in Figure~\ref{fig:prefecture}.
Although both the physical and the human damage were concentrated on Tokyo and Kanagawa prefectures, Chiba prefecture was also considerably affected not only by the main shock but also by the aftershocks.\footnote{\setstretch{0.84}While Tokyo was the largest prefecture with approximately four million inhabitants in 1922, Kanagawa and Chiba were middle-sized prefectures with 1.36 and 1.34 million inhabitants, respectively; Statistics Bureau of the Cabinet, \textit{Jink\=od\=otait\=okei [1922 edition]}, p. 347. The shares of the agricultural, industrial, and commercial sectors in Chiba were 70\%, 10\%, and 10\%, respectively; Statistics Bureau of the Cabinet, \textit{Kokuseich\=osah\=okoku [1920 edition]}, pp.~26--27.}
Overall, 53\% of municipalities suffered physical disruption and roughly one in ten households in Chiba were damaged by the earthquake.
Figure~\ref{fig:spatial_hh} shows the spatial distribution of the percentage distribution of affected housing units.\footnote{\setstretch{0.84}The spatial distribution of victims is similar but shows more regionally smaller distribution patterns. Thus, I prefer to use the percentage distribution of affected housing units in my empirical analysis (Online Appendix Figure~\ref{fig:humanr}).}
The affected municipalities were concentrated on the western coast (\textit{uchi-b\=o}), especially in the counties of Awa and Kimitsu, because this area includes the seismic fault plane named the \textit{Kamogawa-teichi} fault zone.

In affected areas, roads and railways were largely destroyed and the production of newspapers, postal services, and telegraph and telephone services also completely stopped.
Topographical changes that made traveling difficult were observed in many municipalities.
Although railway services had largely restarted by the end of September 1923, passengers still had to walk between heavily damaged sections.
Further, roughly nine in 10 post offices including telegraph and telephone stations were damaged.\footnote{\setstretch{0.84} Division of Social Affairs, Chiba Prefecture, \textit{Taish\=odaishinsai [volume 1]}, p.~151; Chiba Prefecture, \textit{Chibakent\=okeisho [1923 edition, volume 5]}, p.~152.}
Wells, the main water source for people at that time, became contaminated by sand and salt, causing difficulties in obtaining drinking water.
Physical disruption was also observed in the agricultural and industrial sectors.
Because approximately 14,000 hectares of arable land were affected, including damage to reservoirs, embankments, and irrigation equipment as well as the upheaval or depression of ground surfaces, agricultural households struggled to sell their products because trade partners were mainly in Kanagawa and Tokyo.
The industrial sector was affected not only by the physical damage but also by the crisis in the financial system, as approximately 77\% of banks in Tokyo were burnt down by the fire.\footnote{\setstretch{0.84}Descriptions of the agricultural sectors are taken from the Division of Social Affairs, Chiba Prefecture, \textit{Taish\=odaishinsai [volume 1]}, pp.~126--130;~133;~136;~141--143; Chiba Prefecture, \textit{Chibakent\=okeisho [1923 edition, volume 1]}, pp.~11;~103. The number of collapsed banks was obtained from the Bank of Japan, \textit{Nihongink\=o}, p.~48.}

While the earthquake led to a certain physical disruption, evidence suggests that the material losses in Chiba improved within the first few weeks. The municipalities in the devastated area experienced a harsh food shortage for a few days after the first hit and before food began to be distributed through the relief effort.\footnote{Division of Social Affairs, Chiba Prefecture, \textit{Taish\=odaishinsai [volume 2]}, pp.33; 131.}
Anecdotal evidence suggests the temporary losses of jobs and parental deaths in the devastated area.\footnote{Division of Social Affairs, Chiba Prefecture, \textit{Taish\=odaishinsai [volume 2]}, pp.96--97; 104; 113; 126.}
However, the parental extrinsic death rate was less than 0.1\% in 1923, suggesting that parental deaths might have actually been limited in the devastated area (Subsection~\ref{sec:sec22}).\footnote{School enrollment and attendance rates did not decrease after the earthquake; Division of Social Affairs, Chiba Prefecture, \textit{Taish\=odaishinsai [volume 2]}, p.208.}
The official documents from each municipality in Awa county (the most affected county) indicate that most primary schoolchildren were not killed by the earthquake because it occurred after school and thus the children had been playing outside.\footnote{\setstretch{0.84}Division of Social Affairs, Chiba Prefecture, \textit{Taish\=odaishinsai [volume 2]}, pp.93; 109; 115; 151; 228. This is indeed a favorable condition for my empirical analysis because it means that there is no systematic selection in my primary school dataset.}
Moreover, while the primary schools were temporarily closed in the devastated area, many restarted classes within a few weeks of the impact.\footnote{Division of Social Affairs, Chiba Prefecture, \textit{Taish\=odaishinsai [volume 2]}, pp.137--138; 144;150; 202.}

These records suggest evidence that while the people in the devastated area experienced severe hunger for a few days, the earthquake might not have killed many children or their parents.
However, another important fact is that large earthquakes are often followed by a pattern of aftershocks that gradually decrease in intensity over the next few months to a year in the affected region.
The available records of the Choshi Meteorological Observatory in Chiba indicate that 779 aftershocks occurred within the first two weeks after the main shock, and 136 aftershocks within the subsequent four weeks.
This implies that the discernable aftershocks must have lasted for, at least, a couple of months after the first hit.\footnote{Division of Social Affairs, Chiba Prefecture, \textit{Taish\=odaishinsai [volume 1]}, pp.25; 32; 39.}
A public officer of Hokujy\=o town in Awa county stated that ``people felt extreme fear at night because of the aftershocks.''\footnote{Division of Social Affairs, Chiba Prefecture, \textit{Taish\=odaishinsai [volume 2]}, p.31.}
Importantly, even people who lived in areas that did not experience physical devastation were shocked by the resulting strong vibrations.\footnote{Division of Social Affairs, Chiba Prefecture, \textit{Taish\=odaishinsai [volume 2]}, pp.805; 1018.}
This implies that the 1923 earthquake affected people broadly through their mental pathway via the fear of vibrations and anticipation of future aftershocks.

\subsection{Possible Pathways} \label{sec:sec22}

The evidence described in the previous subsection suggests that the earthquake led to adverse shocks on child health through two main pathways.
The first is maternal mental stress in pregnant women.
Medical evidence indicates that strong antenatal stress can increase cortisol concentrations, which leads to abnormal birth outcomes.\footnote{Stein et al., `Effects', p.1812.}
In fact, massive earthquakes can cause prenatal maternal stress,\footnote{Hibino et al., `Health'.} which can increase the risk of adverse pregnancy outcomes.
For instance, Glynn et al. investigate 40 pregnant women who experienced an earthquake of a magnitude of 6.8 in California in 1994 during pregnancy or shortly after, finding that maternal stress experienced in early pregnancy is associated with a shorter gestational period.\footnote{Glynn et al., `When'.}
Torche investigates the influence of acute stress exposure to the large Chilean earthquake of 2005 on birth weight using birth registry data.
She shows that maternal stress results in a decline in birth weight and an increase in the proportion of low birth weight deliveries.\footnote{Torche, `The effect'.}
Kim et al. also provide evidence that psychological maternal stress from exposure to the Northridge earthquake of 1994 increased the likelihood of low birth weight.\footnote{\setstretch{0.84} Kim et al., `Maternal'. The low birth weight can be driven both by prematurity and by reduced fetal growth. To disentangle these factors, data on both birth weight and pregnancy duration, which are usually unavailable as historical statistics, are required. Considering this, the present study focuses on the overall impacts of fetal earthquake exposure on children's heights and weights, but does not delve into the mechanism behind the low birth weight.}
Such low birth weight due to reduced gestational age and intrauterine growth restriction can have adverse effects on the development of children.\footnote{\setstretch{0.84} Stein et al., `Effects'; Au Yeung et al. `Birth'.}

The second pathway is the nutritional deprivation of pregnant women and children.
The fetal origin hypothesis proposed by Barker argues that poor nutritional intake ``programs'' the fetus to have disturbed metabolic characteristics that can permanently affect exposed individuals.\footnote{Barker, \textit{Fetal}; Barker, \textit{Mothers}.}
A growing body of the literature based on this hypothesis has offered causal evidence on the long-run adverse health effects of fetal exposure to nutritional shocks.\footnote{See Almond and Currie, `Killing'; Currie and Vogl, `Early-life'.}
It has also been argued that the nutritional status of children can have instantaneous effects on their subsequent growth.\footnote{Schneider and Ogasawara, `Disease'.}
Therefore, a number of channels could drive the nutritional pathway.

The first channel is pecuniary shocks on children from parental deaths.\footnote{Banerjee et al., `Long-run'.}
The number of extrinsic deaths of people aged 20--54 in Chiba prefecture increased from 172 in 1922 to 572 in 1923, before dropping to 157 in 1924.
Although the increase in 1923 is clearly due to the hit of the earthquake, this total of fewer than 600 people accounts for just 0.1\% of adults aged 20--54.\footnote{There was no gender difference in the rates in 1923. The data on extrinsic deaths and number of people aged 20--54 are from the Statistics Bureau of the Cabinet, \textit{Nihonteikokushiint\=okei [1922--1923 editions]}, pp.168--171; Statistics Bureau of the Cabinet, \textit{Tais\=okyunen}, pp.62; 64.}
Given that eight in 10 men aged 20--54 were married at that time, the rate of parental extrinsic deaths should be roughly 0.08\%.\footnote{Statistics Bureau of the Cabinet, \textit{Tais\=okyunen}, pp.63; 65.}
This implies that while the injury and death of parents due to the earthquake might decrease the welfare of their children in the devastated area, this channel is implausible in non-devastated areas.
The second channel is the nutritional shocks from a shortage of food and/or declines in household income due to shocks on transportation and agricultural production and/or sales.\footnote{Division of Social Affairs, Chiba Prefecture, \textit{Taish\=odaishinsai [volume 2]}.}
As described in Subsection~\ref{sec:sec21}, people in the devastated area might have experienced hunger until the disaster relief started.
Regarding prices, the retail prices of food and other necessities still increased after the earthquake.\footnote{Division of Social Affairs, Chiba Prefecture, \textit{Taish\=odaishinsai [volume 1]}, pp.~293--296.}
However, the increased prices of food and daily commodities reverted to the pre-earthquake trend relatively quickly.\footnote{Hunter and Ogasawara, `Price’.}
This evidence suggests that the nutritional pathway might only matter in the physically devastated area.\footnote{Another related channel may be declining sanitary conditions. Infection can significantly reduce fetal nutrition via inflammation, high fever, lost appetite, vomiting, and complications. However, neither the infant mortality rate nor the crude death rate dramatically increased in 1923 (Online Appendix~\ref{sec:selection}).}

\section{Data} \label{sec:sec3}

\subsection*{\textit{Biological Outcomes}}

Child stunting is considered to be the best overall indicator of the well-being of children.\footnote{de Onis and Branca, `Childhood'.}
Therefore, I consider the biological outcomes of children (i.e., their height and weight) to be their key health indicators.
I regard height as the main measurement of the overall health outcomes of children, as this measure reflects accumulated nutritional status and is associated with cognitive ability and long-term adult health and socioeconomic outcomes.\footnote{\setstretch{0.84}
Fogel, `Economic'; Currie and Vogl, `Early-life'. I do not use the height-for-age z-score of modern WHO standards to control for age effects. The pubertal growth spurt of children in the early 20th century occurred at older ages than in modern healthy children. This causes a rotated shift of the entire growth curve and thus leads to a distorted height-for-age profile at all ages; Schneider and Ogasawara, `Disease'. For example, the mean heights of boys aged six, eight, and 10 years old in the WHO growth reference are 118.6, 129.7, and 140.2 centimeters, respectively (World Health Organization, \textit{2007}), whereas those in 1930 in my sample are 108.5, 118.4, and 127.5 centimeters, respectively. The differences between the two samples range from 10.1 to 12.7 centimeters, showing the distorted change due to the rotation.}
I consider weight to be a secondary measurement of children's health.
Since weight is more likely to be sensitive to instantaneous effects, I regard weight as a subsidiary measurement that can capture accumulated nutritional status.\footnote{The instantaneous nature of the weight implies that the time passed since the earthquake is sufficient for the recent food consumption to dominate the children's weights. In Online Appendix~\ref{sec:alt_spec_bmi}, I confirm that there are no systematic statistically significant effects of fetal earthquake exposure on BMI.}

First, I introduce the historical documents used to build the dataset.
I collected an available set of the annual statistical reports named \textit{Gakk\=o seitojid\=o shintaikensa t\=okei} (Statistics of Physical Examinations for Students; SPES) compiled by Chiba prefecture between 1925 and 1935.
Japan is indeed a good context for my study because of its comprehensive school physical examination records.
Since these physical examinations had to be conducted in April of each year for all schools under the \textit{Gakuseiseito shintaikensa kitei} (Official Regulations for School Physical Examination) from 1897, most schoolchildren undertook one.
Therefore, my dataset on the primary school students from all 434 schools covers approximately 95\% of the juvenile population aged 6--11 in Chiba at that time.\footnote{\setstretch{0.84} See Online~\ref{sec:secb} for a detailed explanation on the comprehensiveness of this dataset.}
This means that the target population in the analyses using the dataset can be regarded as almost the entire child population in the same age range in the prefecture.\footnote{I confirm that the potential selections based on both school enrollment and school attendance do not disturb my main results (Section~\ref{sec:sec6}).}

Second, I describe the structure of my dataset.
The SPES records the average heights and weights of children for each age (from six to 11) and primary school (more than 400 schools) measured between 1925 and 1935.
Thus, my dataset consists of school-age-specific units followed over time and contains the average heights and weights of children aged 6--11 born between 1914 and 1929.\footnote{Online Appendix Figure~\ref{fig:ah} illustrates average height and weight by cohort, gender, and area, with 95\% confidence intervals. Online Appendix Figure~\ref{fig:hw_trend} illustrates the mean heights and weights for each age and measured year.}
As I will explain, in the empirical analysis, I systematically stratified my samples into two subsamples to investigate the heterogeneous stunting effects by age group: 6--8 (early primary) and 9--11 (late primary).\footnote{\setstretch{0.84} This does not mean that children's ages are aggregated into two groups, but that the early (late) primary subsample includes children aged six, seven, and eight (nine, 10, and 11).}
Panel A of Table~\ref{tab:sum} reports the summary statistics on the biological outcomes by age group.

Third, I argue that potential sorting due to migration should not be a threat in my dataset.
To understand this argument, I compute the proportion of children who live in their place of birth to total children at the same age using the 1930 Population Census.\footnote{Statistics Bureau of the Cabinet, \textit{Sh\=owagonen}.}
According to the census, approximately 94\% and 90\% of children aged 0--9 and 10--14 were born in their current municipalities, respectively.\footnote{\setstretch{0.84} The age ranges reported in the census are systematically divided into 0--9 and 10--14 years. Since those figures for boys and girls are similar, I present the average figures herein.}
I regard the figures on children aged 0--9 as plausible for primary schoolchildren since the census figure for children aged 10--14 should be biased downward because of the existence of graduate children aged over 12 who could have jobs in other municipalities.\footnote{\setstretch{0.84}These students were also unlikely to change their primary schools within each municipality: Hijikata, \textit{Kindainihon}, pp. 159--165.}
This indicates that most schoolchildren did not leave their original places until finishing primary school, making potential sorting not an issue in my analyses.\footnote{\setstretch{0.84}
This is consistent with the historical fact that the internal migration of schoolchildren in the interwar period was limited; Nakagawa, `Kokuseish\=osa', p.~42.
I also test whether children's heights and weights were associated with migration status.
Although municipal-level information on migration is unavailable, I confirm that there were no systematic correlations between the prefecture-level data on the non-migrant rate (share of people living in their original municipalities) and prefecture-level average heights and weights of primary schoolchildren (Online Appendix Table~\ref{tab:migration}).}

\subsection*{\textit{Earthquake Intensity}}

I used data on the physical damage from the official report for the Great Kant\=o Earthquake named \textit{Taish\=o shinsaishi} (History of the Taish\=o Earthquake; HTE) published by the Social Welfare Bureau of the Cabinet in 1926.
Since the HTE surveyed all damaged households in Chiba by November 15, 1923, it provides a complete picture of the degree of physical disruption at the municipality level.\footnote{\setstretch{0.84}Division of Social Affairs, Chiba Prefecture, \textit{Taish\=odaishinsai}, pp. 412--420.}

I calculated the physical disruption rate (PDR) as number of affected housing units per 100 households for each municipality, as shown in Figure~\ref{fig:spatial_hh}.
However, one must be careful because the distribution of the PDR is highly skewed, as in the case of the Chernobyl radioactive fallout in Sweden.\footnote{Almond et al., `Chernobyl's'.}
Thus, if the impacts of the earthquake had a stronger magnitude in devastated areas than in marginally affected areas, it is functionally difficult to use the PDR as a continuous intensity variable in the regressions.\footnote{In fact, Torche, `The effect' provides evidence that stresses from vibrations can be positively but heterogeneously correlated with the intensity of earthquakes.}
As I will explain later, the specification using the PDR as a continuous variable indeed fails to capture the heterogeneous impacts of earthquake exposure on the outcome variables (Subsection~\ref{sec:sec42}).

To address this issue, I systematically divided municipalities into three categories according to the Japan Meteorological Agency seismic intensity scale (JMA-SIS): JMA-SIS of 5, 6, and 7.
To do so, I first predicted the collapse rate using the PDR based on the prediction equation and then converted the rates into the seven categories of the JMA-SIS.
Table~\ref{tab:class} summarizes the classification.
The mean PDRs reported in this table suggest that the physical disruption was concentrated in the JMA-SIS7 area with a mean rate of 80\%.
A large number of affected municipalities were in the Awa and Kimitsu counties.
While the municipalities in the SIS6 area were moderately affected (23\%), the municipalities in the JMA-SIS5 area were hardly physically damaged by the earthquake (0.4\%).
This distribution of damage suggests that the negative effects of the earthquake were most obvious in the JMA-SIS7 area.
Figure~\ref{fig:categories} illustrates these regions.

\section{Empirical Analysis} \label{sec:sec4}

\subsection{Identification Strategy}\label{sec:sec41}

I use the quasi-experimental estimation strategy that matches the exogenous shocks due to fetal earthquake exposure with the corresponding birth cohorts.
If the physical disruption enhanced the earthquake stress placed on the fetus, a higher degree of disruption may be associated with stronger stunting effects.
I thus include the product terms between the 1923 birth cohort indicator variable and indicator variable for the JMA-SIS7 area to allow the effects of earthquake stress to vary across areas.\footnote{\setstretch{0.84}I confirm that the effects of fetal earthquake exposure in the JMA-SIS6 area are identical to those in the JMA-SIS5 area in the statistical sense. Online Appendix~\ref{sec:alt_spec_area} presents and discusses these results.}
My baseline specification is given as follows:
\begin{eqnarray}\label{eq1}
\footnotesize{
\begin{split}
y_{sta} = \alpha + \beta_{0} \text{\textit{I}(\textit{YOB}=1923)}_{ta} + \beta_{1} \text{\textit{I}(\textit{YOB}=1923)}_{ta} \times \text{\textit{SIS7}}_{g_{s}} + \vx'_{g^{c}_{s}ta} \vgamma + \mu_{sa} + \lambda_{t} + e_{sta}
\end{split}
}
\end{eqnarray}
where $s$ indexes schools, $t$ indexes survey years, and $a$ indexes ages.\footnote{\setstretch{0.84}The group variables $g_{s}$ and $g^{c}_{s}$ indicate municipalities and counties, respectively.}
The variable $y$ is either height or weight, $\textit{I}(\textit{YOB}=1923)$ is an indicator variable that equals one for children born in 1923, $\text{\textit{SIS7}}$ is an indicator variable for the JMA-SIS7 area, $\vx$ is a vector of the county birth-year-level control variables, $\mu$ is a school age-specific fixed effect, $\lambda$ is a year fixed effect, and $e$ is a random error term.

Since the earthquake hit on September 1, 1923, the physical and human loss mostly occurred in that month.
This timing suggests that children born between September 1923 and July 1924 experienced the earthquake \textit{in utero}.
This means that the 1923 birth cohort includes those children impacted by the earthquake \textit{in utero} because children born in 1923 in my sample were individuals born between April 1923 and March 1924 as the academic year starts in April and ends in March in Japan.\footnote{\setstretch{0.84}The average pregnancy term was nine to 10 months in prewar Japan; Tokyo City Office, \textit{Ny\=ujishib\=och\=osa}. Hereafter, I refer to the 1923 birth cohort instead of the 1923 academic year birth cohort for simplicity. Since the 1923 cohort was nearly all \textit{in utero} during the earthquake, the estimated effects are less likely to be attenuated by the birth cohort cells containing children who were not truly exposed. Conservatively speaking, however, the estimates must be attenuated because the school-year-age-level data are matched with the intensity variable measured at the municipality level.}
Considering the feature of exposure, I expect the estimated coefficients on the affected cohort indicator $\textit{I}(\cdot)$ and the area interaction term $\text{\textit{I}}(\textit{YOB}=1923) \times \text{\textit{SIS7}}$ to be negative.\footnote{\setstretch{0.84} Drixler, `Hidden' highlights the imprecision of birth data in prewar Japan. In light of this study, one must be careful about the potential age heaping in April. For example, children born in March 1923 might have been disadvantaged in their development, perhaps during their early primary school ages, compared with those born in April 1922. This implies that parents might have had an incentive to register their children born in March 1923 as born in April 1923. However, I confirm that there was no such systematic age heaping in April using vital statistics; Statistics Bureau of the Cabinet, \textit{Jink\=od\=otait\=okei [1923 edition]}.}

The timing and spatial distribution of the physical disruption must be exogenous to improve the identification.
The timing of the earthquake was obviously unexpected.
The distribution of the disruption used was also dominated by the distribution of the fault plane found \textit{after} the earthquake, suggesting that the distribution is plausibly exogenous.\footnote{\setstretch{0.84}I also confirm that the distribution of the physical disruption did not depend on that of soil compaction (which can be associated with potential agricultural productivity), measured as the spectral intensity values observed in 2011 (Online~\ref{sec:secb}).}
I also allow the unobserved factor of school to vary over age by introducing a fixed effect ($\mu$) in each school-age cell in all the specifications.\footnote{\setstretch{0.84} Thus, the identification uses the within variation over the measured year. This means that the increasing trends in the growth of child height should be similar over school-age cells. I confirm that the growth patterns of the sampled children over the measured years are similar (Online Appendix~\ref{sec:secc_trend}).}
The advantage of this approach is that the school-specific unobservable preference on improving children's health can be controlled for.\footnote{In addition, my approach can control for the unobservable shocks in measured years such as the 1930 depression and the trends in potential wealth and public health using the year fixed effect ($\lambda$).}
I further use the county birth-year-level variables at baseline, namely the fetal death rate, rice yield, coverage of doctors, and school enrollment rate of the parental generation.
These variables are included to control for the potential mortality selection effects, wealth levels related to agricultural productivity, accessibility to medical care around birth, and changes in parental characteristics across cohorts, respectively.\footnote{Bozzoli et al., `Adult'; Brown and Thomas, `On the long term'.}
See Online~\ref{sec:secb} for the summary statistics and data sources of these control variables.

To assess the potential spatial and school cohort-specific correlations, I intend to cluster the standard errors at the 13-county level.
Thus, my method controls for any autocorrelation and heteroskedasticity within clusters and can deal with heteroskedasticity across clusters.
Since a school is nested in a county, this means that both the dependency across observations of the same birth cohort within a school and the dependency across schools within the same county are allowed in my estimation.
Although a potential threat in the inference is the spatial autocorrelation among clusters, I confirm that there are no systematic spatial dependencies over the counties (Online Appendix~\ref{sec:alt_spatial}).
For the statistical inference, I adopt the wild cluster bootstrap-t method to deal with the issue of the small number of clusters in the cluster-robust variance estimator.\footnote{\setstretch{0.84} Cameron et al., `Bootstrap'. I prefer to cluster at the county level than the municipal level because regressors grouped at the county level are used and the correlation within the same county over time such as the school cohort-specific correlation may be problematic; Bertrand et al. `How'.}
The regression equation is estimated separately for boys and girls for each developmental stage to investigate both gender differences in the effects and the potential catch-up growth against the shocks.\footnote{\setstretch{0.84} Steckel and Ziebarth, `Trader'. Unfortunately, I am unable to use the data on the number of inspected children for each age and school. However, since the number of schools should have been set to reflect the size of municipalities, the number of inspected children in each age cell should be largely similar across schools; Ministry of Education, \textit{Gakuseihyakunenshi}. Thus, my analytical results should be robust to the weighting.}
The age bins are systematically divided into 6--8 (early primary) and 9--11 (late primary) because the growth patterns of children during primary school ages were almost linear.\footnote{\setstretch{0.84}See Online Appendix~\ref{sec:secc_trend}. I also confirm that my main results are robust to the use of alternative definitions of the developmental-stage age bin (Online Appendix~\ref{sec:alt_spec_def}).}

\subsection{Main Results}\label{sec:sec42}

Table~\ref{tab:r_main} presents the results from my preferred specification denoted in equation~(\ref{eq1}).
Panel A-1 lists the estimates of $\vbeta$ for height.
Columns (1) and (2) present the results for the boys, whereas columns (3) and (4) present the results for the girls.
In column (1), primary school boys aged 6--8 born in 1923 are found to be approximately $0.15$ cm shorter than the surrounding cohorts.
Column (2) shows a similar but slightly smaller stunting effect on the boys aged 9--11.
The estimated coefficients on the area interaction terms are statistically insignificant in columns (1) and (2).
When I look at the results for the girls, those stunting effects are greater in their magnitude.
Column (3) indicates that the girls aged 6--8 born in 1923 are $0.28$ cm and $0.59$ cm shorter in the non-JMA-SIS7 and JMA-SIS7 areas, respectively.
Similarly, column (4) indicates that the girls aged 9--11 born in 1923 are $0.20$ cm and $0.85$ cm shorter in each area.
These estimates suggest that both boys and girls in the non-JMA-SIS7 area experienced some catch-up growth.
This implies adaptive responses in relation to the growth of children exposed to the earthquake \textit{in utero} who also experienced better postnatal conditions than those in the prenatal period because they grew up in areas in which the physical damage was relatively limited.\footnote{See Schneider, `Children's' for the mechanism of the adaptive responses. Out-migration by the 9--11 year olds should not ameliorate the earlier effects at age 6--8 because the migration rates were unlikely to be changed throughout the primary school ages (Section~\ref{sec:sec3}; footnote 52).}

Panel A-2 of Table~\ref{tab:r_main} shows the estimates of the effects on weight, which is my secondary measurement of children's health, following the same column layout.
Although the estimate in column (1) suggests that the primary school boys born in 1923 are approximately $0.06$ kg lighter than those in the surrounding cohorts, that listed in column (2) is negative but no longer statistically significant.
This result is consistent with the weaker stunting effects on the boys reported in Panel A-1 and with the fact that weight is more likely to be a rough measurement because it is sensitive to instantaneous effects such as current nutritional status (Subsection~\ref{sec:sec3}).
Similar to the results on the girls' height, the affected birth cohorts among the primary school girls are slightly lighter than those in the surrounding cohorts.
Columns (3) and (4) indicate that the girls born in 1923 are $0.1$ kg and $0.36-0.42$ kg lighter in the non-JMA-SIS7 and JMA-SIS7 areas, respectively.

Overall, these results are consistent with the recent literature on early-life shocks on human growth.\footnote{Rosales-Rueda and Triyana, `The persistent'.}
The Great Kant\=o Earthquake of 1923 had negative lasting effects on girls' growth in the damaged area with a maximum JMA-SIS of seven.
However, while the boys in the exposed cohort were slightly stunted, I find no systematic heterogeneous stunting effects in terms of seismic intensity.
I return to this gender difference in stunting effects in greater detail in Subsection~\ref{sec:sec52}.

Before I move onto the next analyses, I test the validity of the baseline specification denoted in equation~(\ref{eq1}) by comparing the results from the specification using a continuous intensity variable (i.e., the PDR) instead of the seismic indicator variable.
Panel B of Table~\ref{tab:r_main} presents the results.
The estimated coefficients on the 1923 birth cohort dummy are statistically significantly negative, which are consistent with the estimates in Panel A of Table~\ref{tab:r_main}.
However, the estimates of the interaction terms between the 1923 birth cohort dummy and PDR are statistically insignificant in most cases in contrast to those in Panel A of Table~\ref{tab:r_main}.
This means that the specification using the PDR fails to capture the heterogeneous effects of earthquake exposure because of its linear functional form assumption (Subsection~\ref{sec:sec3}), supporting the validity of my preferred specification.
Therefore, I use the logically defined seismic intensity indicator as my main intensity variable in the following analyses.

\subsection{Postnatal Exposure and Instantaneous Effects} \label{sec:sec43}

The regressions discussed thus far have aimed to capture fetal exposure to the earthquake.
However, evidence suggests that postnatal exposure to the Great Earthquake could have also had adverse effects on human capital accumulation.\footnote{Caruso and Miller, `Long run'.}
It has also been shown that exposure to adverse conditions throughout childhood can disturb children's growth.\footnote{Howe and Schiller, `Growth'; Tanner, `With'; Tanner, `Fetus'.}

To test both possibilities, I consider two alternative specifications.
First, I add two types of exposure variables into the baseline specification of equation~(\ref{eq1}).
One is an indicator variable that takes one for children exposed to the earthquake at ages 1--5 and the other is an indicator variable that takes one for children exposed to the earthquake at ages 6--9.
The former variable captures the impacts of early-life postnatal exposure to the earthquake, whereas the latter variable captures the instantaneous effects of the earthquake.
To analyse the potential heterogeneous effects, the area interaction terms for both indicator variables are considered.
Columns (1) and (3) of Table~\ref{tab:r_robustness3} present the results (Panel A for heights and Panel B for weights, respectively).
I limit the sample to children aged 6--9 because no children experienced the earthquake at ages 10 and 11 in my full sample.
As shown, the estimated coefficients on these exposure variables are statistically insignificant in both columns.

Second, to further delve into the potential instantaneous impacts of the earthquake, I limit the sample to children aged 8 and 9 and use an alternative exposure indicator variable that takes one for children exposed to the earthquake at ages 6 and 7.
In this specification, I intend to capture the short-run direct effects of the earthquake on children's health.
Similar to the previous exercise, I consider the area interaction term as well.
Columns (2) and (4) of Table~\ref{tab:r_robustness3} show the results.
The estimates are negative in some cases but not statistically significant in any of the regressions.

These results suggest that neither early-life postnatal exposure nor instantaneous exposure to the earthquake has stunting effects---at least in the statistical sense.
This implies that affected children were less likely to suffer from malnutrition due to the earthquake.
As described in Subsection~\ref{sec:sec22}, although the well-being of children must be low if their parents had been killed, the parental extrinsic death rate in 1923 was estimated to be less than 0.1\%.\footnote{There were no systematic child protection rules for orphans at that time. The child welfare policy had begun to be organized in the late 1920s when the discussion on the establishment of Poor Relief Act of 1929 was initiated; Yoshida 2004, pp.~234--235.}
While children in the devastated area might have suffered severe hunger for a few days after the earthquake first hit, agricultural households could rely on their conservable food to compensate for the temporary market disruption and market functions recovered relatively quickly after the earthquake (Subsection~\ref{sec:sec21}).
In Subsection~\ref{sec:sec511}, I will also provide evidence on the mitigating impacts of the food relief effort.
This evidence suggests that the direct effects of the earthquake on children's nutritional status were not problematic.

Finally, although one can speculate that exposure to the earthquake during an early-life period such as infancy might be problematic, I find little evidence of adverse health effects on the 1922 birth cohort, implying that postnatal exposure within two years of birth does not matter. I further confirm that the observed stunting effects of fetal earthquake exposure are not a systematic decreasing pre-trend in the height or weight of children (Online Appendix~\ref{sec:alt_spec_placebo}).

\section{Mechanisms} \label{sec:sec5}

\subsection{Mental and Nutritional Stress} \label{sec:sec51}

I extend my discussion to the pathways (i.e., maternal mental and nutritional stress) through which fetal earthquake exposure may affect children's health.
First, I test whether disaster relief spending mitigated the adverse effects of the earthquake on the development of children.
Second, I use the market disruption due to damage to the railway transportation network to identify the potential stress behind these adverse health effects.
I do not intend to estimate any causal effects in this subsection. Rather, I try to present suggestive evidence on the pathways using data on relief expenses and damage to railway transportation networks.

\subsubsection{Relief Effects} \label{sec:sec511}

Soon after the earthquake, Chiba prefecture organized a headquarters for emergency disaster control and decided to provide disaster relief to affected counties, namely, Awa, Kimitsu, Ichihara, Higashikatsushika, Ch\=osei, and Isumi; approximately 12\% of the population in these counties received relief.
Total relief expenditure was $490,837$ yen for the impacted year and most was used within the few first months.\footnote{\setstretch{0.84} Division of Social Affairs, Chiba Prefecture, \textit{Taish\=odaishinsai [volume 1]}, p. 420. Given that the average expenditure for food in peasant households was roughly 8 yen per family member (footnote 79), this figure is considered to be sufficiently large. Total relief expenditure in 1924 dropped to $2,790$ yen. This means that relief was considered to be a temporal and one-shot investment for affected counties in 1923.}
Of relief expenditure, 66\% was used to provide food, 31\% was for temporary housing, and 0.3\% was for medical treatment.\footnote{\setstretch{0.84}Although the remaining 2.7\% was used for the other miscellaneous goods, I do not use this category for the regression analysis because the purpose of the relief is unclear and thus it is difficult to interpret the results.}
Medical and food relief started within a few days of the first hit and provided first-aid treatment, set up soup kitchens, and distributed rice in devastated areas.
Both types of relief were therefore stopped relatively quickly, that is, by the end of October 1923.\footnote{Social Welfare Bureau of the Cabinet, \textit{Taish\=oshinsaishi}, pp. 1035--1047; Division of Social Affairs, Chiba Prefecture, \textit{Taish\=odaishinsai [volume 1]}, pp.434--438.}
Housing relief was used to build temporary barracks for people who had lost their homes.
Given the scarcity of construction materials in the aftermath, a large part of these expenses started to be incurred from late October 1923. Such temporary housing protected people in the subsequent winter.\footnote{Division of Social Affairs, Chiba Prefecture, \textit{Taish\=odaishinsai [volume 1]}, pp.307--308.}
To investigate the potential compensating effects of the disaster relief, I use statistics on the county-level relief expenses by expense item in the SRCP.\footnote{Chiba Prefecture, \textit{Chibaken [1923 edition, volume 1]}.}
Panel B of Table~\ref{tab:sum} shows the summary statistics of the relief expense variables.

To assess the relief effects, I introduce the interaction term between the measures of relief expenses and indicators of affected cohorts into equation~(\ref{eq1}) as follows:
\begin{eqnarray}\label{eq2} 
\footnotesize{
\begin{split}
y_{sta} = \kappa + \delta_{0} \text{\textit{I}(\textit{YOB}=1923)}_{ta} + \delta_{1} \text{\textit{I}(\textit{YOB}=1923)}_{ta} \times \text{\textit{SIS7}}_{g_{s}} + \delta_{2} \text{\textit{I}(\textit{YOB}=1923)}_{ta} \times \text{\textit{Relief}}_{g^{c}_{s}}\\
 + \vx'_{g^{c}_{s}ta} \vtheta + \eta_{sa} + \nu_{t} + \eps_{sta}
 \end{split}
}
\end{eqnarray}
where $\text{\textit{Relief}}$ is the county-level per capita relief expenses (total, food, shelter, or medical treatment).
While medical treatment is obviously necessary for injured people, immediate hunger relief intervention is the principal emergency response to sustain people in a severe situation after a crisis.
Temporary shelter is also an important emergency facility for personal safety, climate protection, security, and resistance to infectious diseases after large-scale disasters.
Therefore, one can expect the estimate of the parameter of interest ($\hat{\delta}_{2}$) to be positive when the disaster relief had the capacity to compensate for the negative effects of the earthquake.
The regression equation is estimated separately for each gender and age bin to investigate how relief spending ameliorated the negative health impacts of the earthquake in detail.

Table~\ref{tab:r_relief} presents the results.\footnote{\setstretch{0.84}The estimated coefficients on the cohort dummies and area interaction terms ({\scriptsize $\hat{\delta}_{0}; \hat{\delta}_{1}$}) in equation (\ref{eq2}) are not reported in the same table because these estimates are largely unchanged from those in Table~\ref{tab:r_main}.}
Panel A (Panel B) lists the estimates of the relief effects on height (weight).
Columns (1)--(4) present the results for total, food, shelter, and medical treatment expenses, respectively.
Overall, I find that the disaster relief had compensating effects on child stunting to a certain degree.
Column (1) of Panel A shows that while the relief ameliorates the stunting effects on the primary school boys aged 6--8 born in 1923, it does not have such effects on the girls' stunting.
Furthermore, column (1) of Panel B indicates that the relief also ameliorates the weight loss of the primary school boys aged 6--8, while such effects are not observed for the girls.

To investigate what type of disaster relief can ameliorate the adverse effects of the earthquake on children, I decompose relief expenses into three subcategories in columns (2)--(4), suggesting that food, shelter, and medical treatment had positive effects on both height and weight.
To compare the impacts, I next calculate the magnitudes of these relief effects estimated on the height of the primary school boys.
The estimate in column (1) of Panel A suggests that a one standard deviation increase in disaster relief (Table~\ref{tab:sum}) might have increased the affected primary school boys' height by approximately $0.07$ ($0.722 \times 0.095$) cm at ages 6--8 on average.
Given that the estimated stunting effect on the early primary school boys is $0.146$ cm as discussed, this magnitude implies that disaster relief might have ameliorated roughly half of the stunting effects.
Similar calculations applied to food (column 2), shelter (column 3), and medical treatment (column 4) indicate that the potential compensating effects are approximately $0.07$cm for food, $0.07$cm for shelter, and $0.06$ cm for medical treatment, respectively.\footnote{These magnitudes are considered to be reasonable. Average relief expenditure on food for treated people in Chiba prefecture was $3.8$ yen, whereas average expenditure for food in peasant households in September 1926 was $7.7$ yen per family member; Statistical Bureau of the Cabinet, \textit{Kakeich\=osah\=okoku}, p.~420. Although this is the most conservative calculation, it suggests that the relief might have provided nearly half of the monthly food needed.}
While the magnitude of medical treatment was slightly smaller than that of the other types of relief, this seems to be reasonable given that relief interventions for food and shelter were provided more broadly, whereas medical treatment was only provided to injured people in devastated areas.\footnote{See Panel B of Table~\ref{tab:sum} for the summary statistics. One might think, however, that the magnitude of medical treatment had an even larger effect than expected.
This is because medical treatment, despite its small proportion of total expenses, was used for first aid, which could have rescued parents, thereby reducing the number of orphans in devastated areas (Subsection~\ref{sec:sec22}).}

Since I use county-level relief expenses as the relief variable, the estimated effects may underestimate the actual effects of the relief.
Despite this, the present study provides suggestive evidence that remediation effects through disaster relief can mitigate stunting effects.
I also find that the relief effects might be biased by gender, as discussed in detail in Subsection~\ref{sec:sec52}.

\subsubsection{Mental Stress} \label{sec:sec512}

The disaster relief results suggest that the ameliorating effects of relief might stem from reducing the risk of maternal nutritional deprivation in the physically devastated area after the crisis.
In this subsection, I test whether the effects of mental and nutritional stress are more remarkable in the area that experienced little physical disruption and received no disaster relief.

Mental stress is caused not only by the experience of strong vibrations and a stressful life in the aftermath of the earthquake but also the physical disruption, whereas nutritional stress is concentrated in physically damaged and market disrupted areas.\footnote{Harada et al. `Mental'.}
Considering this feature, I begin my analysis by trimming the sample as follows.
First, I exclude all municipalities in the JMA-SIS6 and JMA-SIS7 areas and leave the JMA-SIS5 area alone because it is difficult to disentangle mental and nutritional stress in those areas that experienced physical disruption.
Second, I drop municipalities that received any disaster relief to eliminate its compensating effects.
The sample reduces to 222 municipalities that include exposed children who experienced strong vibrations \textit{in utero} in the limited area with a JMA-SIS of five but received no disaster relief.\footnote{\setstretch{0.84}The JMA-SIS ranges from one to seven, meaning that a JMA-SIS of five is still a very strong vibration. At a scale of five, people are frightened and feel the need to hold onto something stable. See the Japan Meteorological Agency, \textit{Tables}.}

For the identification, I sort the locations which suffered market disruption due to damage to railway transportation networks.
Regarding railway transport in Chiba prefecture, there were considerable losses of freight arrivals on the Noda line in Higashikatsushika county and Kururi line in Kimitsu county.\footnote{\setstretch{0.84}Both lines served several populated provincial towns and villages such as Noda town, Funabashi town, Chiyoda village, Kisarazu town, Makuta village, Obitsu village, and Kururi town; Chiba Prefecture, \textit{Chibaken [1923 edition, volume 1]}, pp. 52--53. As described in Subsection~\ref{sec:sec21}, the other lines recovered relatively soon after the earthquake in Chiba prefecture; Ministry of Railways, \textit{Kokuy\=utetsud\=oshinsaishi}.}
In fact, the annual tonnages of freight arrivals on both lines dropped by more than half of those in previous years.\footnote{Chiba Prefecture, \textit{Chibaken [1923 edition, volume 1]}, pp.~125--126.}
Since the main freight items were food, wood, and fuels, this suggests that the regional markets near both lines were more likely to have been disrupted than other areas and thus the alimentary deficiency might have occurred after the earthquake hit.\footnote{Chiba Prefecture, \textit{Chibaken [1923 edition, volume 5]}, pp.~138--145.}
While railway transport suffered physical damage, however, marine transport was not impacted by the earthquake. Indeed, Ch\=oshi port, a famous export port for the fish industry located at the western end of the prefecture, restarted to export goods to Tokyo soon after the earthquake.\footnote{\setstretch{0.84} Chiba Prefecture, \textit{Chibaken [1923 edition, volume 1]}, pp.~150--151.}

Exploiting these facts, I test the pathways by estimating the following regression:
 \begin{eqnarray}\label{eq3}
\footnotesize{
\begin{split}
y_{sta} = \varphi + \zeta_{0} \text{\textit{I}(\textit{YOB}=1923)}_{ta} + \zeta_{1} \text{\textit{I}(\textit{YOB}=1923)}_{ta} \times \text{\textit{Rail}}_{g_{s}} + \vx'_{g^{c}_{s}ta}\vphi  + \iota_{sa} + \tau_{t} + \varepsilon_{sta}
\end{split}
}
\end{eqnarray}
where \textit{Rail} is an indicator variable for municipalities within 10 km of any municipality including either the Noda line or the Kururi line.\footnote{\setstretch{0.84}I set the radius to cover neighboring municipalities: 10 km is roughly the median of the 5th percentile of the distances from the origins (i.e., municipalities including railways) to the other municipalities. Most municipalities within the radius are far from other railway networks, making it difficult for them to access other regional markets. Given the mean distance to the nearest municipality is approximately 3 km, I check the sensitivity of my results by changing the diameter to 3 km range and confirm that my results remain similar (Online Appendix~\ref{sec:alt_spec_def}).}
The estimate ($\hat{\zeta}_{0}$) is expected to be statistically significantly negative when maternal mental stress to vibrations was associated with adverse health effects.
In addition, the estimate of the coefficient on the area interaction term ($\hat{\zeta}_{1}$) is statistically significantly negative when maternal nutritional stress due to disrupted transportation networks had adverse effects on children's health.
The combination of the estimates provides evidence of which pathway might have played an important role in disturbing children's growth in the limited JMA-SIS5 area.
The regression equation is estimated separately for each gender and age bin.

Panel A of Table~\ref{tab:r_rail} presents the results for height.
Columns (1)--(2) list the estimates for the boys, whereas columns (3)--(4) list the estimates for the girls.
While the estimated coefficients on the 1923 birth cohort dummies are statistically significantly negative in most cases, those on the interaction terms are not statistically significant.
In the same columns in Panel B, I find similar effects for the girls' weight.
While these results may provide evidence of the role of mental stress, my estimates offer less clear arguments on the stunting effects of nutritional stress.

One might worry about the proxy measure of market disruption used herein.
A market survey reported that the retail prices of daily food such as rice, wheat, miso, and soy source increased after the earthquake in the regional markets of Chiba, which might have temporarily reduced the nutritional intake of mothers.\footnote{\setstretch{0.84}Systematic statistics on retail prices around the time of the earthquake are unavailable; Hunter and Ogasawara, `Price'. Although my data on retail prices are also insufficient for use in the regression analysis, the retail prices in some regional markets on September 20 were reported; Division of Social Affairs, Chiba Prefecture, \textit{Taish\=odaishinsai}, pp.~293--296. For instance, in the Matsudo market in Higashikatsushika, the maximum price of rice involuntarily increased from 39 to 45 sen per sh\=o (1.8 liter) after the earthquake hit. Another example shows that in the Kururi market in Kimitsu county, the maximum price of rice increased from 36 to 41 sen per sh\=o and the maximum price of miso rose from 80 to 85 sen per kan (3.75 kg).}
However, despite such market responses, rural households might have benefitted from rice and conservable vegetables for self-consumption to compensate for the temporary market disruption.\footnote{\setstretch{0.84}Moreover, in the initial stage after the earthquake, there might have been a greater supply of food in rural areas because of the damaged transport links in Tokyo and Kanagawa prefectures (Subsection~\ref{sec:sec21}).}
If this kind of unobservable self-protection behavior by households played a role, my estimates would understate the adverse effects of the disruption.
While I acknowledge these potential issues, the result obtained herein is consistent with the growing body of the medical and economic literature that shows the maternal mental stress caused by the earthquake is associated with the adverse health outcomes at birth (Subsection~\ref{sec:sec22}).

\subsection{Gender bias in the stunting effects}\label{sec:sec52}

The main results suggest that fetal earthquake exposure had stronger stunting effects on the girls than the boys.
I discuss the potential factors behind this gender bias from the timing dimension: prenatal selection and postnatal remediation.

\begin{figure}[]
\centering
\subfloat[Aug. vs nine-month average from Sept.]{\label{fig:scatter_mbs}\includegraphics[width=0.45\textwidth]{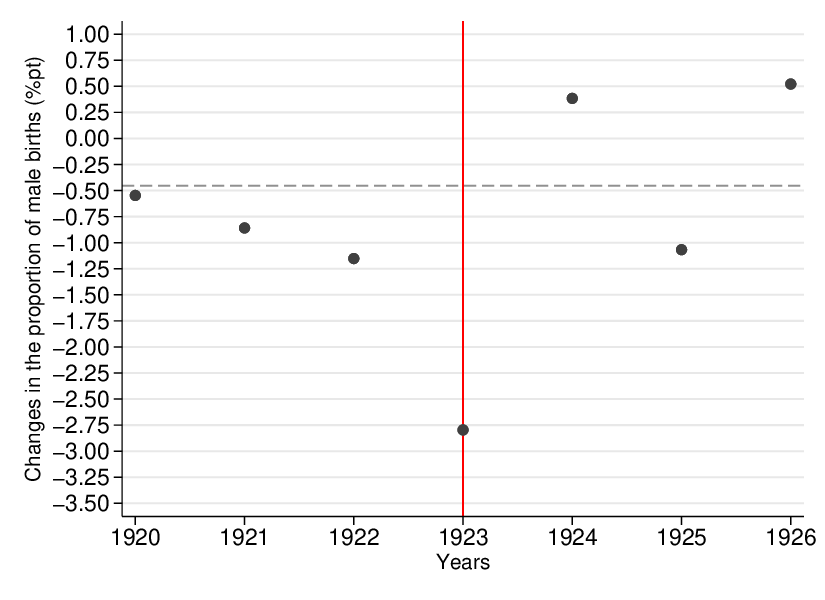}}
\subfloat[Tests based on regressions]{\label{fig:scatter_mbs_test}\includegraphics[width=0.45\textwidth]{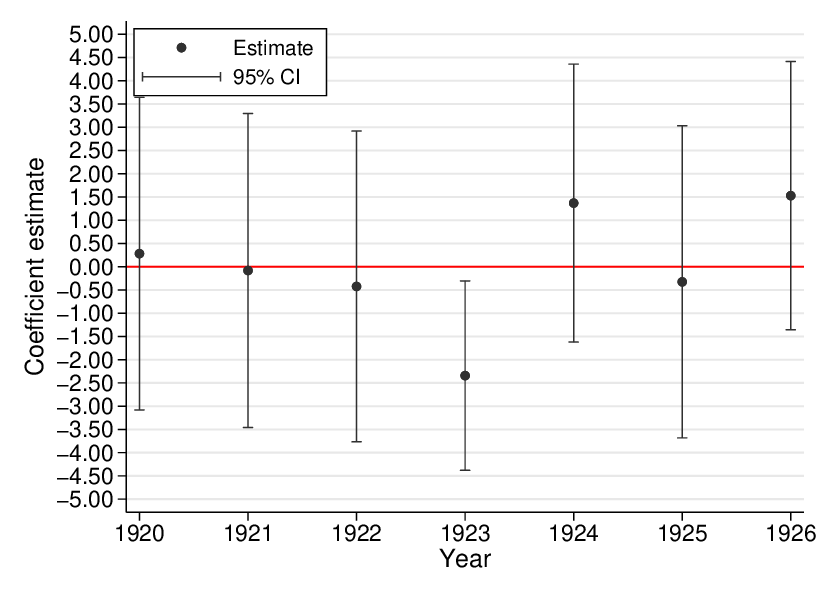}}
\captionsetup{justification=centering}
\caption{Changes in the proportion of boys at birth: \\Pre- and post-earthquake comparisons}
\label{fig:spatial}
\scriptsize{\begin{minipage}{400pt}
\setstretch{0.85}Notes: 
Each circle in Figure~\ref{fig:scatter_mbs} indicates the changes in the proportion of boys at birth (i.e., male births as a percentage of all births) between August and nine-month average of the rates from September to May.
For example, the circle on 1923 in Figure~\ref{fig:scatter_mbs} indicates the difference between the rate in August 1923 and the nine-month average rate from September 1923 (i.e., the earthquake month) to May 1924 (nine months after the earthquake).
Each circle in Figure~\ref{fig:scatter_mbs_test} indicates the estimate from a regression on each year dummy using the seven observations in Figure~\ref{fig:scatter_mbs}.
For example, the circle on 1923 in Figure~\ref{fig:scatter_mbs} indicates the estimated coefficient on the 1923 dummy.
Seasonality is controlled for in both figures by comparing the same months in each year.\\
Sources: Calculated by the author from the Statistics Bureau of the Cabinet, \textit{Nihonteikoku}.
\end{minipage}}
\end{figure}

One plausible explanation of prenatal selection is the growing body of the literature supporting evidence that male fetuses are more vulnerable to ambient stress \textit{in utero} than female fetuses and thus positive selection into birth is larger for boys than girls.\footnote{\setstretch{0.84}
Kraemer, `The fragile'. The selection mechanism indicates a shift of the survival threshold to the right. Given a certain distribution of fetal health endowments, this rightward shift tends to kill unhealthy fetuses before birth. Since male fetuses are more likely to be distributed on the lower side of the distribution than female fetuses, the number of male fetal deaths becomes greater than that of girls, leading to the reduced share of boys at birth.}
Under this mechanism, the proportion of boys at birth could have reduced after the earthquake hit.
To test this possibility, I digitize the monthly-level time-series information on live births by gender using the official vital statistics.\footnote{Unfortunately, systematic municipal- and/or county-level panel data on live births by gender are not available in Chiba.}
Figure~\ref{fig:scatter_mbs} illustrates the differences in the proportion of male births in August and the nine-month average rates from September in each year between 1920 and 1926.\footnote{\setstretch{0.84} The post-earthquake figure uses the nine-month average because the average pregnancy period was about nine to 10 months at that time. Seasonality in the changes in the proportion of male births is controlled for because the same range of months is used for all the sampled years. In fact, the time-series plots in Figure~\ref{fig:scatter_mbs} are mostly unchanged if I use the time-series plots of residuals after removing seasonality using month dummies.}
Figure~\ref{fig:scatter_mbs_test} provides a set of statistical tests by regressing the values plotted in Figure~\ref{fig:scatter_mbs} on each year dummy.
Clearly, both figures provide suggestive evidence that the earthquake of September 1923 reduced male births relative to female births.

My result on disaster relief suggested that while such relief might have played a role in nourishing the primary school boys to a certain extent, it had little compensatory effect on the primary school girls (Subsection~\ref{sec:sec511}).
The relief effects should be equal for both female and male fetuses because parents cannot identify the gender of their children before birth.
Therefore, one plausible explanation of the biased impacts of the relief is the biased postnatal remediation under the strong son preference in the family system (\textit{ie seido}) in prewar Japan.
The prewar Civil Code regarded the first-born son as the next head of the family (\textit{koshu}) who had the right to inherit property and dictatorial power to allocate household resources.\footnote{\setstretch{0.84}Ramseyer, \textit{Odd}. In fact, Article 970 of the Civil Code prioritized the of wedlock boy than the legitimate girl (within the same degree of relationship) in inheritance if the boy is recognized by his father; Cabinet Official Gazette Bureau, \textit{Houreizensho}, pp. 148--149. While Edict No.162 of the Grand Council of the State (\textit{dajy\=okan}) allows wives to access courts to see a divorce, women were not able to be awarded custody and were not always supported by their ex-husbands; Fuess, `Meijiki', p.179.}
Consequently, a gender bias against girls in intrahousehold resource allocations was widely observed in prewar Japan.\footnote{\setstretch{0.84}Hijikata, \textit{Kindainihon}. The most extreme case of biased resource allocation is that daughters were often sold by their parents (\textit{miuri}) in depression periods given the higher return of boys in the labor market; Nakamura, \textit{Sh\=owaky\=ok\=o}, pp.~115--116.}
From this perspective, one must be careful about the fact that the 1923 birth cohorts include children born after the earthquake hit but before or during the distribution of the relief.
Given that the institutions and customs in family systems favored boys, postnatal biased remediation via the relief might explain the gender-biased stunting effects against girls.

\section{Robustness}\label{sec:sec6}

Thus far, I have documented the adverse effects of fetal earthquake exposure on children's health. Before discussing the findings of the present study, I conduct a set of exercises to test the sensitivity of the main results in terms of the following concerns: sorting issues owing to internal migration; selection issues because of absenteeism, mortality, and fertility; and the potential impacts of other historical events.
I provide evidence that these concerns do not confound my findings.

First, the potential selection issue might have been arisen from children absent from school on the date of the physical examination.
If children exposed to the earthquake \textit{in utero} had lingering health issues and thus were more likely to be absent from school for health reasons than other children, my estimates may suffer from downward bias.
In addition, if the absentees came from poor households and thus were unhealthier than other children, my estimates would also understate the effects because of systematic positive selection.
However, the absenteeism rates for primary schools were approximately 0.5\% at that time, which is considered to be negligible.\footnote{Chiba Prefecture, \textit{Chibaken [1930 edition, volume 2]}, p.~21.}
Moreover, the school enrollment and school attendance rates were stable at around 99\% and 95\%, respectively (Online Appendix Table~\ref{tab:sum_cov}). I confirm that my baseline results are unchanged if I control for both rates as proxies for the proportion of children with health issues (Online Appendix Table~\ref{tab:r_robustness1}, Panel A).

Second, mortality selection after birth is another potentially important issue to be discussed.\footnote{I also confirm that positive fertility selection, which suggests declines in fertility rates in the aftermath of the disaster, did not occur (Online Appendix~\ref{sec:selection}).}
If unhealthy infants were less likely to survive into the sample, my estimates would understate the impacts of the earthquake because survivors should have been healthier than those who died.
As discussed, I control for a part of mortality selection before birth by including the fetal death rate in the regressions.
However, since unhealthy babies might also die within the first 12 months of birth, I further check whether infant deaths involuntarily increased after the earthquake hit.
Although no systematic statistics on infant mortality are available, a time-series dataset of annual prefecture-average infant mortality rates between 1917 and 1926 is available from the official vital statistics.
The time-series plots of the rates indicate no systematic increasing trend in infant mortality rates after the earthquake in Chiba prefecture (Online Appendix Figure~\ref{fig:tsimr}).\footnote{To test the instantaneous impacts of the earthquake on infants' health, I also regress the first-differenced infant mortality rates on the 1923 year dummy. I find that the estimated coefficient on the earthquake year dummy is statistically insignificant (Online Appendix~\ref{sec:selection}). This result is considered to be consistent with the evidence that the number of children growing up as orphans was small (Subsection~\ref{sec:sec22}).}

Finally, our sample period includes two other important historical events: the First World War (1915--1919) and influenza pandemic (1918--1920). However, I confirm that my baseline results remain unchanged if I control for the two indicator variables for these potentially affected cohorts (Panel B of Table~\ref{tab:r_robustness1} in Online Appendix~\ref{sec:selection}).

\section{Discussion} \label{sec:sec7}

As discussed thus far, the Great Kant\=o Earthquake had negative consequences for children's health.
I find that the primary school girls aged 9--11 (6--8) who were exposed to the earthquake \textit{in utero} in the JMA-SIS7 area were approximately $0.9$ cm ($0.6$ cm) shorter than those in surrounding cohorts.
While this decline in height accounts for approximately a $0.19$ ($0.14$) standard deviation change in the average height of the primary school girls aged 9--11 (6--8) (Panel A of Table~\ref{tab:sum}), one must be careful not to attach too much importance to the small stunting effects.
Mazumder at al. found that the U.S. birth cohort exposed to the 1919 pandemic influenza experienced a $0.1$ cm decline in final height on average, which is associated with a 5\% higher risk of cardiovascular disease in old age.\footnote{Mazumder at al. `Lingering'.}
Although they did not ascertain the causal links between such stunting and the later-life risk of contracting diseases, a number of medical and epidemiological studies including randomized intervention trials have found similar associations between shorter stature in early life and suboptimal function later in life.\footnote{Dewey and Begum, `Long-term'.}

Furthermore, a growing body of the literature provides evidence on the links between child stunting and lower earnings later in life.
The comprehensive review by McGovern et al. finds that $0.5$ cm child stunting is associated with a decrease in wages of 2--3\% on average.\footnote{McGovern et al. `A review'.}
A recent height premium-based calculation also suggests that almost $2$ cm lower height translates into 1.6\% lower earnings for the affected population.\footnote{Rosales-Rueda and Triyana, `The persistent'.}
Another important fact is that the medical evidence revealed an epigenetic effect that the girls' stunting has a risk of passing lasting adverse health effects to their offspring.
A growing body of evidence has shown that maternal stunting restricts the growth of the uterus and placenta and thus increases the risk of intrauterine growth restriction, which causes insufficient neurological and intellectual development and the shorter stature of their infants.\footnote{Black et al. `Maternal'.}

This study provides suggestive evidence that the disaster relief might have played a role in nourishing exposed primary school boys.
The estimates suggest that relief provisions ameliorated roughly half of the stunting effects at early primary school ages.
This result thus adds evidence to the recent literature on the optimal timing of the remediation of early disadvantage.\footnote{Heckman, `The development'; Vellore, `The dust'. Few studies have also suggested the possibility that these interventions may mask other long-term effects; Wells, `The metabolic'; Harris, `Anthropometric'.}
My results further show that relief effects can be biased by gender depending on the parents' resource allocation based on the strong preference for a son regulated by the patriarchal institution.\footnote{Hayashi and Prescott, `The depressing'. This kind of biased resource allocation may also be in line with those of Parman, who shows that parents with a child exposed to the 1918 influenza pandemic \textit{in utero} reallocated their resources to the child's older siblings, which led to higher educational attainment for these siblings; Parman, `Childhood'.}

Given the unavoidable mental stress of the enormous exogenous shock, the role of disaster relief indicates that nutritional stress is a plausible pathway of the adverse health effects in the devastated area.
In addition, among municipalities that experienced little physical damage and disaster relief, I find suggestive evidence that maternal mental stress due to vibrations was more likely to be associated with stunting than nutritional stress due to market disruption.
That the earthquake can lead to maternal mental stress is consistent with previous studies.\footnote{E.g., Kim et al. `Maternal'.}
My results further shed light on the role of the different pathways behind the adverse health effects of fetal exposure to the earthquake.
That is, while mental stress is an important pathway underlying the adverse health effects, nutritional stress can be another pathway in the physically devastated area, thus enhancing the adverse health effects on exposed children.

\section{Conclusion}\label{sec:sec8}

This study used a catastrophic earthquake from the 1920s to analyse the long-term effects of a natural disaster on children's health.
I found that fetal exposure to the Great Kant\=o Earthquake had stunting effects on primary schoolchildren and that the magnitude of such effects increased with the degree of earthquake stress.
Disaster relief was found to have compensating effects on stunting among boys by late primary school age. Although mental stress was associated with adverse health effects, nutritional stress also enhanced the effects in the physically devastated area.

My evidence from industrializing Japan is, however, not without its limitations.
First, given the scarcity of individual-level data with the date and place of births in Japan, I predominantly used school year-age-level datasets.
Since this assignment can attenuate the estimates, those obtained in this study should be considered to be the lower bounds of the average treatment effects.
Relatedly, it was difficult to sufficiently account for the unobservable events in the children's lives in the years following the earthquake.
Second, I used data that cover one prefecture in the Kant\=o region, Chiba prefecture.
Since there was no systematic rule on editing the school physical examination records in prefectures, obtaining systematic statistics on school-level physical examinations across prefectures is difficult.
Investigating the potential compensating effects of teenage growth spurt on the early stunting effects found in this preset paper must also be an important future research direction.
While I acknowledge these limitations, this study does use a localized measure of the severity of the 1923 earthquake.
Municipal-level geospatial variations in the physical devastation could improve the assignment by detecting the most impacted area in the prefecture.

This study contributes to our understanding of the long-term effects of the great earthquake on children's health given that the lingering health impacts of earthquakes on children have been neglected in the literature.
It also offers suggestive evidence of the importance of the remediation of early disadvantage via disaster relief and of the potentially significant impacts of maternal mental stress on children \textit{in utero}.

\begin{spacing}{1.0}

\renewcommand{\refname}{{\large Documents, Statistical Reports, and Database}}

\end{spacing}
\clearpage
\begin{landscape}
\begin{table}[]
\def\arraystretch{0.91}
\begin{center}
\caption{Summary Statistics}
\label{tab:sum}
\footnotesize
\scalebox{0.96}[1]{
\begin{tabular}{lccD{.}{.}{2}D{.}{.}{2}D{.}{.}{2}D{.}{.}{2}D{.}{.}{2}D{.}{.}{2}}
\toprule
&\multirow{2}{*}{Unit}&Age&\multicolumn{3}{c}{Boys}&\multicolumn{3}{c}{Girls}\\
\cmidrule(rrr){4-6}\cmidrule(rrr){7-9}
Panel A: Height and weight&&Intervals&\multicolumn{1}{c}{Mean}&\multicolumn{1}{c}{Std. Dev.}&\multicolumn{1}{c}{Obs.}
&\multicolumn{1}{c}{Mean}&\multicolumn{1}{c}{Std. Dev.}&\multicolumn{1}{c}{Obs.}\\\hline
All cohorts											&&&&&&&&\\
\hspace{10pt}Height (cm)							&School-year-age		&6--8		&113.46	&4.37	&14,139	&112.26	&4.33	&14,145\\
													&School-year-age		&9--11		&127.48	&4.13	&14,133	&126.89	&4.65	&14,138\\
\hspace{10pt}Weight (kg)							&School-year-age		&6--8		&20.07		&1.72	&14,139	&19.40		&1.68	&14,145\\
													&School-year-age		&9--11		&26.49		&2.15	&14,133	&26.27		&2.58	&14,138\\
Exposed cohorts										&&&&&&&&\\
\hspace{10pt}Height (cm)							&School-year-age		&6--8		&113.33	&4.42	&1,295		&111.94	&4.38	&1,294\\
													&School-year-age		&9--11		&127.72	&4.18	&1,284		&127.09	&4.71	&1,278\\
\hspace{10pt}Weight (kg)							&School-year-age		&6--8		&19.96		&1.69	&1,295		&19.25		&1.69	&1,294\\
													&School-year-age		&9--11		&26.66		&2.21	&1,284		&26.37		&2.66	&1,278\\
Unexposed cohorts									&&&&&&&&\\
\hspace{10pt}Height (cm)							&School-year-age		&6--8		&113.48	&4.36	&12,844	&112.30	&4.32	&12,851\\
													&School-year-age		&9--11		&127.45	&4.12	&12,849	&126.87	&4.64	&12,860\\
\hspace{10pt}Weight (kg)							&School-year-age		&6--8		&20.08		&1.73	&12,844	&19.42		&1.68	&12,851\\
													&School-year-age		&9--11		&26.47		&2.14	&12,849	&26.26		&2.57	&12,860\\
&&&&&&&&\\\hline
Panel B: Exposure variables and relief expenses~~~~~~~~ 
&Unit&&\multicolumn{1}{c}{Mean}&\multicolumn{1}{c}{Std. Dev.}&\multicolumn{1}{c}{Min}&\multicolumn{1}{c}{Max}&\multicolumn{1}{c}{$N$}&\\\hline
Exposure variables									&&&&&&&\\
\hspace{10pt}Physical disruption rate [\textit{PDR}]	&Municipality			&&8.70		&21.26						&0	&98.81	&344	&\\
\hspace{10pt}JMA-SIS7 [\textit{SIS7}]				&Municipality			&&0.07		&\multicolumn{1}{r}{$-$}	&0	&1		&344	&\\
\hspace{10pt}Railway disruption [\textit{Rail}]		&Municipality			&&0.14		&\multicolumn{1}{r}{$-$}	&0	&1		&222	&\\
Relief expenses per capita (yen)						&&&&&&&\\
\hspace{10pt}Total									&County				&&0.253	&0.722						&0	&2.619	&13	&\\
\hspace{10pt}Food									&County				&&0.165	&0.500						&0	&1.816	&13	&\\
\hspace{10pt}Shelter								&County				&&0.083	&0.208						&0	&0.740	&13	&\\
\hspace{10pt}Medical treatment						&County				&&0.001	&0.002						&0	&0.007	&13	&\\\bottomrule
\end{tabular}
}
{\scriptsize
\begin{minipage}{640pt}
\setstretch{0.84}Notes:
Panel A reports the summary statistics for the school-year-age-level average values of height (cm) and weight (kg).
The exposed cohort includes the 1923 birth cohort, whereas unexposed cohorts include all birth cohorts other than the exposed cohort.
Panel B reports the summary statistics for the exposure variables and relief expenses per capita.
\textit{PDR} is the number of affected housing units per 100 households (Figure~\ref{fig:spatial_hh}).
\textit{SIS7} is an indicator variable for municipalities in the JMA-SIS7 area (Table~\ref{tab:class}).
\textit{Rail} is an indicator variable for municipalities satisfying the following conditions: (a) located in the JMA-SIS7 area, (b) did not receive any disaster relief, and (c) located within 10 km of any municipality including either the Noda line or the Kururi line.
Each relief variable is defined as the relief expense divided by the number of people (yen).\\
Sources: Biological outcome data are from the SPES (1925--1935 editions).
See Table~\ref{tab:class} for the data used to calculate the earthquake intensity measure. Data on the railway disruption and relief expenses per capita are from the SRCP (1923 edition).
\end{minipage}
}
\end{center}
\end{table}
\end{landscape}
\begin{landscape}
\begin{table}[]
\def\arraystretch{1.0}
\begin{center}
\captionsetup{justification=centering}
\caption{Classification of Municipalities by JMA-SIS (shindo)}
\label{tab:class}
\footnotesize
\scalebox{1.0}[1]{
\begin{tabular}{lrrrr}
\toprule
							&Number of		&Counties mainly						&Mean of the damaged	\\
Seismic intensity scale		&municipalities	&included								&households (PDR in \%)	\\\hline
JMA-SIS7 (shindo nana)		&23			&Awa, Kimitsu							&79.8					\\
JMA-SIS6 (shindo roku)		&46			&Awa, Kimitsu, Ichihara, Ch\=osei		&22.8					\\
JMA-SIS5 (shindo go)		&275			&Chiba, Higashikatsushika,				&0.39					\\
							&				&Imba, Isumi, Ichihara, Kaijy\=o,			&						\\
							&				&Katori, Sanbu, Sousa					&						\\\bottomrule
\end{tabular}
}
{\scriptsize
\begin{minipage}{450pt}
\setstretch{0.85}Notes:
The physical disruption rate (PDR) is the number of affected housing units per $100$ households.
The JMA-SIS ranges from one (minimum) to seven (maximum).
It is calculated based on the collapse rate predicted using the prediction equation of Moroi and Takemura, `1995nen': {\scriptsize $\text{Collapse Rate} = -1.61+0.46 \times \text{PDR}+0.0051 \times \text{PDR}^{2}$}.
The JMA-SIS is classified based on Takemura and Moroi, `Chishitsuch\=osajyo': a rate greater than 30\% is defined as JMA-SIS7 (shindo nana), greater than 1\% and less than 30\% is defined as JMA-SIS6 (shindo roku), and greater than 0.1\% and less than 1\% is defined as upper JMA-SIS5 (shindo go).
No municipalities experienced JMA-SIS1--4.
Following the official classification, JMA-SIS6 includes both 6-lower (roku jyaku) and 6-upper (roku ky\=o); JMA-SIS5 includes both 5-lower (go jyaku) and 5-upper (go ky\=o).\\
Sources: Data used to calculate the earthquake intensity measures are from the HTE and Statistics Bureau of the Cabinet, \textit{Taish\=ojyuyonen}.
\end{minipage}}
\end{center}
\end{table}
\end{landscape}
\begin{table}[]
\def\arraystretch{1.0}
\begin{center}
\captionsetup{justification=centering}
\caption{Effects of Fetal Earthquake Exposure on Height (cm) and Weight (kg)}
\label{tab:r_main}
\footnotesize
\scalebox{1.0}[1]{
\begin{tabular}{lD{.}{.}{-2}D{.}{.}{-2}D{.}{.}{-2}D{.}{.}{-2}D{.}{.}{-2}D{.}{.}{-2}D{.}{.}{-2}D{.}{.}{-2}}
\toprule
&\multicolumn{2}{c}{Boys}&\multicolumn{2}{c}{Girls}\\
\cmidrule(rr){2-3}\cmidrule(rr){4-5}
&\multicolumn{1}{c}{(1)}&\multicolumn{1}{c}{(2)}&\multicolumn{1}{c}{(3)}&\multicolumn{1}{c}{(4)}\\
&\multicolumn{1}{c}{Ages 6--8}&\multicolumn{1}{c}{Ages 9--11}&\multicolumn{1}{c}{Ages 6--8}&\multicolumn{1}{c}{Ages 9--11}\\\hline
Panel A-1: Effects on height							&&&&\\
\hspace{5pt}1923 birth cohort						&-0.146$**$	&-0.137$**$	&-0.280$***$	&-0.203$***$	\\
													&[0.024]		&[0.038]		&[0.002]		&[0.003]		\\
\hspace{5pt}1923 birth cohort $\times$ SIS7		&0.043			&0.113			&-0.311$***$	&-0.643$***$	\\
													&[0.750]		&[0.482]		&[0.004]		&[0.007]		\\
Panel A-2: Effects on weight							&&&&\\
\hspace{5pt}1923 birth cohort						&-0.057$**$	&-0.031	&-0.101$***$	&-0.114$**$	\\
													&[0.044]		&[0.276]	&[0.004]		&[0.011]		\\
\hspace{5pt}1923 birth cohort $\times$ SIS7		&0.006			&0.102		&-0.258$***$	&-0.307$**$	\\
													&[0.918]		&[0.480]	&[0.002]		&[0.019]		\\
\hspace{5pt}Observations							
&\multicolumn{1}{c}{14,139}
&\multicolumn{1}{c}{14,133}
&\multicolumn{1}{c}{14,145}
&\multicolumn{1}{c}{14,138}		\\
&&&&\\
Panel B-1: Effects on height							&&&&\\
\hspace{5pt}1923 birth cohort						&-0.158$**$	&-0.156$**$	&-0.268$***$	&-0.188$***$	\\
													&[0.020]		&[0.024]		&[0.004]		&[0.003]		\\
\hspace{5pt}1923 birth cohort $\times$ PDR		&0.002			&0.003			&-0.004		&-0.007$***$	\\
													&[0.104]		&[0.290]		&[0.296]		&[0.007]		\\
Panel B-2: Effects on weight							&&&&\\
\hspace{5pt}1923 birth cohort						&-0.062$**$	&-0.040	&-0.096$**$	&-0.118$***$	\\
													&[0.022]		&[0.188]	&[0.012]		&[0.009]		\\
\hspace{5pt}1923 birth cohort $\times$ PDR		&0.001			&0.002		&-0.002		&-0.002		\\
													&[0.568]		&[0.136]	&[0.388]		&[0.315]		\\
\hspace{5pt}Observations							
&\multicolumn{1}{c}{14,139}
&\multicolumn{1}{c}{14,133}
&\multicolumn{1}{c}{14,145}
&\multicolumn{1}{c}{14,138}		\\\bottomrule
\end{tabular}
}
{\scriptsize
\begin{minipage}{420pt}
\setstretch{0.85}
***, **, and * represent statistical significance at the 1\%, 5\%, and 10\% levels based on the $p$-values from the wild cluster bootstrap resampling method in brackets, respectively.
The data are clustered at the 13-county level in the bootstrap procedure.
The number of replications is fixed to 1,000 for all the specifications.\\
Notes:
The numbers of observations in columns (1)--(4) are $14,139$, $14,133$, $14,145$, and $14,138$, respectively.
The data on the school-year-age-level average values of height (cm) or weight (kg) are used in the regressions.
All the regressions include controls for the rice yield in the birth year; fetal death rate in the birth year; school enrollment rate of the parental generation; school-age-specific fixed effects; and year fixed effects.
\end{minipage}
}
\end{center}
\end{table}
\begin{table}[]
\def\arraystretch{1.0}
\begin{center}
\captionsetup{justification=centering}
\caption{Effects of Fetal Earthquake Exposure on Height (cm) and Weight (kg): Testing the Potential Effects of Early-life and Instantaneous Exposure}
\label{tab:r_robustness3}
\footnotesize
\scalebox{1.0}[1]{
\begin{tabular}{lcccccccc}
\toprule
&\multicolumn{2}{c}{Boys}&\multicolumn{2}{c}{Girls}\\
\cmidrule(rr){2-3}\cmidrule(rr){4-5}
&\multicolumn{1}{c}{(1)}&\multicolumn{1}{c}{(2)}&\multicolumn{1}{c}{(3)}&\multicolumn{1}{c}{(4)}\\
&\multicolumn{1}{c}{Ages 6--9}&\multicolumn{1}{c}{Ages 8--9}&\multicolumn{1}{c}{Ages 6--9}&\multicolumn{1}{c}{Ages 8--9}\\\hline
Panel A: Effects on height							&&&&\\
\hspace{5pt}1923 birth cohort						&-0.103*		&			&-0.261**	&\\
													&[0.078]		&			&[0.016]	&\\
\hspace{5pt}1923 birth cohort $\times$ SIS7		&-0.031		&			&-0.362***	&\\
													&[0.982]		&			&[0.002]	&\\
\hspace{5pt}Early-life exposure (ages 1-5)			&0.069			&			&-0.076	&\\
													&[0.134]		&			&[0.474]	&\\
\hspace{5pt}Early-life exposure $\times$ SIS7		&-0.134		&			&-0.093	&\\
													&[0.762]		&			&[0.930]	&\\
\hspace{5pt}Instantaneous exposure (ages 6-9)		&0.279			&			&-0.014	&\\
													&[0.186]		&			&[0.936]	&\\
\hspace{5pt}Instantaneous exposure $\times$ SIS7	&-0.563		&			&-0.395	&\\
													&[0.508]		&			&[0.410]	&\\
\hspace{5pt}Exposure at ages 6--7					&				&0.392		&			&-0.028\\
													&				&[0.104]	&			&[0.910]\\
\hspace{5pt}Exposure at ages 6--7 $\times$ SIS7	&				&-0.520	&			&-0.224\\
													&				&[0.450]	&			&[0.542]\\
\hspace{5pt}Observations							
&\multicolumn{1}{c}{18,851}
&\multicolumn{1}{c}{9,424}
&\multicolumn{1}{c}{18,857}
&\multicolumn{1}{c}{9,422}		\\
&&&&\\
Panel B: Effects on weight							&&&&\\
\hspace{5pt}1923 birth cohort						&-0.020		&			&-0.087**		&\\
													&[0.442]		&			&[0.012]		&\\
\hspace{5pt}1923 birth cohort $\times$ SIS7		&-0.000		&			&-0.274*		&\\
													&[1.000]		&			&[0.060]		&\\
\hspace{5pt}Early-life exposure (ages 1-5)			&0.039			&			&-0.002		&\\
													&[0.282]		&			&[0.982]		&\\
\hspace{5pt}Early-life exposure $\times$ SIS7		&-0.021		&			&-0.015		&\\
													&[1.000]		&			&[0.866]		&\\
\hspace{5pt}Instantaneous exposure (ages 6-9)		&0.048			&			&-0.002		&\\
													&[0.620]		&			&[0.908]		&\\
\hspace{5pt}Instantaneous exposure $\times$ SIS7	&-0.292		&			&-0.131		&\\
													&[0.504]		&			&[0.460]		&\\
\hspace{5pt}Exposure at ages 6--7					&				&0.100		&				&0.039\\
													&				&[0.268]	&				&[0.730]\\
\hspace{5pt}Exposure at ages 6--7 $\times$ SIS7	&				&-0.273	&				&-0.081\\
													&				&[0.430]	&				&[0.498]\\
\hspace{5pt}Observations							
&\multicolumn{1}{c}{18,851}
&\multicolumn{1}{c}{9,424}
&\multicolumn{1}{c}{18,857}
&\multicolumn{1}{c}{9,422}		\\\bottomrule
\end{tabular}
}
{\scriptsize
\begin{minipage}{370pt}
\setstretch{0.85}
***, **, and * represent statistical significance at the 1\%, 5\%, and 10\% levels based on the $p$-values from the wild cluster bootstrap resampling method in brackets, respectively.
The data are clustered at the 13-county level in the bootstrap procedure.
The number of replications is fixed to 1,000 for all the specifications.\\
Notes:
The numbers of observations for each regression reported in columns (1)--(4) are $18,851$, $9,424$, $18,857$, and $9,422$, respectively.
The data on the school-year-age-level average values of height (cm) or weight (kg) are used in the regressions.
All the regressions include controls for the rice yield in the birth year; fetal death rate in the birth year; school enrollment rate of the parental generation; school-age-specific fixed effects; and year fixed effects.
\end{minipage}
}
\end{center}
\end{table}
\begin{landscape}
\begin{table}[]
\def\arraystretch{1.0}
\begin{center}
\captionsetup{justification=centering}
\caption{Effects of Disaster Relief on Height (cm) and Weight (kg) by Type of Relief Expense}
\label{tab:r_relief}
\footnotesize
\scalebox{1.0}[1]{
\begin{tabular}{llccccc}
\toprule
&&&\multicolumn{4}{c}{Relief expenses per capita}\\
\cmidrule(rrrr){4-7}
&&&\multicolumn{1}{c}{Total}&\multicolumn{1}{c}{Food}&\multicolumn{1}{c}{Shelter}&\multicolumn{1}{c}{Medical treatment}\\
\cmidrule(r){4-4}\cmidrule(r){5-5}\cmidrule(r){6-6}\cmidrule(r){7-7}
&&&\multicolumn{1}{c}{(1) Expenses $\times$}&\multicolumn{1}{c}{(2) Expenses $\times$}&\multicolumn{1}{c}{(3) Expenses $\times$}&\multicolumn{1}{c}{(4) Expenses $\times$}\\
&&Observations&\multicolumn{1}{c}{1923 birth cohort}&\multicolumn{1}{c}{1923 birth cohort}&\multicolumn{1}{c}{1923 birth cohort}&\multicolumn{1}{c}{1923 birth cohort}\\\hline

Panel A: 		&Effects on height	&&&&&\\
Boys			&Ages 6--8			&14,139	&0.095 [0.030]**	&0.138 [0.018]**	&0.318 [0.050]**	&32.144 [0.100]*	\\
				&Ages 9--11		&14,133	&0.076 [1.000]		&0.117 [1.000]		&0.215 [1.000]		&28.207 [1.000]	\\
Girls			&Ages 6--8			&14,145	&-0.008 [0.954]	&-0.005 [0.954]	&-0.070 [0.946]	&8.124 [1.000]		\\
				&Ages 9--11		&14,138	&0.075 [0.504]		&0.114 [0.480]		&0.225 [0.556]		&37.606 [0.468]	\\
		 		&					&&&&&\\
Panel B: 		&Effects on weight	&&&&&\\
Boys			&Ages 6--8			&14,139	&0.026 [0.060]*	&0.037 [0.094]*	&0.089 [0.032]**	&15.050 [0.032]**	\\
				&Ages 9--11		&14,133	&0.047 [0.938]		&0.069 [0.834]		&0.154 [0.986]		&19.161 [0.744]	\\
Girls			&Ages 6--8			&14,145	&0.004 [1.000]		&0.006 [1.000]		&0.018 [1.000]		&11.109 [0.900]	\\
				&Ages 9--11		&14,138	&0.056 [0.458]		&0.079 [0.450]		&0.204 [0.466]		&29.698 [0.460]	\\
\bottomrule
\end{tabular}
}
{\scriptsize
\begin{minipage}{550pt}
\setstretch{0.85}
** and * represent statistical significance at the 5\% and 10\% levels based on the $p$-values from the wild cluster bootstrap resampling method in brackets.
The data are clustered at the 13-county level in the bootstrap procedure.
The number of replications is fixed to 1,000 for all the specifications.\\
Notes: 
Estimated coefficient on $\textit{I(YOB=1923)}\times\textit{Relief}$ in equation~(\ref{eq2}) are reported in the table.
The number of observations for boys (girls) aged 6--8 and 9--11 are 14,139 (14,145) and 14,133 (14,138), respectively.
The data on the school-year-age-level average values of height (cm) or weight (kg) are used in the regressions.
All the regressions include controls for the rice yield in the birth year; fetal death rate in the birth year; school enrollment rate of the parental generation; school-age-specific fixed effects; and year fixed effects.
\end{minipage}
}
\end{center}
\end{table}
\end{landscape}
\begin{table}[h!]
\def\arraystretch{1.0}
\begin{center}
\captionsetup{justification=centering}
\caption{Effects of Fetal Earthquake Exposure on Height (cm) and Weight (kg) in the Limited JMA-SIS5 Area}
\label{tab:r_rail}
\footnotesize
\scalebox{0.96}[1]{
\begin{tabular}{lD{.}{.}{-2}D{.}{.}{-2}D{.}{.}{-2}D{.}{.}{-2}}
\toprule
&\multicolumn{2}{c}{Boys}&\multicolumn{2}{c}{Girls}\\
\cmidrule(rr){2-3}\cmidrule(rr){4-5}
&\multicolumn{1}{c}{(1)}&\multicolumn{1}{c}{(2)}&\multicolumn{1}{c}{(3)}&\multicolumn{1}{c}{(4)}\\
&\multicolumn{1}{c}{Ages 6--8}&\multicolumn{1}{c}{Ages 9--11}&\multicolumn{1}{c}{Ages 6--8}&\multicolumn{1}{c}{Ages 9--11}\\\hline

Panel A: Effects on height									&&&&\\
\hspace{5pt}1923 birth cohort								&-0.135$**$	&-0.126		&-0.290$***$	&-0.186$**$	\\
															&[0.032]		&[0.230]		&[0.004]		&[0.017]		\\
\hspace{5pt}1923 birth cohort $\times$ Railway disruption	&0.135			&0.236			&0.179			&0.343			\\
															&[0.776]		&[0.554]		&[0.252]		&[0.767]		\\
\hspace{5pt}Observations							
&\multicolumn{1}{c}{9,352}
&\multicolumn{1}{c}{9,348}
&\multicolumn{1}{c}{9,353}
&\multicolumn{1}{c}{9,347}		\\
															&&&&\\
Panel B: Effects on weight									&&&&\\
\hspace{5pt}1923 birth cohort								&-0.056$*$		&-0.010		&-0.108$***$	&-0.094$*$	\\
															&[0.054]		&[0.854]		&[0.010]		&[0.067]	\\
\hspace{5pt}1923 birth cohort $\times$ Railway disruption	&0.071			&-0.103		&0.055			&-0.066	\\
															&[0.756]		&[0.482]		&[0.432]		&[0.711]	\\
\hspace{5pt}Observations							
&\multicolumn{1}{c}{9,352}
&\multicolumn{1}{c}{9,348}
&\multicolumn{1}{c}{9,353}
&\multicolumn{1}{c}{9,347}		\\\bottomrule

\end{tabular}
}
{\scriptsize
\begin{minipage}{440pt}
\setstretch{0.85}
***, **, and * represent statistical significance at the 1\%, 5\%, and 10\% levels based on the $p$-values from the wild cluster bootstrap resampling method in brackets, respectively.
The data are clustered at the 13-county level in the bootstrap procedure.
The number of replications is fixed to 1,000 for all the specifications.\\
Notes:
The samples include municipalities receiving no disaster relief in the JMA-SIS5 area.
The numbers of observations in columns (1)--(4) are 9352, 9348, 9353, and 9347, respectively.
The data on the school-year-age-level average values of height (cm) or weight (kg) are used in the regressions.
All the regressions include controls for the rice yield in the birth year; fetal death rate in the birth year; school enrollment rate of the parental generation; school-age-specific fixed effects; and year fixed effects.
\end{minipage}
}
\end{center}
\end{table}
\clearpage
\begin{figure}[]
\centering
\subfloat[Physical disruption rates (\%)]{\label{fig:spatial_hh}\includegraphics[width=0.35\textwidth]{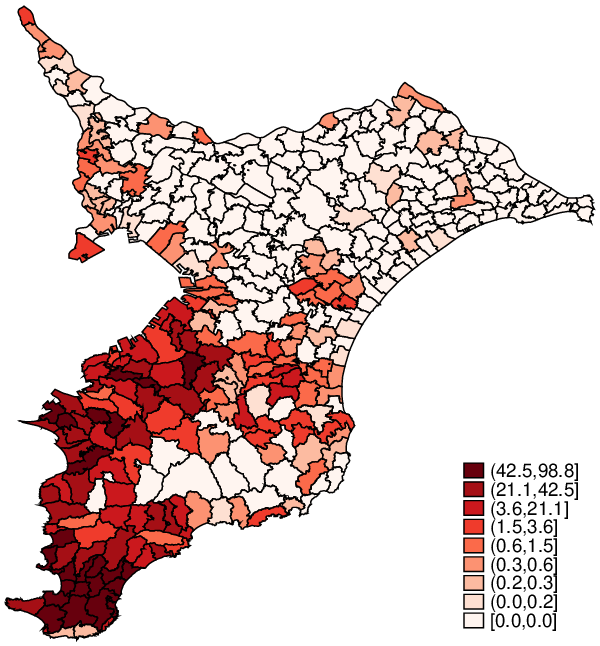}}
\subfloat[Intensity classification]{\label{fig:categories}\includegraphics[width=0.35\textwidth]{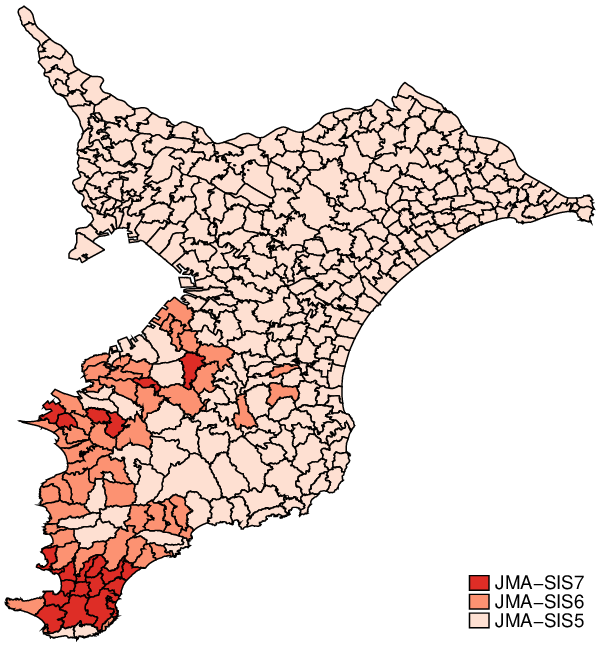}}
\caption{Spatial distribution of physical disruptions in Chiba prefecture}
\label{fig:spatial}
\scriptsize{\begin{minipage}{400pt}
\setstretch{0.85}Notes: 
Figure~\ref{fig:spatial_hh} illustrates the physical disruption rate, defined as the number of affected housing units (collapsed or semi-collapsed due to the earthquake) per 100 households.
The classifications of JMA-SIS5, JMA-SIS6, and JMA-SIS7 in Figure~\ref{fig:categories} include municipalities exposed to a JMA-SIS of 5 (shindo go), 6 (shindo roku), and 7 (shindo nana), respectively.\\
Sources: Calculated by the author from the Division of Social Affairs, Chiba Prefecture, \textit{Taish\=odaishinsai [volume 2]}; Statistics Bureau of the Cabinet, \textit{Taish\=okyunen}. Shapefile is based on the database of the Ministry of Land, Infrastructure, Transport and Tourism of Japan, \textit{Kokudos\=uchijy\=oh\=o}.
\end{minipage}}
\end{figure}
\clearpage
\thispagestyle{empty}

\begin{center}
\qquad

\qquad

\qquad

\qquad

\qquad

\qquad

{\Large \textbf{
Appendices
}}
\end{center}
\clearpage
\appendix
\def\thesection{Appendix~\Alph{section}}
\def\thesubsection{\Alph{section}.\arabic{subsection}}
\setcounter{page}{1}
\setcounter{footnote}{0}
\begin{spacing}{1.05}
\section{Background appendix}\label{sec:seca}
\setcounter{figure}{0} \renewcommand{\thefigure}{A.\arabic{figure}}

\begin{figure}[h]
\centering
\subfloat[Number of extrinsic deaths]{\label{fig:tsed}\includegraphics[width=0.33\textwidth]{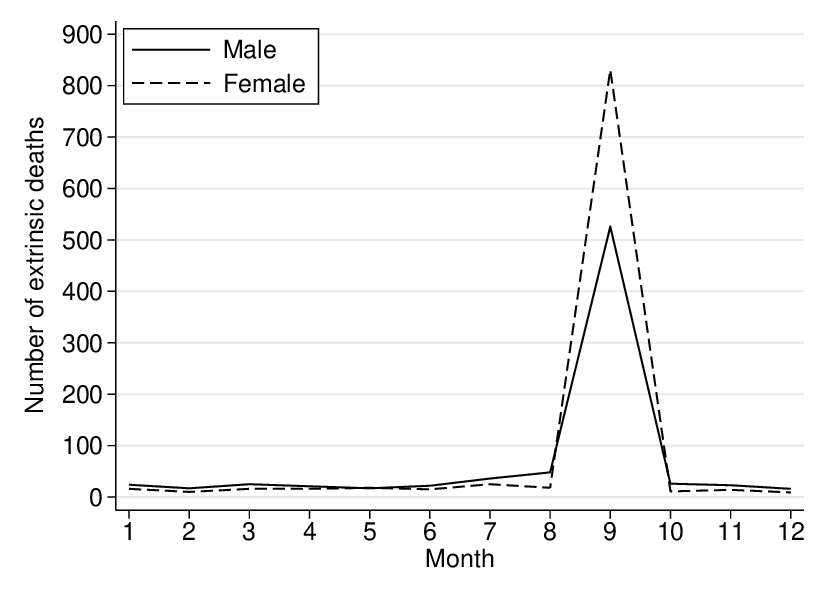}}
\subfloat[Extrinsic deaths (\%)]{\label{fig:tsedp}\includegraphics[width=0.33\textwidth]{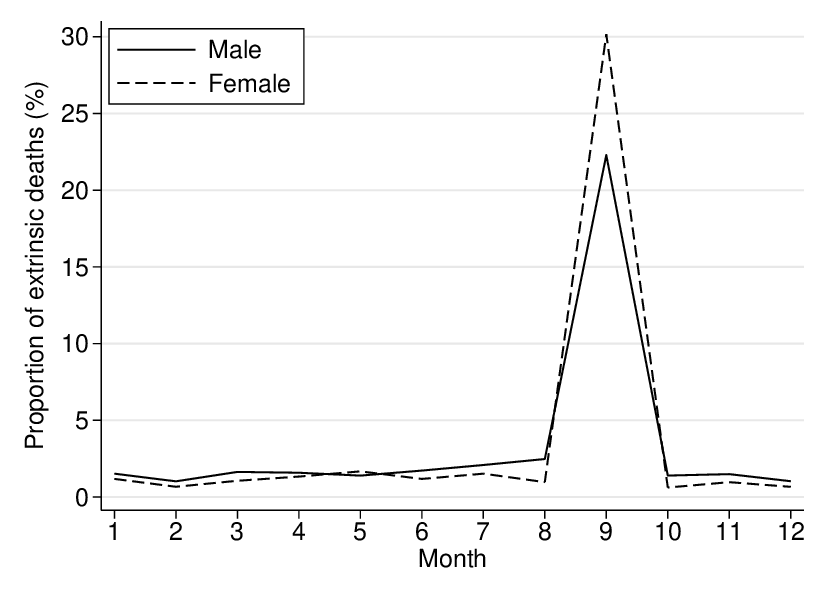}}
\subfloat[Extrinsic death rate]{\label{fig:edr}\includegraphics[width=0.33\textwidth]{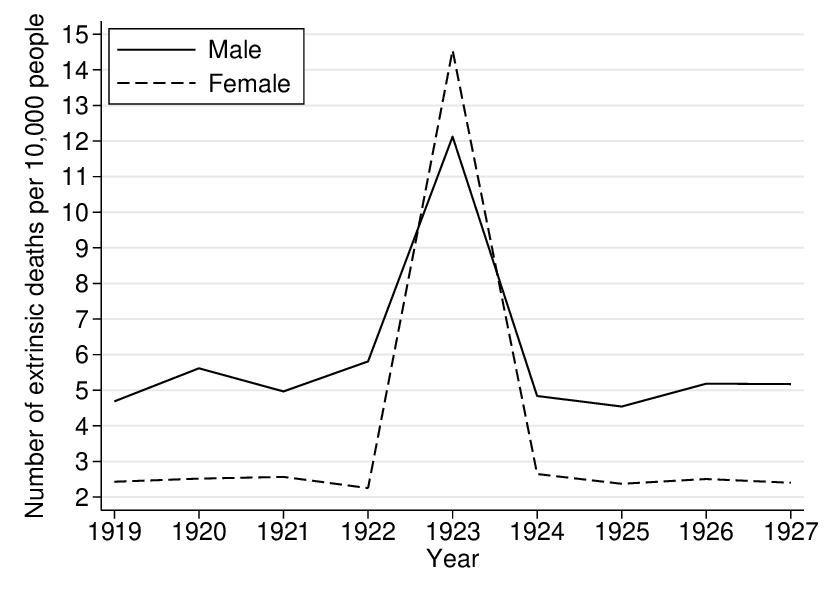}}
\caption{Extrinsic deaths in Chiba around the Great Kant\=o Earthquake}
\label{fig:ts}
\scriptsize{\begin{minipage}{450pt}
\setstretch{0.85}Notes: 
The proportion of extrinsic deaths shown in Figure~\ref{fig:tsedp} is defined as the number of extrinsic deaths relative to total deaths.
The extrinsic death rate shown in Figure~\ref{fig:edr} is defined as the number of extrinsic deaths per 10,000 people.\\
Sources: All figures are calculated by the author from the Statistics Bureau of the Cabinet, \textit{Nihonteikoku shiint\=okei [1919--1927 editions]}.
\end{minipage}}
\end{figure}

\begin{figure}[h]
 \centering
  \includegraphics[width=5cm]{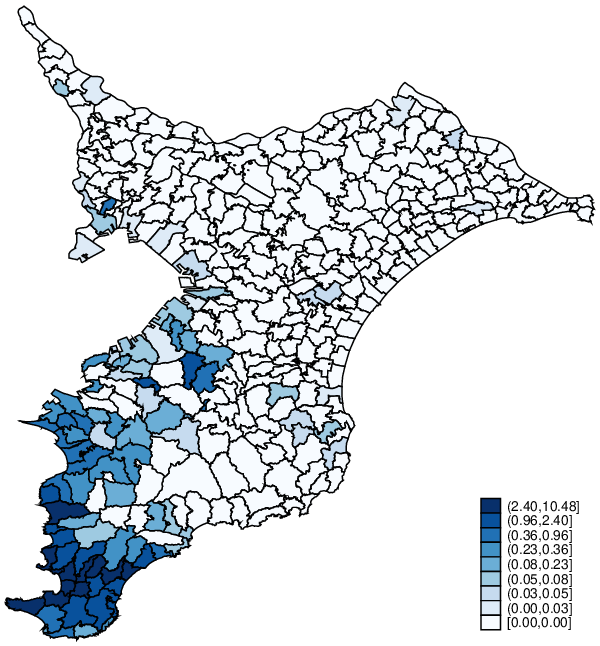}
  \caption{Spatial distribution of victims (per 100 people)\\ in Chiba prefecture}
 \label{fig:humanr}
\scriptsize{\begin{minipage}{350pt}
\setstretch{0.85}Note: 
The number of victims is defined as the number of deaths, missing people, and injured people due to the earthquake. \\
Source: Calculated by the author from Division of Social Affairs, Chiba Prefecture, \textit{Taish\=odaishinsai [volume.2]}.
\end{minipage}}
\end{figure}

\begin{figure}[h]
\centering
\includegraphics[width=0.6\textwidth]{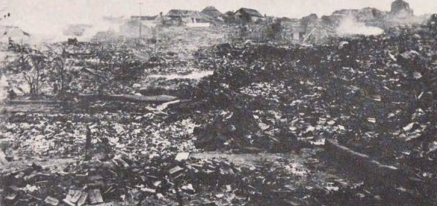}
\caption{Devastation in Funagata town in Awa county}
\label{fig:photo}
\scriptsize{\begin{minipage}{350pt}
Source: Division of Social Affairs, Chiba Prefecture, \textit{Taish\=odaishinsai [volume.1]}.
\end{minipage}}
\end{figure}

Figure~\ref{fig:tsed} confirms the dramatic increase in extrinsic deaths in September 1923. 
Figure~\ref{fig:edr} shows the extrinsic death rate, defined as the number of extrinsic deaths per 10,000 people, in Chiba between 1919 and 1927. 
The small disparity in the male and female death rates of 1923 suggests that women were more likely to have been affected by the earthquake, which struck households around lunchtime when most wives had cooked their family's meal at home.
Figure~\ref{fig:humanr} shows the spatial distribution of the number of deaths, missing people, and injured people due to the earthquake per 100 people (Online~\ref{sec:secb}).
This figure shows a similar but spatially modest distribution compared with that of affected housing units in Figure~\ref{fig:spatial_hh}.
Indeed, while 28\% of municipalities (97 out of 349) had victims, 53\% suffered household damage.
The example picture in Figure~\ref{fig:photo} illustrates the devastation in Awa, confirming the validity of the measure of earthquake stress.


\section{Data appendix}\label{sec:secb}
\setcounter{figure}{0} \renewcommand{\thefigure}{B.\arabic{figure}}
\setcounter{table}{0} \renewcommand{\thetable}{B.\arabic{table}}

The school-level primary school dataset used in this study is constructed from the SPES published by Chiba prefecture between 1925 and 1936.\footnote{\setstretch{0.84}The SPES originally consisted of two reports that contained similar information: the \textit{Gakk\=o seitojid\=o shintaikensa t\=okei} published in 1925 and from 1927 to 1934 (data for 1925--1934) and the \textit{Seitojid\=o shintaikensa t\=okei} published in 1936 (data for 1935). I uniformly refer to these publications as the SPES (1925--1935 editions) for simplicity. Some errata in these documents were corrected. For instance, if the sequence of height (in cm) from ages 6 to 11 was 109, \textit{218}, 123, 125, 128, and 130, I corrected the second observation to \text{118} because a height of 218 cm is clearly unrealistic for a seven-year-old and thus can be regarded as a typo. I also excluded some observations because of outliers, missing values, and consolidations of the municipalities. Consequently, 99\% of my school-year panels are balanced. I confirm that trimming my sample to keep a balanced panel structure produces results virtually identical to those presented in the main results in Table~\ref{tab:r_main} (not reported). This means that the particular subsets of my school panels do not disturb my main results.}
Since the academic term runs from April to March in Japan, children in the first grade of primary school are aged six and seven and those in the final grade are aged 11 and 12.
To ensure the consistency of the data structure, however, I refer to the range of ages in my primary students sample as 6--11 years throughout this paper.
The systematic data on the secondary schools are unfortunately unavailable.
The SPES indicates that the number of secondary schools in the physically devastated areas was indeed too small to conduct any statistical inferences.
The number of primary schools from 1925 to 1935 was 429, 429, 429, 433, 434, 433, 433, 432, 428, 427, and 422, respectively.
According to the SPES and Population Census in 1935, approximately 95\% of primary school-aged children in Chiba were covered in the SPES datasets.\footnote{\setstretch{0.84}The number of examinees in the primary schools is taken from Physical Education Bureau, Ministry of Education, \textit{Seitojid\=o [1938 edition]}, p.~7;~23. The number of children is taken from Statistics Bureau of the Cabinet, \textit{Sh\=owajy\=unen}, pp.~34--35.}
Although earthquake severity is measured at the municipal level, I use data on the biological outcomes at the school-year-age level to control for the unobserved factors of each school using the bilateral-specific fixed effect and exploit the full information around the JMA-SIS7 area. Despite this design, I find that the main results in Table~\ref{tab:r_main} are mostly unchanged if I use the municipal-year-age-level datasets (not reported).

\begin{figure}[h]
\centering
\captionsetup{justification=centering}
\subfloat[Height: PS-boys]{\label{fig:phb}\includegraphics[width=0.45\textwidth]{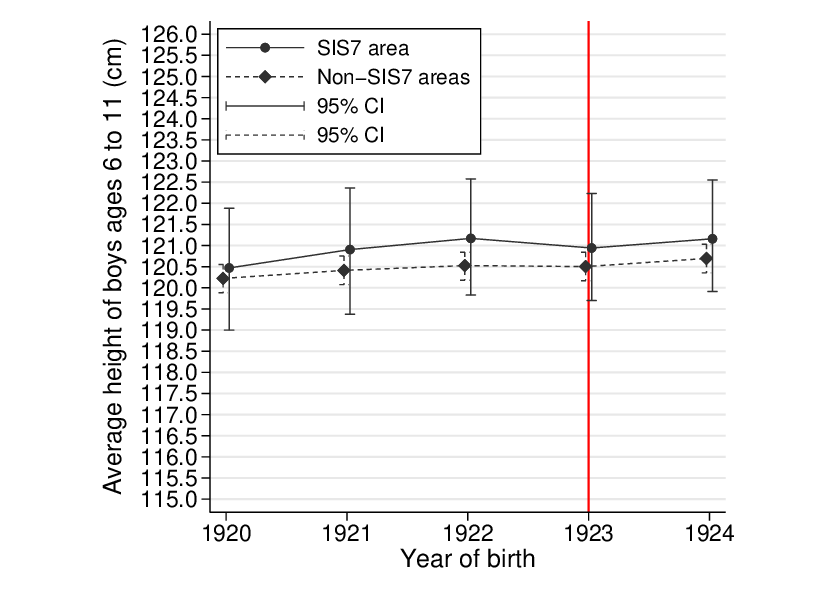}}
\subfloat[Height: PS-girls]{\label{fig:phg}\includegraphics[width=0.45\textwidth]{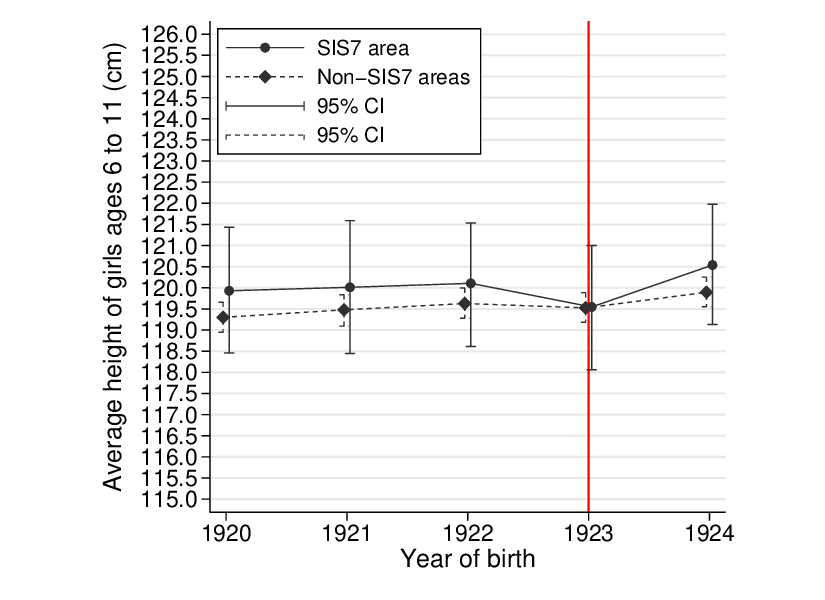}}\\
\subfloat[Weight: PS-boys]{\label{fig:pwb}\includegraphics[width=0.45\textwidth]{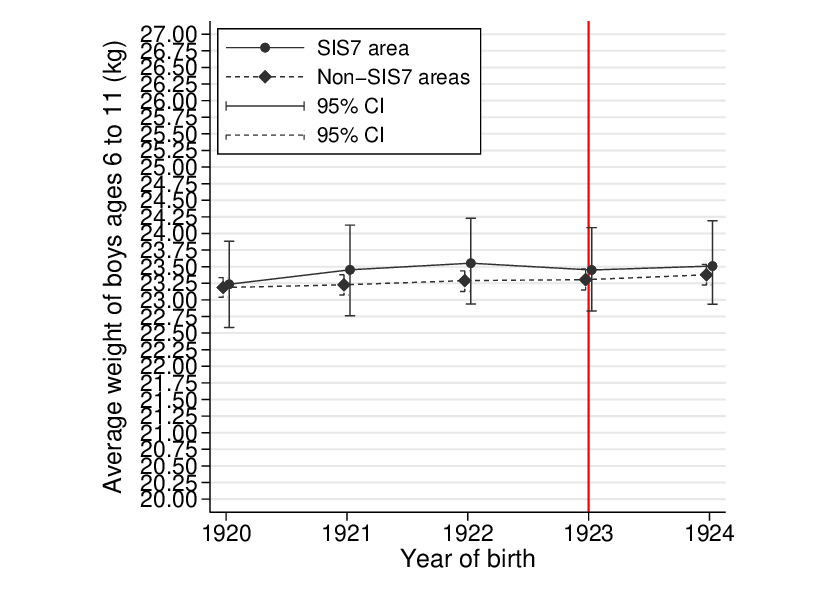}}
\subfloat[Weight: PS-girls]{\label{fig:pwg}\includegraphics[width=0.45\textwidth]{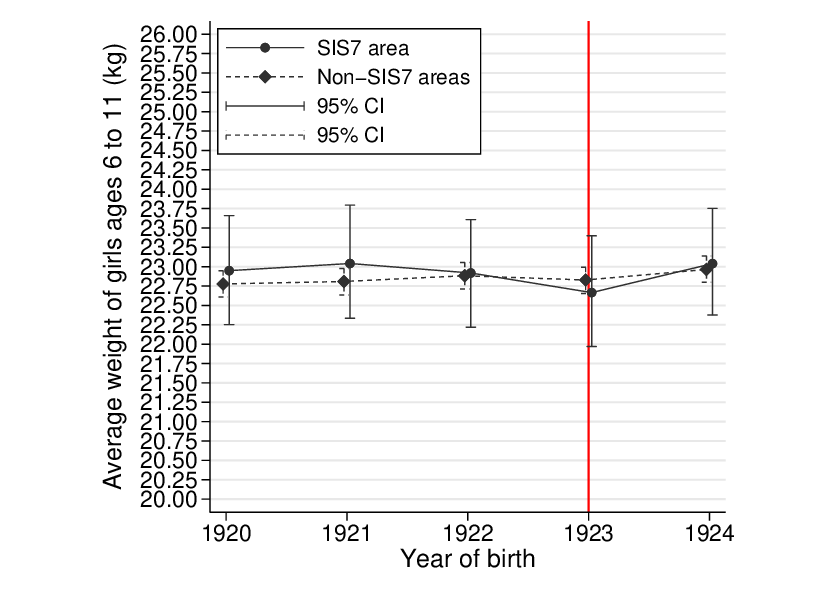}}
\caption{Average height (in cm) and weight (in kg) of the primary school (PS) students by area and gender}
\label{fig:ah}
\scriptsize{\begin{minipage}{450pt}
\setstretch{0.85}Notes: 
Figure~\ref{fig:phb} and \ref{fig:phg} show the average height of primary boys and girls, respectively.
Figure~\ref{fig:pwb} and \ref{fig:pwg} show the average weight of primary boys and girls, respectively.
The SIS7 area refers to the area extremely affected by the earthquake with the maximum seismic intensity scale (JMA-SIS7), whereas the non-SIS7 area refers to the JMA-SIS5--6 areas reported in Table~\ref{tab:class}.
Bootstrap percentile confidence intervals are illustrated.
The number of replications is fixed to 1,000.\\
Sources: Calculated by the author from the SPES (1925--1935 editions).
\end{minipage}}
\end{figure}

Figure~\ref{fig:ah} illustrates the raw relationship between schoolchildren's average height and weight by their year of birth.\footnote{\setstretch{0.84}I used Bunmeid\=o, \textit{Kaitei}, pp.~1--27 to match the schools with the municipalities. The confidence intervals for the non-JMA-SIS7 areas are systematically smaller than those for the JMA-SIS7 area because of the greater number of observations in the former areas than in the latter area.}
While there is a general increasing trend in height, reduced growth is observed in the 1923 birth cohort, especially in the JMA-SIS7 area (Figure~\ref{fig:phb} and~\ref{fig:phg}).
Regarding the weight of children, reduced growth might be observed in the 1923 birth cohort in the JMA-SIS7 area (Figures~\ref{fig:pwb} and~\ref{fig:pwg}).

In Subsection~\ref{sec:sec3}, I confirm that the internal migration rates of primary schoolchildren were substantially low, less than 5\% of all children. Here, I further test the potential correlations between the migration rates and children's heights and weights.
While no systematic municipal-level information on migrants is available, the 1930 Population Census documented the proportion of non-migrants, namely, the ratio of the number of people living in their original municipality to the entire population.
Table~\ref{tab:migration} presents the correlation coefficients between this ``non-migrant rate'' and prefecture-average heights and weights of primary schoolchildren obtained from the 1930 edition of the SSPE.
Clearly, all the coefficients are close to zero and statistically insignificant in most cases, suggesting no systematic correlations between migration and children's heights and weights.

\begin{table}[h!]
\def\arraystretch{1.0}
\begin{center}
\captionsetup{justification=centering}
\caption{Correlation coefficients between non-migrant rates and children's heights and weights in 1930 Japan}
\label{tab:migration}
\footnotesize
\scalebox{1.0}[1]{
\begin{tabular}{lD{.}{.}{-1}D{.}{.}{-1}c}
\toprule
&\multicolumn{1}{c}{Boys}&\multicolumn{1}{c}{Girls}&Obs.\\\hline
Heights										&&&\\
\hspace{10pt}Average heights (ages 6--11)		&-0.2249 [0.1329]&-0.2049 [0.1719]&46\\
\hspace{10pt}Age 6							&-0.1508 [0.3171]&-0.2531 [0.0896]&46\\
\hspace{10pt}Age 7							&-0.1613 [0.2841]&-0.1610 [0.2851]&46\\
\hspace{10pt}Age 8							&-0.1937 [0.1972]&-0.1238 [0.4126]&46\\
\hspace{10pt}Age 9							&-0.2214 [0.1392]&-0.2867 [0.0534]&46\\
\hspace{10pt}Age 10							&-0.1968 [0.1900]&-0.1746 [0.2459]&46\\
\hspace{10pt}Age 11							&-0.3102 [0.0359]&-0.1319 [0.3822]&46\\
&&&\\
Weights											&&&\\
\hspace{10pt}Average weights (ages 6--11)		&-0.1619 [0.2824]&-0.0683 [0.6521]&46\\
\hspace{10pt}Age 6							& 0.0497 [0.7431]&-0.1111 [0.4623]&46\\
\hspace{10pt}Age 7							&-0.0682 [0.6523]& 0.0054 [0.9714]&46\\
\hspace{10pt}Age 8							&-0.2070 [0.1675]&-0.0284 [0.8516]&46\\
\hspace{10pt}Age 9							&-0.2220 [0.1381]&-0.0934 [0.5370]&46\\
\hspace{10pt}Age 10							&-0.1959 [0.1919]&-0.0698 [0.6449]&46\\
\hspace{10pt}Age 11							&-0.1149 [0.4470]&-0.0685 [0.6513]&46\\\bottomrule
\end{tabular}
}
{\scriptsize
\begin{minipage}{420pt}
\setstretch{0.85}Notes: 
The correlation coefficients between the non-migrant rates and average heights or weights of primary schoolchildren are listed in this table.
$p$-values are in brackets.
Prefecture-level data on the average heights and weights of primary schoolchildren in 1930 are used.
The non-migrant rate is defined as the number of people living in their original municipality per 100 people in 1930.
Okinawa prefecture, a Southern most island prefecture, is excluded as an outlier, as it has an extremely high non-migrant rate.
The number of prefectures is 46 for all the calculations.\\
Sources: Data on the prefecture-level average heights and weights of primary schoolchildren in 1930 are obtained from the 1930 edition of the SSPE (Physical Education Bureau, Ministry of Education, \textit{Seitojid\=o}.
Data on the non-migrant rates are taken from the 1930 Population Census; Statistics Bureau of the Cabinet, \textit{Sh\=owagonen}.
\end{minipage}
}
\end{center}
\end{table}

As described in Subsection~\ref{sec:sec3}, the data on physical damage are obtained from \textit{Taish\=o shinsaishi} (History of the Taish\=o Earthquake; HTE) published by the Social Welfare Bureau of the Cabinet in 1926.
The HTE surveyed all damaged households in Chiba by November 15, 1923.
Since more than two months had passed after the earthquake, the number of affected housing units cumulatively documented in the report should be accurate.
One observation (Nakagawa village in Kimitsu county) showed an unrealistically high value in the number of affected housing units. This was considered to be a misprint and was replaced by the figure based on the survey conducted by the prefecture on October 3, 1923 reported in the Division of Social Affairs, Chiba Prefecture, \textit{Taish\=odaishinsai}, pp.~2--3.
The physical disruption rate is defined as the number of affected housing units per 100 households.
The number of households in each municipality is based on the 1920 Population Census.
According to the 1925 Population Census, the number of households increased from 259,026 in 1920 to 270,796 in 1925; Statistics Bureau of the Cabinet, \textit{Taish\=ojy\=unen}.
Assuming this increasing trend, total households around 1923 might have increased by roughly 5,900, accounting for 17 households more per municipality from 1920.
This figure accounts for a 2\% increase per municipality based on 1920 values. Thus, the data from the 1920 Population Census are considered to be plausible to use as a proxy for the number of households in municipalities around the time of the earthquake hitting.

\begin{figure}[]
\centering
\includegraphics[width=5cm]{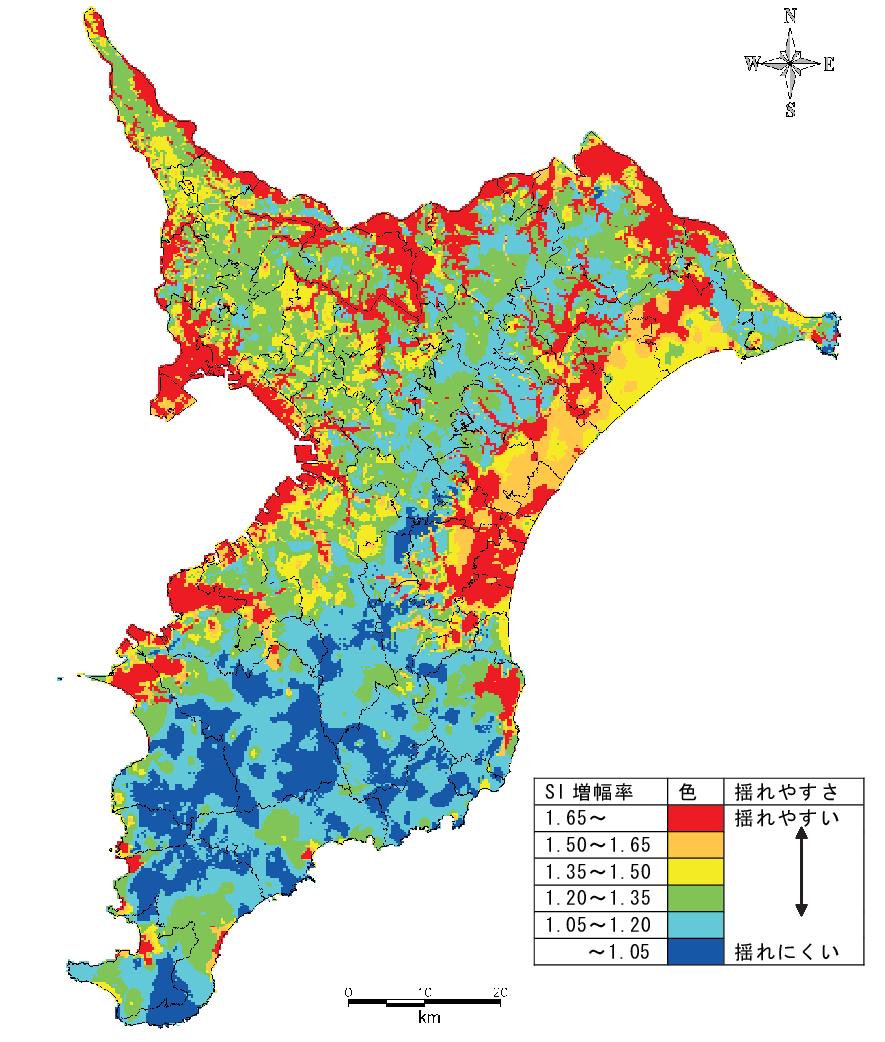}
\caption{Soil compaction in Chiba prefecture}
\label{fig:soil}
\scriptsize{\begin{minipage}{370pt}
\setstretch{0.85}Notes: 
Areas with the red (blue) meshes are more (less) likely to be affected by the earthquake. 
This map was created based on boring data in 2011 (approximately 50,000 observations).
The spectral intensity values are shown in 250-m meshes. Although the data are investigated in 2011, such wide meshes mean that the map could show nearly the same distribution of soil compaction in the early 20th century.\\
Source: Chiba prefecture \textit{Yureyasusa map}.
\end{minipage}}
\end{figure}

Figure~\ref{fig:soil} shows a map of soil compaction in Chiba prefecture, based on boring data.
In this map, the areas with red (blue) meshes were more (less) likely to be affected by the earthquake.
The interesting fact is that the physical disruption in Figure~\ref{fig:spatial_hh} shows the opposite distribution of affected housing units. 
This fact implies that my key measure of earthquake stress does not depend on soil compaction, which can correlate with the potential spatial distribution of industries and agricultural production.
As discussed in Subsection~\ref{sec:sec21}, the distribution of physical disruption used is more likely to be dominated by the distribution of the fault plane (\textit{Kamogawa teichi} active fault), which snakes across Awa county.

Finally, I list the sources of the documents used to construct the control variables.
The birth year fetal death rate, defined as the number of fetal deaths per 1,000 births.
Data on fetal deaths as well as live births between 1913--1921 are taken from the Statistical Report of Chiba Prefecture (SRCP) (1913--1921 editions) published by Chiba prefecture between 1915 and 1922.
The missing values for 1922--1924 and 1926--1929 are linearly interpolated by using the values of 1921, 1925, and 1930 given that the time-series plots of the rates in Chiba prefecture at that time were nearly linear (Figure~\ref{fig:tsfdr} in Online Appendix~\ref{sec:selection}).
Data on fetal deaths and live births in 1925 and 1930 are taken from the \textit{Shich\=osonbetsu jink\=od\=otai t\=okei} (Vital Statistics for Municipalities, 1925 and 1930 editions) published by the Statistics Bureau of the Cabinet in 1927 and 1933, respectively.
The rice yield is the volume of rice yield per hectare. The coverage of doctors is the number of doctors per 100 people.
Data on rice yield (hectoliter per 0.1 ha) and the number of doctors are taken from the SRCP (1913--1930 editions) published by Chiba prefecture between 1915 and 1931.
The school enrollment rate of the parental generation is the 17-year lagged primary school enrollment rate from the year of birth.
Data on the primary school enrollment rate of the parental generation are obtained from the SRCP (1897--1912 editions) published between 1899 and 1914.\footnote{\setstretch{0.84} As described by Hijikata, the school enrollment rate is defined as $\text{School Enrollment Rate} = 100\times (\text{Students aged 6--13}+\text{Graduates aged 6--13})/{\text{Children aged 6--13}}$; Hijikata, \textit{Kindainihon}, p.~13. The Vital Statistics of 1923 report that the average age at first marriage in Chiba was 25 years, implying that parents might have had their first child at 26 on average; Statistics Bureau of the Cabinet, \textit{Nihonteikoku [1923 edition]}, pp.~18--21. Therefore, one can guess that the average age of parents is roughly 26 years. Given the year of birth of the sampled children described above and primary school entrance age of 6, one may want to use the enrollment rates from 1894--1909, namely the 20-year lagged rate from the year of birth. Since the data are severely limited before 1896, however, I use the 1897--1912 editions of the SRCP (17-year lagged rate from the year of birth) to establish the primary school entrance rate of the parental generation.}
In the sensitivity analysis, I added the municipal-level primary school enrollment rate and county-level primary school attendance rate.
Data on these rates are taken from the SRCP (1925--1935 editions) published by the Chiba prefecture between 1927 and 1937.

\begin{table}[h!]
\def\arraystretch{1.0}
\begin{center}
\caption{Summary statistics of the control variables}
\label{tab:sum_cov}
\scriptsize
\scalebox{1.0}[1]{
\begin{tabular}{llD{.}{.}{-1}D{.}{.}{-1}D{.}{.}{-1}}
\toprule
&Unit&\multicolumn{1}{c}{Mean}&\multicolumn{1}{c}{Std. Dev.}&\multicolumn{1}{c}{$N$}\\\hline
Baseline control variables											&&&&\\
\hspace{10pt}Fetal death rate in the birth year						&County-birth year	&72.05	&15.16	&208\\
\hspace{10pt}Rice yield in the birth year								&County-birth year	&30.91	&4.36	&208\\
\hspace{10pt}Coverage of doctors									&County-birth year	&0.08	&0.07	&208\\
\hspace{10pt}School enrollment rate of the parental generation		&County-birth year	&87.19	&12.31	&208\\
Additional control variables											&&&&\\
\hspace{10pt}Primary school enrollment rate in the measured year	&Municipal-year		&99.62	&0.38	&3769\\
\hspace{10pt}Primary school attendance rate in the measured year	&County-year		&96.09	&0.93	&143\\\bottomrule
\end{tabular}
}
{\scriptsize
\begin{minipage}{420pt}
\setstretch{0.85}Notes: 
The fetal death rate is the number of still births per 1,000 births.
Rice yield is the volume of rice yield per 1 hectare. 
Coverage of doctors is the number of doctors per 100 people.
The primary school enrollment rate and attendance rate are the shares of enrolled and attended children relative to total school-aged children.\\
Sources: See Online~\ref{sec:secb}.
\end{minipage}
}
\end{center}
\end{table}
\section{Empirical Analysis Appendix}\label{sec:secc}
\setcounter{figure}{0} \renewcommand{\thefigure}{C.\arabic{figure}}
\setcounter{table}{0} \renewcommand{\thetable}{C.\arabic{table}}

\subsection{Trends in Height and Weight}\label{sec:secc_trend}

\begin{figure}[]
\centering
\subfloat[Boys' height (6--11)]{\label{fig:heightb_trend}\includegraphics[width=0.45\textwidth]{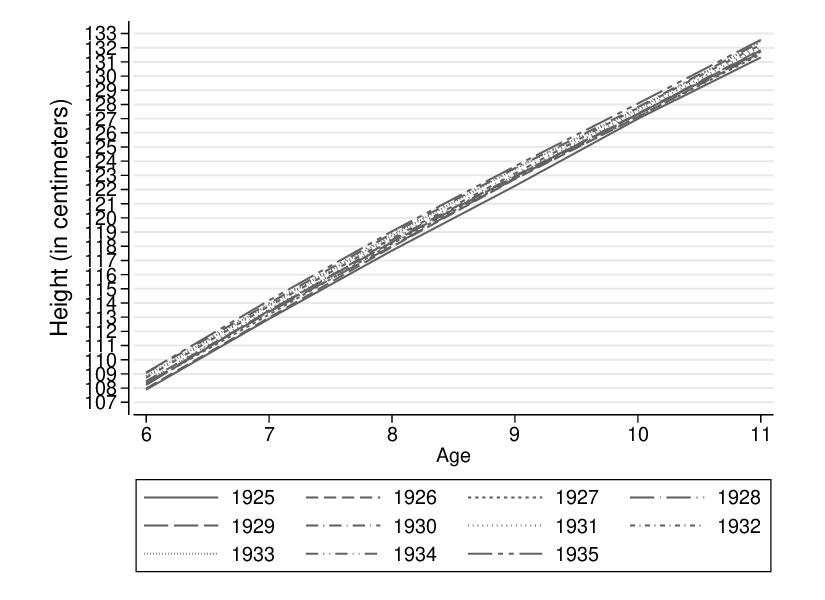}}
\subfloat[Girls' height (6--11)]{\label{fig:heightg_trend}\includegraphics[width=0.45\textwidth]{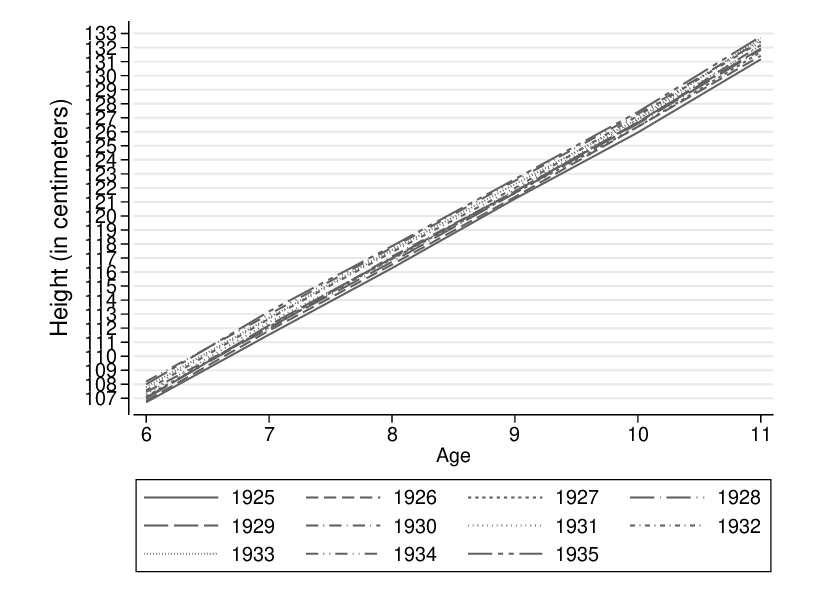}}\\
\subfloat[Boys' weight (6--11)]{\label{fig:weightb_trend}\includegraphics[width=0.45\textwidth]{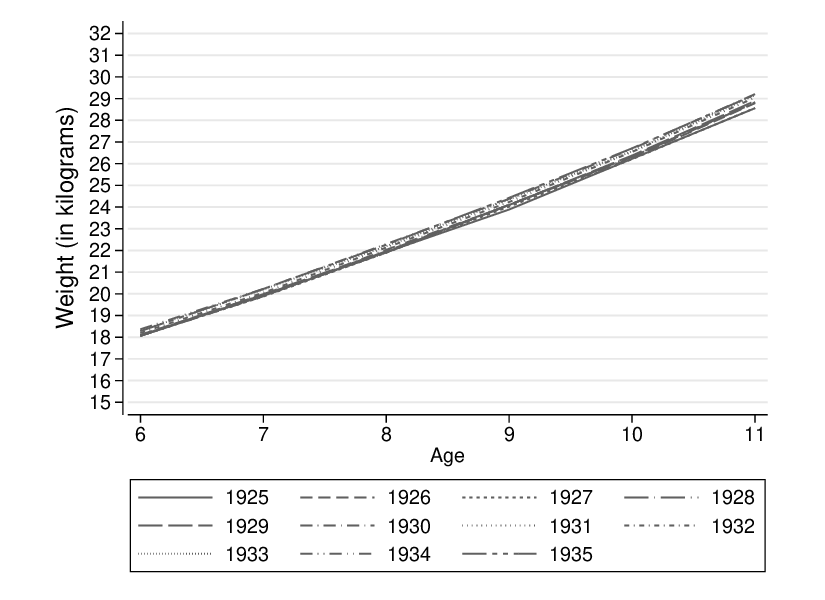}}
\subfloat[Girls' weight (6--11)]{\label{fig:weightg_trend}\includegraphics[width=0.45\textwidth]{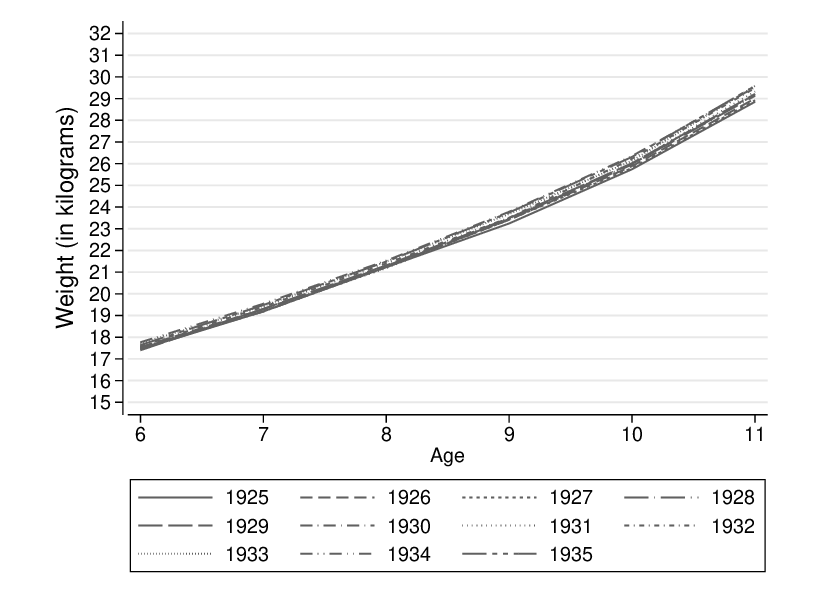}}
\caption{Trends in the Height and Weight of Children}
\label{fig:hw_trend}
\scriptsize{\begin{minipage}{400pt}
\setstretch{0.85}Note: Each figure shows the average height or weight of children in each measured year.\\
Sources: Calculated by author from the SPES (1925--1935 editions).
\end{minipage}}
\end{figure}

Figures~\ref{fig:heightb_trend} and \ref{fig:heightg_trend} present the heights of boys and girls by age and measured year, respectively.
Similarly, Figures~\ref{fig:weightb_trend} and \ref{fig:weightg_trend} present the weights of boys and girls, respectively. 
The trends of height and weight show near parallel translation over the measured year for both genders, suggesting that the trends in child development are similar during my sample periods.

\subsection{Robustness Exercises}\label{sec:alt_spec}

\subsubsection{Including Additional Area Interactions}\label{sec:alt_spec_area}

Table~\ref{tab:r_alt_sis6} shows the results from the alternative specification of equation~(\ref{eq1}) including the additional interaction terms between the affected cohort dummies and indicator variable for the JMA-SIS6 area.
The estimated coefficients on the interaction terms with respect to the JMA-SIS6 area are statistically insignificant in all cases.
This means that the effects of the earthquake on the affected cohorts are similar in the JMA-SIS5 and JMA-SIS6 areas.
This makes sense because the physical disruption in the JMA-SIS7 area was massive relative to the JMA-SIS5 and JMA-SIS6 areas (Table~\ref{tab:class}).

\begin{table}[]
\def\arraystretch{1.0}
\begin{center}
\captionsetup{justification=centering}
\caption{Effects of Fetal Earthquake Exposure on Height (cm) and Weight (kg):\\ Including the Additional Area Interaction Terms}
\label{tab:r_alt_sis6}
\footnotesize
\scalebox{1.0}[1]{
\begin{tabular}{lD{.}{.}{-2}D{.}{.}{-2}D{.}{.}{-2}D{.}{.}{-2}}
\toprule
&\multicolumn{2}{c}{Boys}&\multicolumn{2}{c}{Girls}\\
\cmidrule(rr){2-3}\cmidrule(rr){4-5}
&\multicolumn{1}{c}{(1)}&\multicolumn{1}{c}{(2)}&\multicolumn{1}{c}{(3)}&\multicolumn{1}{c}{(4)}\\
&\multicolumn{1}{c}{Ages 6--8}&\multicolumn{1}{c}{Ages 9--11}&\multicolumn{1}{c}{Ages 6--8}&\multicolumn{1}{c}{Ages 9--11}\\\hline
Panel A: Height									&&&&\\
\hspace{5pt}1923 birth cohort              			&-0.178$**$	&    -0.156$**$	&  -0.287$***$	&  -0.221$***$\\
                     										&[0.010]		&   [0.026]   	& [0.004]		& [0.003]\\
\hspace{5pt}1923 birth cohort $\times$ SIS6 	&0.228			&     0.128   		&   0.054		&   0.126\\
                       									&[0.192]		&   [0.585]		& [0.662] 		& [0.375]\\
\hspace{5pt}1923 birth cohort $\times$ SIS7 	&0.075			&     0.132		&  -0.303$**$  	&  -0.625$***$\\
                        									&[0.312]		&   [0.482]		& [0.034]   		& [0.007]\\
\hspace{5pt}Observations							
&\multicolumn{1}{c}{14,139}
&\multicolumn{1}{c}{14,133}
&\multicolumn{1}{c}{14,145}
&\multicolumn{1}{c}{14,138}\\
Panel B: Weight									&&&&\\
\hspace{5pt}1923 birth cohort            			&-0.065$**$	&    -0.042   &  -0.111$***$ 		&  -0.141$***$\\
                       		 							&[0.016]		&   [0.136]   & [0.006]			& [0.009]\\
\hspace{5pt}1923 birth cohort $\times$ SIS6 	&0.059			&     0.075   &   0.073   			&   0.188\\
                       		 							&[0.200]		&   [0.244]   & [0.340]  			& [0.235]\\
\hspace{5pt}1923 birth cohort $\times$ SIS7 	&0.015			&     0.113   &  -0.247$***$ 		&  -0.280$**$\\
                        									&[0.694]		&   [0.442]  & [0.002]   			& [0.025]\\
\hspace{5pt}Observations							
&\multicolumn{1}{c}{14,139}
&\multicolumn{1}{c}{14,133}
&\multicolumn{1}{c}{14,145}
&\multicolumn{1}{c}{14,138}\\\bottomrule
\end{tabular}
}
{\scriptsize
\begin{minipage}{410pt}
\setstretch{0.85}***, **, and * represent statistical significance at the 1\%, 5\%, and 10\% levels based on the $p$-values from the wild cluster bootstrap resampling method in brackets, respectively.
The data are clustered at the 13-county level in the bootstrap procedure.
The number of replications is fixed to 1,000 for all the specifications.\\
Notes:
The numbers of observations in columns (1)--(4) are $14,139$, $14,133$, $14,145$, and $14,138$, respectively.
The data on the school-year-age-level average values of height (cm) or weight (kg) are used in the regressions.
All the regressions include controls for the rice yield in the birth year; fetal death rate in the birth year; school enrollment rate of the parental generation; school-age-specific fixed effects; and year fixed effects.
\end{minipage}
}
\end{center}
\end{table}

\subsubsection{Testing the Spatial Correlations among Counties}\label{sec:alt_spatial}

The regressions reported in the main text are robust to the potential spatio-temporal autocorrelations within each county (i.e., cluster), heteroskedasticity within counties, and heteroskedasticity across counties.
In this subsection, I further test the spatial autocorrelations among clusters.
Table~\ref{tab:r_alt_spatial} presents the chi-square statistics from Moran tests for spatial dependencies across 13 counties.
As shown in the table, the obtained statistics are small and close to zero in most cases.
While a few cases exceed the reasonable criteria for a small sample size, they do not show systematic patterns and should not disturb my variance estimation in the regression.
Moreover, all the regressions presented in the main text include the school-age-specific fixed effects and year fixed effects as well as the control variables, which should further reduce potential spatial dependencies.
Considering this evidence, I can conclude that there are no systematic spatial autocorrelations across counties.

\begin{table}[]
\def\arraystretch{1.0}
\begin{center}
\captionsetup{justification=centering}
\caption{Moran Tests for Potential Spatial Autocorrelations}
\label{tab:r_alt_spatial}
\scriptsize
\scalebox{0.92}[1]{
\begin{tabular}{lcccccccccccc}
\toprule
&\multicolumn{6}{c}{Boys}&\multicolumn{6}{c}{Girls}\\
\cmidrule(rrrrrr){2-7}\cmidrule(rrrrrr){8-13}
Year&Age 6&Age 7&Age 8&Age 9&Age 10&Age 11&Age 6&Age 7&Age 8&Age 9&Age 10&Age 11\\\hline
Panel A: Height&&&&&&&&&&&&\\
1925 &	0.32 &	0.56 &	0.47 &	0.98 &	1.62  &	0.50 &	0.52 &	0.52 &	1.97 &	0.74 &	2.85 &	1.42  \\
1926 &	2.95 &	1.47 &	1.78 &	0.72 &	1.87  &	0.23 &	0.24 &	0.82 &	2.65 &	2.98 &	2.03 &	2.02  \\
1927 &	0.60 &	3.55 &	0.55 &	1.96 &	3.02  &	1.37 &	0.54 &	4.36 &	0.74 &	3.18 &	1.45 &	2.58  \\
1928 &	0.28 &	0.31 &	11.42$\dag$&	5.27 &	10.87$\dag$ &	8.09 &	0.09 &	0.41 &	3.50 &	1.45 &	14.8 &	9.35  \\
1929 &	0.28 &	0.32 &	0.22 &	4.86 &	5.04  &	7.58 &	0.67 &	0.33 &	0.21 &	6.23 &	9.15 &	10.92$\dag$ \\
1930 &	0.30 &	0.48 &	0.72 &	0.72 &	1.40  &	6.45 &	0.88 &	0.56 &	0.61 &	0.67 &	5.71 &	5.52  \\
1931 &	0.23 &	0.19 &	0.65 &	0.55 &	0.57  &	2.23 &	0.12 &	0.39 &	0.78 &	0.23 &	0.59 &	5.84  \\
1932 &	0.01 &	0.21 &	0.43 &	0.46 &	0.25  &	0.69 &	0.20 &	0.04 &	0.53 &	0.62 &	0.24 &	0.62  \\
1933 &	0.05 &	0.07 &	0.21 &	0.21 &	0.51  &	0.44 &	0.06 &	0.55 &	0.24 &	0.31 &	0.35 &	0.21  \\
1934 &	0.23 &	0.40 &	0.35 &	0.33 &	0.38  &	0.56 &	0.01 &	0.32 &	0.84 &	0.01 &	0.61 &	0.57  \\
1935 &	0.08 &	0.49 &	0.09 &	0.06 &	0.61  &	0.29 &	0.01 &	0.15 &	0.34 &	0.00 &	0.41 &	0.61  \\
&&&&&&&&&&&&\\
Panel B: Weight&&&&&&&&&&&&\\
1925 &	0.03 &	2.04 &	1.86 &	1.30 &	0.45 &	0.42 &	0.09 &	1.77 &	3.23 &	1.44 &	1.44 &	0.53\\
1926 &	1.96 &	0.09 &	0.43 &	0.89 &	0.05 &	0.20 &	1.73 &	0.41 &	2.23 &	1.78 &	0.96 &	4.31\\
1927 &	0.15 &	0.71 &	2.51 &	1.39 &	0.40 &	3.49 &	0.05 &	7.29 &	2.09 &	2.40 &	0.92 &	2.02\\
1928 &	0.38 &	0.11 &	8.34 &	0.68 &	9.89 &	4.87 &	0.13 &	0.43 &	0.35 &	0.29 &	9.69 &	2.91\\
1929 &	0.13 &	0.13 &	0.25 &	4.06 &	5.27 &	1.27 &	0.40 &	0.16 &	0.29 &	1.59 &	0.42 &	5.67\\
1930 &	0.27 &	0.88 &	0.99 &	0.01 &	7.26 &	6.34 &	0.23 &	0.45 &	0.93 &	0.69 &	2.65 &	4.73\\
1931 &	0.19 &	0.06 &	0.55 &	0.32 &	0.31 &	0.85 &	0.25 &	0.03 &	0.45 &	0.27 &	0.49 &	3.59\\
1932 &	0.06 &	0.20 &	0.47 &	0.44 &	0.47 &	0.45 &	0.16 &	0.55 &	0.42 &	0.51 &	0.53 &	0.53\\
1933 &	0.59 &	0.28 &	0.01 &	0.40 &	0.15 &	0.36 &	0.08 &	0.87 &	0.04 &	0.38 &	0.19 &	0.39\\
1934 &	0.65 &	0.40 &	0.33 &	0.35 &	0.39 &	0.80 &	0.65 &	0.45 &	0.49 &	0.27 &	0.41 &	0.59\\
1935 &	0.21 &	0.12 &	0.29 &	0.36 &	0.10 &	0.63 &	0.14 &	0.32 &	0.22 &	0.30 &	0.25 &	0.59\\\bottomrule
\end{tabular}
}
{\scriptsize
\begin{minipage}{450pt}
\setstretch{0.85}
The value in each cell indicates the chi-square statistic from the Moran test for spatial autocorrelation across 13 counties in Chiba prefecture.
$\dag$ represents statistical significance at the 0.1\% level.\\
Notes:
The data on the county-level average values of height (cm) or weight (kg) are used in the regressions.
The test is based on the residuals calculated from the specification that regresses either height or weight on the indicator variables for SIS6 and SIS7 areas and the control variables used in the main text.
Those controls include the rice yield in the birth year, fetal death rate in the birth year, and school enrollment rate of the parental generation.
The number of observations in each regression is 13 (counties).
\end{minipage}
}
\end{center}
\end{table}

\subsubsection{Alternative Definitions of the Developmental-Stage Age Bin and Railway Disruption}\label{sec:alt_spec_def}

In the main text, I systematically divided the age bins into 6--8 and 9--11 to estimate the impact of the earthquake on the (conditional) mean height at each age bin.
This is because the observed growth patterns of height are mostly straight in the primary school ages, as shown in Figure~\ref{fig:hw_trend}.
Although one must be careful while simply applying the modern growth reference to historical child growth, these bins are also consistent with the fact that the height-for-age z-scores of $-2$ in the modern WHO growth reference show similar growth curves to those children.\footnote{de Onis et al., `Development'.}
In Table~\ref{tab:r_alt_age_rail}, I check the sensitivity to this definition of developmental-stage age bins by slightly changing the bins to 6--9 and 8--11.
The estimated coefficients are similar to the results reported in Table~\ref{tab:r_main}, supporting the robustness of my baseline results.

In Table~\ref{tab:r_rail_alt}, I check the sensitivity of my results from equation~(\ref{eq3}) using the same sample used in Panel A of Table~\ref{tab:r_rail} but changing the radius to a 1.5 km range.
In Panels A and B of Table~\ref{tab:r_rail_alt}, I set \textit{Rail} in equation~(\ref{eq3}) as an indicator variable for municipalities within 8.5 km and 11.5 km of any municipality including either the Noda line or the Kururi line, respectively.
The results are largely unchanged from those reported in Table~\ref{tab:r_rail}, supporting the robustness of my baseline results.

\begin{table}[]
\def\arraystretch{1.0}
\begin{center}
\captionsetup{justification=centering}
\caption{Effects of Fetal Earthquake Exposure on Height (cm) and Weight (kg): \\Alternative Definitions of the Developmental-Stage Age Bins}
\label{tab:r_alt_age_rail}
\footnotesize
\scalebox{1.0}[1]{
\begin{tabular}{lD{.}{.}{-2}D{.}{.}{-2}D{.}{.}{-2}D{.}{.}{-2}}
\toprule
&\multicolumn{2}{c}{Boys}&\multicolumn{2}{c}{Girls}\\
\cmidrule(rr){2-3}\cmidrule(rr){4-5}
&\multicolumn{1}{c}{(1)}&\multicolumn{1}{c}{(2)}&\multicolumn{1}{c}{(3)}&\multicolumn{1}{c}{(4)}\\
&\multicolumn{1}{c}{Ages 6--9}&\multicolumn{1}{c}{Ages 8--11}&\multicolumn{1}{c}{Ages 6--9}&\multicolumn{1}{c}{Ages 8--11}\\\hline

Panel A: Effects on height			&&&&\\
1923 birth cohort					&-0.133$**$	&-0.116$**$	&-0.222$***$	&-0.199$***$	\\
									&[0.034]		&[0.037]		&[0.002]		&[0.002]		\\
1923 birth cohort $\times$ SIS7		&0.066			&0.129			&-0.291$**$	&-0.631$*$		\\
									&[0.244]		&[0.481]		&[0.012]		&[0.072]		\\
\hspace{5pt}Observations							
&\multicolumn{1}{c}{18,851}
&\multicolumn{1}{c}{18,845}
&\multicolumn{1}{c}{18,857}
&\multicolumn{1}{c}{18,848}\\
&&&&\\
Panel B: Effects on weight			&&&&\\
1923 birth cohort					&-0.041$*$		&-0.034	&-0.087$**$	&-0.100$***$	\\
									&[0.064]		&[0.229]	&[0.002]		&[0.002]		\\
1923 birth cohort $\times$ SIS7		&0.031$*$		&0.096		&-0.257$*$		&-0.266$*$		\\
									&[0.056]		&[0.427]	&[0.022]		&[0.084]		\\
\hspace{5pt}Observations							
&\multicolumn{1}{c}{18,851}
&\multicolumn{1}{c}{18,845}
&\multicolumn{1}{c}{18,857}
&\multicolumn{1}{c}{18,848}\\\bottomrule
\end{tabular}
}
{\scriptsize
\begin{minipage}{410pt}
\setstretch{0.85}
***, **, and * represent statistical significance at the 1\%, 5\%, and 10\% levels based on the $p$-values from the wild cluster bootstrap resampling method in brackets, respectively.
The data are clustered at the 13-county level in the bootstrap procedure.
The number of replications is fixed to 1,000 for all the specifications.\\
Notes:
The numbers of observations in columns (1)--(4) are $18,851$, $18,845$, $18,857$, and $18,848$, respectively.
The data on the school-year-age-level average values of height (cm) or weight (kg) are used in the regressions.
All the regressions include controls for the rice yield in the birth year; fetal death rate in the birth year; school enrollment rate of the parental generation; school-age-specific fixed effects; and year fixed effects.
\end{minipage}
}
\end{center}
\end{table}
\begin{table}[h!]
\def\arraystretch{1.0}
\begin{center}
\captionsetup{justification=centering}
\caption{Effects of Fetal Earthquake Exposure on Height (cm) and Weight (kg) in the Limited JMA-SIS5 Area: Alternative Definitions of Railway Network Disruption}
\label{tab:r_rail_alt}
\footnotesize
\scalebox{0.91}[1]{
\begin{tabular}{lD{.}{.}{-2}D{.}{.}{-2}D{.}{.}{-2}D{.}{.}{-2}}
\toprule
&\multicolumn{2}{c}{Boys}&\multicolumn{2}{c}{Girls}\\
\cmidrule(rr){2-3}\cmidrule(rr){4-5}
&\multicolumn{1}{c}{(1)}&\multicolumn{1}{c}{(2)}&\multicolumn{1}{c}{(3)}&\multicolumn{1}{c}{(4)}\\
&\multicolumn{1}{c}{Ages 6--8}&\multicolumn{1}{c}{Ages 9--11}&\multicolumn{1}{c}{Ages 6--8}&\multicolumn{1}{c}{Ages 9--11}\\\hline

Panel A-1: Effects on height								&&&&\\
1923 birth cohort										&-0.114$*$	&-0.141	&-0.290$***$	&-0.202$**$\\
														&[0.076]	&[0.198]	&[0.004]		&[0.013]	\\
1923 birth cohort $\times$ Railway disruption (8.5km)	&-0.117	&0.525		&0.224			&0.668		\\
														&[0.942]	&[0.546]	&[0.150]		&[0.529]	\\
\hspace{5pt}Observations							
&\multicolumn{1}{c}{9,352}
&\multicolumn{1}{c}{9,348}
&\multicolumn{1}{c}{9,353}
&\multicolumn{1}{c}{9,347}\\
Panel A-2: Effects on weight								&&&&\\
1923 birth cohort										&-0.053$*$	&-0.021	&-0.108$***$	&-0.108$**$\\
														&[0.094]	&[0.736]	&[0.010]		&[0.023]	\\
1923 birth cohort $\times$ Railway disruption (8.5km)	&0.044		&0.018		&0.073			&0.126		\\
														&[0.644]	&[0.890]	&[0.330]		&[0.693]	\\
\hspace{5pt}Observations							
&\multicolumn{1}{c}{9,352}
&\multicolumn{1}{c}{9,348}
&\multicolumn{1}{c}{9,353}
&\multicolumn{1}{c}{9,347}\\
														&&&&\\
Panel B-1: Effects on height								&&&&\\
1923 birth cohort										&-0.134$**$&-0.120	&-0.289$***$	&-0.181$**$\\
														&[0.038]	&[0.234]	&[0.004]		&[0.017]	\\
1923 birth cohort $\times$ Railway disruption (11.5km)	&0.115		&0.154		&0.153			&0.261		\\
														&[0.818]	&[0.590]	&[0.316]		&[0.767]	\\
\hspace{5pt}Observations							
&\multicolumn{1}{c}{9,352}
&\multicolumn{1}{c}{9,348}
&\multicolumn{1}{c}{9,353}
&\multicolumn{1}{c}{9,347}\\
Panel B-2: Effects on weight								&&&&\\
1923 birth cohort										&-0.052$*$	&-0.006	&-0.109$***$	&-0.093$*$	\\
														&[0.062]	&[0.916]	&[0.010]		&[0.075]	\\
1923 birth cohort $\times$ Railway disruption (11.5km)	&0.022		&-0.135	&0.058			&-0.061	\\
														&[0.960]	&[0.208]	&[0.384]		&[0.737]	\\
\hspace{5pt}Observations							
&\multicolumn{1}{c}{9,352}
&\multicolumn{1}{c}{9,348}
&\multicolumn{1}{c}{9,353}
&\multicolumn{1}{c}{9,347}\\\bottomrule

\end{tabular}
}
{\scriptsize
\begin{minipage}{440pt}
\setstretch{0.85}
***, **, and * represent statistical significance at the 1\%, 5\%, and 10\% levels based on the $p$-values from the wild cluster bootstrap resampling method in brackets, respectively.
The data are clustered at the 13-county level in the bootstrap procedure.
The number of replications is fixed to 1,000 for all specifications.\\
Notes:
Samples include municipalities receiving no disaster relief in the SIS5 area.
The numbers of observations in columns (1)--(4) are 9352, 9348, 9353, and 9347, respectively.
The data on the school-year-age-level average values of height (cm) or weight (kg) are used in the regressions.
All the regressions include controls for the rice yield in the birth year; fetal death rate in the birth year; school enrollment rate of the parental generation; school-age-specific fixed effects; and year fixed effects.
\end{minipage}
}
\end{center}
\end{table}

\subsubsection{Results for the BMI}\label{sec:alt_spec_bmi}

Table~\ref{tab:alt_spec_bmi} presents the results from the specification using the BMI as a dependent variable in the baseline specification of equation~(\ref{eq1}).
The estimates are statistically insignificant in all the regressions.
This result is considered to be consistent with my main result presented in Table~\ref{tab:r_main}, indicating that both heights and weights are reduced by fetal exposure to the earthquake.
In addition, I have not used body mass index (BMI) as the primary measure of child obesity because child growth spurts disturbs the measurement of obesity at different ages, making it difficult to identify whether the observed child stunting is due to fetal shocks or just the timing of children's growth.\footnote{See Schneider, `Sample' for a detailed explanation of this mechanism.}

\begin{table}[]
\def\arraystretch{1.0}
\begin{center}
\captionsetup{justification=centering}
\caption{Effects of Fetal Earthquake Exposure on the Body Mass Index (BMI)}
\label{tab:alt_spec_bmi}
\footnotesize
\scalebox{1.0}[1]{
\begin{tabular}{lD{.}{.}{-2}D{.}{.}{-2}D{.}{.}{-2}D{.}{.}{-2}D{.}{.}{-2}D{.}{.}{-2}D{.}{.}{-2}D{.}{.}{-2}}
\toprule
&\multicolumn{2}{c}{Boys}&\multicolumn{2}{c}{Girls}\\
\cmidrule(rr){2-3}\cmidrule(rr){4-5}
&\multicolumn{1}{c}{(1)}&\multicolumn{1}{c}{(2)}&\multicolumn{1}{c}{(3)}&\multicolumn{1}{c}{(4)}\\
&\multicolumn{1}{c}{Ages 6--8}&\multicolumn{1}{c}{Ages 9--11}&\multicolumn{1}{c}{Ages 6--8}&\multicolumn{1}{c}{Ages 9--11}\\
Mean BMI&\multicolumn{1}{c}{15.6}&\multicolumn{1}{c}{16.3}&\multicolumn{1}{c}{15.4}&\multicolumn{1}{c}{16.3}\\\hline
\hspace{5pt}1923 birth cohort						&-0.001		&0.013		&-0.005	&-0.024		\\
												&[0.930]		&[0.362]		&[0.858]	&[0.127]		\\
\hspace{5pt}1923 birth cohort $\times$ SIS7			&-0.009		&0.031		&-0.129	&-0.032		\\
												&[0.784]		&[0.458]		&[0.182]	&[0.765]		\\
\hspace{5pt}Observations							
&\multicolumn{1}{c}{14,139}
&\multicolumn{1}{c}{14,133}
&\multicolumn{1}{c}{14,145}
&\multicolumn{1}{c}{14,138}		\\\bottomrule
\end{tabular}
}
{\scriptsize
\begin{minipage}{350pt}
\setstretch{0.85}
***, **, and * represent statistical significance at the 1\%, 5\%, and 10\% levels based on the $p$-values from the wild cluster bootstrap resampling method in brackets, respectively.
The data are clustered at the 13-county level in the bootstrap procedure.
The number of replications is fixed to 1,000 for all the specifications.\\
Notes:
The numbers of observations in columns (1)--(4) are $14,139$, $14,133$, $14,145$, and $14,138$, respectively.
The data on the school-year-age-level average values of height (cm) or weight (kg) are used in the regressions.
All the regressions include controls for the rice yield in the birth year; fetal death rate in the birth year; school enrollment rate of the parental generation; school-age-specific fixed effects; and year fixed effects.
\end{minipage}
}
\end{center}
\end{table}

\subsubsection{Including Additional Control Variables}\label{sec:alt_spec_add}

First, I include two additional variables that proxy for the proportion of children with health issues in my baseline specification.
Panel A of Table~\ref{tab:r_robustness1} presents the results from the specification that includes both the municipal-year-level primary school enrollment rate and the county-year-level primary school attendance rate.
The results are largely unchanged, which is consistent with the fact that the school enrollment rate and school attendance rate were stable around 99\% and 95\%, respectively (Table~\ref{tab:sum_cov}).

Second, I consider the potential impacts of other historical events in my baseline specification.
In Japan, Spanish influenza cases occurred between August 1918 and July 1920 and its intensity (measured by the number of deaths) spiked in November 1918 and January 1920.\footnote{\setstretch{0.84}Representative historical studies investigating Spanish influenza in Japan include Rice and Palmer, `Pandemic' and Hayami, \textit{Spanish}. Although the rice riots of 1918 might also have affected regional food prices, the riots in Chiba prefecture were small and thus had a negligible impact on the regional food prices; Shoji, \textit{Study}, p. 36).}
This feature of pandemics implies that a large proportion of 1919--1920 birth cohorts in my sample might have been affected by pandemic flu \textit{in utero}.
Another important event during the sample period that might have influenced children's health is the First World War.
In Panel B of Table~\ref{tab:r_robustness1}, I include two indicator variables for these potentially affected cohorts.\footnote{\setstretch{0.84}
I include a 1919--1920 birth cohort dummy to control for the potential pandemic cohort effects. Regarding the wartime cohort effects, I include an indicator variable for the children \textit{in utero} in wartime between 1915 and 1919 to control for the potential long-term impacts of wartime shocks on health outcomes. When I include these indicators, the coefficients on the 1923 and 1924 birth cohort variables are all measured relative to the small number of reference cohorts, which loses the advantage of averaging across many reference cohorts. In other words, if the coefficients are estimated relative to few reference cohorts, one loses the ability to see whether outliers are generated because the earthquake cohort looks unique or because the small set of cohorts serves as a baseline. Given this result, I do not include the indicator variables for the pandemic and wartime cohorts in my baseline specifications (equations~(\ref{eq1})—(\ref{eq3})).}
As shown, the estimated coefficients on the influenza pandemic and wartime birth cohorts dummies are statistically insignificant in all cases.
This result is consistent with the fact that the influenza epidemics in Chiba prefecture had not been severe during the epidemics in 1918--1920; Statistics Bureau of the Cabinet, \textit{Nihonteikoku shiint\=okei [1918--1920 editions]}.
As for the impacts of the First World War, it is widely accepted in Japanese history that the war did not change the daily lives of children in Japan and thus did not influence their growth patterns; Kudo et al. `Growth'.
These findings suggest that there is no significant difference in these cohorts and reference birth cohorts in this setting.

In the same way, I show the results from specifications including the same additional control variables in equations~(\ref{eq2}) and (\ref{eq3}) in Tables~\ref{tab:rob_relief} and \ref{tab:rob_rail}, respectively.
In Table~\ref{tab:rob_relief}, I report the results for the boys aged 6--8 given that all the estimates for the other subsamples are statistically insignificant (Table~\ref{tab:r_relief}).
As shown in Table~\ref{tab:rob_relief}, the results are largely unchanged from those reported in Tables~\ref{tab:r_relief}.
Similarly, Table~\ref{tab:rob_rail} confirms that the results are unchanged from those reported in Table~\ref{tab:r_rail}.

\subsubsection{Time Series Plots of the Birth and Mortality Measures}\label{sec:selection}

\begin{figure}[h]
\centering
\subfloat[Fetal death rate]{\label{fig:tsfdr}\includegraphics[width=0.32\textwidth]{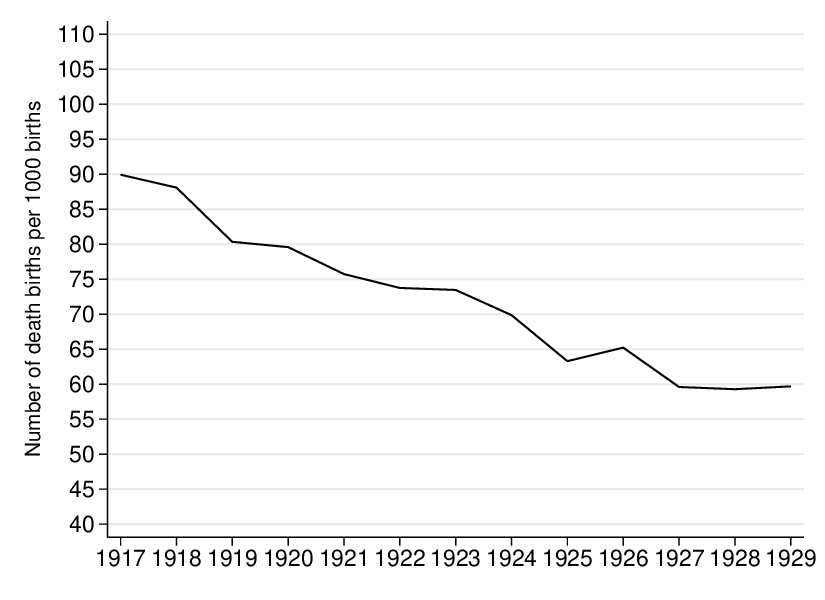}}
\subfloat[Infant mortality rate]{\label{fig:tsimr}\includegraphics[width=0.32\textwidth]{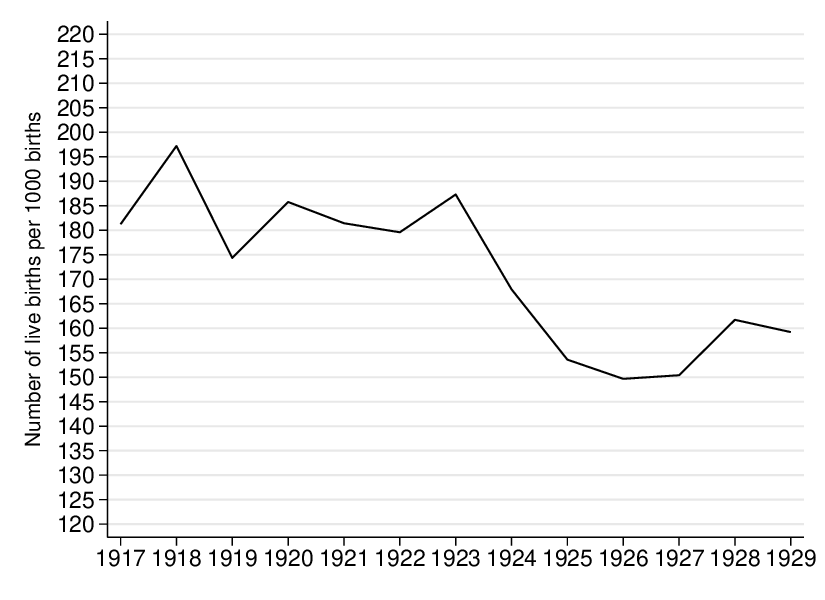}}
\subfloat[Crude birth rate]{\label{fig:tscbr}\includegraphics[width=0.32\textwidth]{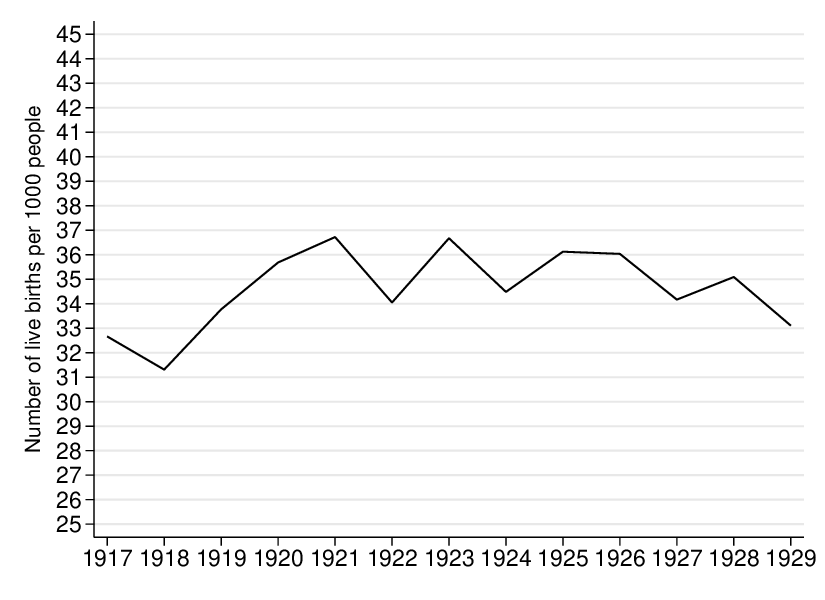}}
\caption{Fetal Death, Infant Mortality, and Fertility Rates in Chiba Prefecture}
\label{fig:ts_selection}
\scriptsize{\begin{minipage}{450pt}
\setstretch{0.85}Notes:
The fetal death rate is the number of deaths births per 1000 births.
The infant mortality rate is the number of infant deaths (i.e., deaths within 12 months of birth) per 1000 live births.
The crude birth rate is the number of live births per 1000 people.\\
Sources: Statistics Bureau of the Cabinet, \textit{Nihonteikoku jink\=od\=otai [1917--1929 editions]}.
\end{minipage}}
\end{figure}

Figure~\ref{fig:ts_selection} shows the time series plots of the fetal death, infant mortality, and crude death rates between 1917 and 1929.
The data on these rates are obtained from the 1917--1929 editions of the Vital Statistics of Japan; Statistics Bureau of the Cabinet, \textit{Nihonteikoku jink\=od\=otai [1917--1929 editions]}.
Figure~\ref{fig:tsfdr} and~\ref{fig:tsimr} show the decreasing trend of the risk of fetal deaths and infant mortality during this period, as also shown by Schneider and Ogasawara, `Disease'.

Regarding Figure~\ref{fig:tsimr}, although the infant mortality rate seems to increase slightly in 1923, the increment was not greater as that observed in 1918, suggesting it is not systematic mortality selection due to the earthquake.
If we apply a simple regression for the first differenced infant mortality rates (IMR) between 1917 and 1929, the estimated coefficient on the 1923 year dummy is positive but insignificant (standard errors in parentheses; $N=12$):
 \begin{eqnarray}\label{eq3}
\footnotesize{
\Delta \text{IMR}_{t} = -2.7(3.8) + 10.4(13.1) I(\text{Year}=1923)_{t}.
}
\end{eqnarray}
The deaths of infants before September 1 (i.e., between January 1 and August 31) were unrelated to the earthquake, whereas those in and after September (i.e., between September 1 and December 31) should have been exposed to the earthquake when they were alive. Considering this, the obtained result is considered to be consistent with the findings that the direct nutritional effects on children might have been less influential on average (Subsections~\ref{sec:sec22} and~\ref{sec:sec512}).

Another important phenomenon in terms of sample selection may be fertility selection after the disaster; Dehejia and Lleras-Muney, `Booms'.
The fertility rate may involuntarily decrease during the aftermath of the disaster because some of the population could not afford to have children in such abnormal circumstances.
In other words, children born during the aftermath of the disaster might be more likely to belong in households that can provide sufficient compensation for any adverse health effects in the circumstances.
Although my estimates would understate the adverse health effects if such fertility selection behavior had been common among parents, Figure~\ref{fig:tscbr} does not show any systematic reduction in the crude birth rate during or after the year in which the earthquake hit.

\subsubsection{Testing the Cohort Effects of Surrounding Cohorts}\label{sec:alt_spec_placebo}

Evidence of the long-term effects of earthquake exposure suggests that postnatal exposure, especially by the second year of life, to the Great Earthquake could also have adverse effects on human capital accumulation; Caruso and Miller, `Long run'.
In Table~\ref{tab:r_robustness2}, I include an indicator of the 1922 birth cohort, which includes children who experienced the earthquake roughly between six months old and two years old, in my baseline specification to test the potential effects of postnatal exposure.\footnote{\setstretch{0.84}Children born in January 1922 were exposed to the earthquake at one years and nine months old, whereas children born in March 1923 were exposed at seven months old.}
While I find robust negative effects on the 1923 birth cohort, I find little evidence of adverse health effects on the 1922 birth cohort, implying that postnatal exposure to the earthquake might not matter.
This finding is consistent with the evidence of the recent study by Rosales-Rueda and Triyana, who show that in-utero exposure to the Indonesian forest fires of 1997 had stunting effects on children aged 10 and 17 years, while postnatal exposure at two years did not have such persistent stunting effects.\footnote{Rosales-Rueda and Triyana, `The persistent'.}
As discussed, children born between April and August 1923 included in the 1923 birth cohort experienced postnatal earthquake exposure in infancy.
These data may thus complicate the interpretation of the estimates of the 1923 birth cohort variables, which could include potential negative effects via the postnatal health shock.
However, my results on the negligible influence of postnatal exposure support the evidence that the estimated effects on the 1923 birth cohort variables in my baseline results should capture the adverse effects of prenatal exposure to the earthquake rather than postnatal exposure.

In Table~\ref{tab:r_robustness2}, I also include an indicator of the 1924 birth cohort to run a placebo test.
Since most of the 1924 birth cohort was not exposed to the earthquake of 1923, the estimated coefficient on this indicator variable should be statistically insignificant.
If the estimates were statistically significantly negative, the observed stunting effects could be systematic decreasing trends in the height or weight of children.
As shown in Table~\ref{tab:r_robustness2}, the estimates are all statistically insignificant, indicating that the estimated stunting effects do not capture pre-trends in height and weight.
One can also argue that if the estimated stunting effects do not capture such trends and can be regarded as the consequences of fetal earthquake exposure, the mean heights and weights of children born in the surrounding years should be similar after excluding the 1923 birth cohort from the sample.
Each estimate of the cohort effect in Table~\ref{tab:r_placebo} is obtained from a regression for the sample excluding the 1923 birth cohort, which means that I run 56 regressions to obtain these estimates.
As shown, the estimated coefficients are not statistically significant in most cases, which suggests that the mean heights and weights of children born between 1918 and 1925 (excluding the 1923 birth cohort) are indeed similar in the statistical sense using the specification including each surrounding birth cohort dummy.

\begin{table}[]
\def\arraystretch{0.95}
\begin{center}
\captionsetup{justification=centering}
\caption{Effects of Fetal Earthquake Exposure on Height (cm) and Weight (kg): \\Robustness of including Additional Control Variables}
\label{tab:r_robustness1}
\footnotesize
\scalebox{1.0}[1]{
\begin{tabular}{lD{.}{.}{-2}D{.}{.}{-2}D{.}{.}{-2}D{.}{.}{-2}}
\toprule
&\multicolumn{2}{c}{Boys}&\multicolumn{2}{c}{Girls}\\
\cmidrule(rr){2-3}\cmidrule(rr){4-5}
&\multicolumn{1}{c}{(1)}&\multicolumn{1}{c}{(2)}&\multicolumn{1}{c}{(3)}&\multicolumn{1}{c}{(4)}\\
&\multicolumn{1}{c}{Ages 6--8}&\multicolumn{1}{c}{Ages 9--11}&\multicolumn{1}{c}{Ages 6--8}&\multicolumn{1}{c}{Ages 9--11}\\\hline
Panel A-1: Effects on height									&&&&\\
\hspace{5pt}1923 birth cohort								&-0.146$**$	&-0.137$**$	&-0.279$***$	&-0.203$***$	\\
															&[0.024]		&[0.038]		&[0.002]		&[0.003]		\\
\hspace{5pt}1923 birth cohort $\times$ SIS7				&0.042			&0.113			&-0.308$**$	&-0.646$**$	\\
															&[0.750]		&[0.482]		&[0.012]		&[0.005]		\\
\hspace{5pt}Primary school enrollment rate					&-0.080		&0.056			&-0.036		&0.003			\\
															&[0.126]		&[0.464]		&[0.506]		&[0.951]		\\
\hspace{5pt}Primary school attendance rate					&-0.028		&-0.007		&0.042			&-0.028		\\
															&[0.452]		&[0.802]		&[0.174]		&[0.379]		\\
\hspace{5pt}Observations							
&\multicolumn{1}{c}{14,139}
&\multicolumn{1}{c}{14,133}
&\multicolumn{1}{c}{14,145}
&\multicolumn{1}{c}{14,138}		\\
Panel A-2: Effects on weight									&&&&\\
\hspace{5pt}1923 birth cohort								&-0.057$**$	&-0.031	&-0.101$***$	&-0.114$**$	\\
															&[0.044]		&[0.278]	&[0.004]		&[0.011]		\\
\hspace{5pt}1923 birth cohort $\times$ SIS7				&0.006			&0.105		&-0.257$***$	&-0.308$**$	\\
															&[0.908]		&[0.480]	&[0.002]		&[0.019]		\\
\hspace{5pt}Primary school enrollment rate					&-0.019		&-0.019	&0.016			&-0.003		\\
															&[0.552]		&[0.340]	&[0.500]		&[0.941]		\\
\hspace{5pt}Primary school attendance rate					&-0.004		&0.029$*$	&0.012			&-0.005		\\
															&[0.806]		&[0.078]	&[0.264]		&[0.777]		\\
\hspace{5pt}Observations							
&\multicolumn{1}{c}{14,139}
&\multicolumn{1}{c}{14,133}
&\multicolumn{1}{c}{14,145}
&\multicolumn{1}{c}{14,138}		\\\hline
&\multicolumn{2}{c}{Boys}&\multicolumn{2}{c}{Girls}\\
\cmidrule(rr){2-3}\cmidrule(rr){4-5}
&\multicolumn{1}{c}{(1)}&\multicolumn{1}{c}{(2)}&\multicolumn{1}{c}{(3)}&\multicolumn{1}{c}{(4)}\\
&\multicolumn{1}{c}{Ages 6--8}&\multicolumn{1}{c}{Ages 9--11}&\multicolumn{1}{c}{Ages 6--8}&\multicolumn{1}{c}{Ages 9--11}\\\hline
Panel B-1: Effects on height									&&&&\\
\hspace{5pt}1923 birth cohort								&-0.149$**$	&-0.138$**$	&-0.280$***$	&-0.202$***$	\\
															&[0.022]		&[0.040]		&[0.002]		&[0.003]		\\
\hspace{5pt}1923 birth cohort $\times$ SIS7				&0.044			&0.114			&-0.311$***$	&-0.645$***$	\\
															&[0.734]		&[0.482]		&[0.004]		&[0.007]		\\
\hspace{5pt}Wartime birth cohorts							&0.035			&-0.005		&0.068			&0.156$*$		\\
															&[0.302]		&[0.898]		&[0.446]		&[0.075]		\\
\hspace{5pt}Influenza pandemic birth cohorts				&-0.096		&-0.022		&-0.020		&0.051			\\
															&[0.260]		&[0.784]		&[0.784]		&[0.565]		\\
\hspace{5pt}Observations							
&\multicolumn{1}{c}{14,139}
&\multicolumn{1}{c}{14,133}
&\multicolumn{1}{c}{14,145}
&\multicolumn{1}{c}{14,138}		\\
Panel B-2: Effects on weight									&&&&\\
\hspace{5pt}1923 birth cohort								&-0.057$**$	&-0.032	&-0.101$***$	&-0.113$**$	\\
															&[0.038]		&[0.280]	&[0.004]		&[0.011]		\\
\hspace{5pt}1923 birth cohort $\times$ SIS7				&0.007			&0.102		&-0.257$***$	&-0.310$**$	\\
															&[0.912]		&[0.476]	&[0.002]		&[0.019]		\\
\hspace{5pt}Wartime birth cohorts							&-0.041$*$		&0.050$*$	&-0.031		&0.015			\\
															&[0.066]		&[0.090]	&[0.386]		&[0.781]		\\
\hspace{5pt}Influenza pandemic birth cohorts				&-0.012		&-0.011	&-0.010		&0.078			\\
															&[0.682]		&[0.520]	&[0.764]		&[0.129]		\\
\hspace{5pt}Observations							
&\multicolumn{1}{c}{14,139}
&\multicolumn{1}{c}{14,133}
&\multicolumn{1}{c}{14,145}
&\multicolumn{1}{c}{14,138}		\\\bottomrule
\end{tabular}
}
{\scriptsize
\begin{minipage}{430pt}
\setstretch{0.85}
***, **, and * represent statistical significance at the 1\%, 5\%, and 10\% levels based on the $p$-values from the wild cluster bootstrap resampling method in brackets, respectively.
The data are clustered at the 13-county level in the bootstrap procedure.
The number of replications is fixed to 1,000 for all the specifications.\\
Notes:
The numbers of observations in columns (1)--(4) are $14,139$, $14,133$, $14145$, and $14138$, respectively.
The data on the school-year-age-level average values of height (cm) or weight (kg) are used in the regressions.
All the regressions include controls for the rice yield in the birth year; fetal death rate in the birth year; school enrollment rate of the parental generation; school-age-specific fixed effects; and year fixed effects.
\end{minipage}
}
\end{center}
\end{table}
\begin{landscape}
\begin{table}[h!]
\def\arraystretch{1.0}
\begin{center}
\captionsetup{justification=centering}
\caption{Effects of Disaster Relief on the Height (cm) and Weight (kg) of Boys Aged 6--8: \\Robustness of Including Additional Control Variables}
\label{tab:rob_relief}
\footnotesize
\scalebox{0.90}[1]{
\begin{tabular}{lcccccccc}
\toprule
&\multicolumn{4}{c}{Height}&\multicolumn{4}{c}{Weight}\\
\cmidrule(rrrr){2-5}\cmidrule(rrrr){6-9}
&\multicolumn{1}{c}{(1) Total}&\multicolumn{1}{c}{(2) Food}&\multicolumn{1}{c}{(3) Shelter}&\multicolumn{1}{c}{(4) Medical}
&\multicolumn{1}{c}{(5) Total}&\multicolumn{1}{c}{(6) Food}&\multicolumn{1}{c}{(7) Shelter}&\multicolumn{1}{c}{(8) Medical}\\\hline

Panel A: School enrollment and attendance rates		&&&&&&&&\\
\hspace{5pt}Expenses $\times$ 1923 birth cohort	&0.095$**$&0.139$**$	&0.320$**$	&32.050$*$&0.026$*$	&0.037$*$	&0.089$**$	&15.070$**$	\\
				 									&[0.032]	&[0.020]	&[0.046]	&[0.096]	&[0.058]	&[0.096]	&[0.032]	&[0.026]		\\
\hspace{5pt}Primary school enrollment rate 			&-0.080	&-0.080	&-0.080	&-0.080	&-0.019	&-0.019	&-0.019	&-0.019		\\
				 									&[0.148]	&[0.148]	&[0.148]	&[0.148]	&[0.524]	&[0.524]	&[0.522]	&[0.520]		\\
\hspace{5pt}Primary school attendance rate 			&-0.028	&-0.028	&-0.028	&-0.027	&-0.004	&-0.004	&-0.004	&-0.003		\\
				 									&[0.426]	&[0.426]	&[0.426]	&[0.428]	&[0.750]	&[0.748]	&[0.756]	&[0.766]		\\
\hspace{5pt}Observations							
&\multicolumn{1}{c}{14,145}
&\multicolumn{1}{c}{14,145}
&\multicolumn{1}{c}{14,145}
&\multicolumn{1}{c}{14,145}
&\multicolumn{1}{c}{14,138}
&\multicolumn{1}{c}{14,138}
&\multicolumn{1}{c}{14,138}
&\multicolumn{1}{c}{14,138}\\
													&&&&&&&&\\
Panel B: Wartime and flu pandemic effects			&&&&&&&&\\
\hspace{5pt}Expenses $\times$ 1923 birth cohort	&0.096$**$&0.140$**$&0.322$**$&32.685$*$&0.026$*$	&0.037$*$	&0.090$**$&15.128$**$	\\
				 									&[0.030]	&[0.014]	&[0.046]	&[0.096]	&[0.060]	&[0.086]	&[0.034]	&[0.036]		\\
\hspace{5pt}Wartime birth cohorts 					&0.035		&0.035		&0.035		&0.035		&-0.041$*$	&-0.041$*$	&-0.041$*$	&-0.041$*$		\\
				 									&[0.336]	&[0.336]	&[0.338]	&[0.338]	&[0.076]	&[0.076]	&[0.076]	&[0.076]		\\
\hspace{5pt}Pandemic influenza birth cohorts 		&-0.097	&-0.098	&-0.097	&-0.097	&-0.012	&-0.012	&-0.012	&-0.012		\\
				 									&[0.276]	&[0.276]	&[0.276]	&[0.276]	&[0.682]	&[0.678]	&[0.684]	&[0.676]		\\
\hspace{5pt}Observations							
&\multicolumn{1}{c}{14,145}
&\multicolumn{1}{c}{14,145}
&\multicolumn{1}{c}{14,145}
&\multicolumn{1}{c}{14,145}
&\multicolumn{1}{c}{14,138}
&\multicolumn{1}{c}{14,138}
&\multicolumn{1}{c}{14,138}
&\multicolumn{1}{c}{14,138}\\\bottomrule
\end{tabular}
}
{\scriptsize
\begin{minipage}{600pt}
\setstretch{0.85}
***, **, and * represent statistical significance at the 1\%, 5\%, and 10\% levels based on the $p$-values from the wild cluster bootstrap resampling method in brackets, respectively.
The data are clustered at the 13-county level in the bootstrap procedure.
The number of replications is fixed to 1,000 for all the specifications.\\
Notes: 
The number of observations in columns (1)--(4) and (5)--(8) are 14,145 and 14,138, respectively.
The data on the school-year-age-level average values of height (cm) or weight (kg) are used in the regressions.
The estimated coefficient on $\textit{I(YOB=1923)}\times\textit{Relief}$ in equation~(\ref{eq2}) is reported in the table.
All the regressions include controls for the rice yield in the birth year; fetal death rate in the birth year; school enrollment rate of the parental generation; school-age-specific fixed effects; and year fixed effects.
\end{minipage}
}
\end{center}
\end{table}
\end{landscape}
\begin{table}[h!]
\def\arraystretch{0.98}
\begin{center}
\captionsetup{justification=centering}
\caption{Effects of Fetal Earthquake Exposure on Height (cm) and Weight (kg) in the Limited JMA-SIS5 Area: Robustness of Including Additional Variables in the Railway Disruption Regressions}
\label{tab:rob_rail}
\footnotesize
\scalebox{0.97}[1]{
\begin{tabular}{lD{.}{.}{-2}D{.}{.}{-2}D{.}{.}{-2}D{.}{.}{-2}}
\toprule
&\multicolumn{2}{c}{Boys}&\multicolumn{2}{c}{Girls}\\
\cmidrule(rr){2-3}\cmidrule(rr){4-5}
&\multicolumn{1}{c}{(1)}&\multicolumn{1}{c}{(2)}&\multicolumn{1}{c}{(3)}&\multicolumn{1}{c}{(4)}\\
Panel A&\multicolumn{1}{c}{Ages 6--8}&\multicolumn{1}{c}{Ages 9--11}&\multicolumn{1}{c}{Ages 6--8}&\multicolumn{1}{c}{Ages 9--11}\\\hline
Panel A-1: Effects on height&&&&\\
1923 birth cohort											&-0.135$**$&-0.126	&-0.289$***$	&-0.186$**$	\\
															&[0.034]	&[0.232]	&[0.004]		&[0.017]	\\
1923 birth cohort $\times$ Railway disruption				&0.139		&0.233		&0.172			&0.341		\\
															&[0.774]	&[0.556]	&[0.288]		&[0.767]	\\
Primary school enrollment rate								&-0.027	&0.059		&-0.056		&-0.049	\\
															&[0.578]	&[0.662]	&[0.452]		&[0.653]	\\
Primary school attendance rate								&-0.038	&0.011		&0.069			&-0.062	\\
															&[0.552]	&[0.904]	&[0.256]		&[0.129]	\\
\hspace{5pt}Observations							
&\multicolumn{1}{c}{9,352}
&\multicolumn{1}{c}{9,348}
&\multicolumn{1}{c}{9,353}
&\multicolumn{1}{c}{9,347}\\
Panel A-2: Effects on weight&&&&\\
1923 birth cohort											&-0.056$*$	&-0.011	&-0.108$***$	&-0.094$*$	\\
															&[0.058]	&[0.856]	&[0.010]		&[0.073]	\\
1923 birth cohort $\times$ Railway disruption				&0.071		&-0.096	&0.052			&-0.063	\\
															&[0.760]	&[0.514]	&[0.464]		&[0.727]	\\
Primary school enrollment rate								&-0.017	&-0.035	&0.000			&-0.069$**$	\\
															&[0.562]	&[0.314]	&[1.000]		&[0.047]	\\
Primary school attendance rate								&0.001		&0.050		&0.021			&-0.005	\\
															&[0.952]	&[0.228]	&[0.304]		&[0.863]	\\
\hspace{5pt}Observations							
&\multicolumn{1}{c}{9,352}
&\multicolumn{1}{c}{9,348}
&\multicolumn{1}{c}{9,353}
&\multicolumn{1}{c}{9,347}\\\hline
&\multicolumn{2}{c}{Boys}&\multicolumn{2}{c}{Girls}\\
\cmidrule(rr){2-3}\cmidrule(rr){4-5}
&\multicolumn{1}{c}{(1)}&\multicolumn{1}{c}{(2)}&\multicolumn{1}{c}{(3)}&\multicolumn{1}{c}{(4)}\\
Panel B&\multicolumn{1}{c}{Ages 6--8}&\multicolumn{1}{c}{Ages 9--11}&\multicolumn{1}{c}{Ages 6--8}&\multicolumn{1}{c}{Ages 9--11}\\\hline
Panel B-1: Effects on height									&&&&\\
\hspace{5pt}1923 birth cohort								&-0.138$**$	&-0.128	&-0.292$***$	&-0.184$**$	\\
															&[0.040]		&[0.220]	&[0.004]		&[0.017]		\\
\hspace{5pt}1923 birth cohort $\times$ Railway disruption	&0.136			&0.238		&0.181			&0.342			\\
															&[0.770]		&[0.554]	&[0.248]		&[0.767]		\\
\hspace{5pt}Wartime birth cohorts 							&0.066			&0.070		&0.185			&0.112			\\
				 											&[0.314]		&[0.412]	&[0.186]		&[0.147]		\\
\hspace{5pt}Pandemic influenza birth cohorts 				&-0.113		&-0.083	&-0.024		&0.131			\\
				 											&[0.302]		&[0.554]	&[0.850]		&[0.217]		\\
\hspace{5pt}Observations							
&\multicolumn{1}{c}{9,352}
&\multicolumn{1}{c}{9,348}
&\multicolumn{1}{c}{9,353}
&\multicolumn{1}{c}{9,347}\\
Panel B-2: Effects on weight									&&&&\\
\hspace{5pt}1923 birth cohort								&-0.057$*$		&-0.011	&-0.109$***$	&-0.091$*$\\
															&[0.066]		&[0.844]	&[0.010]		&[0.077]	\\
\hspace{5pt}1923 birth cohort $\times$ Railway disruption	&0.071			&-0.102	&0.055			&-0.067	\\
															&[0.756]		&[0.486]	&[0.432]		&[0.713]	\\
\hspace{5pt}Wartime birth cohorts 							&-0.043		&0.065		&-0.003		&0.046		\\
				 											&[0.296]		&[0.102]	&[0.994]		&[0.557]	\\
\hspace{5pt}Pandemic influenza birth cohorts 				&-0.015		&-0.013	&-0.026		&0.129		\\
				 											&[0.696]		&[0.692]	&[0.606]		&[0.135]	\\
\hspace{5pt}Observations							
&\multicolumn{1}{c}{9,352}
&\multicolumn{1}{c}{9,348}
&\multicolumn{1}{c}{9,353}
&\multicolumn{1}{c}{9,347}
\\\bottomrule

\end{tabular}
}
{\scriptsize
\begin{minipage}{440pt}
\setstretch{0.85}
***, **, and * represent statistical significance at the 1\%, 5\%, and 10\% levels based on the $p$-values from the wild cluster bootstrap resampling method in brackets, respectively.
The data are clustered at the 13-county level in the bootstrap procedure.
The number of replications is fixed to 1,000 for all the specifications.\\
Notes:
Samples include municipalities receiving no disaster relief in the SIS5 area.
The numbers of observations in columns (1)--(4) are 9352, 9348, 9353, and 9347, respectively.
The data on the school-year-age-level average values of height (cm) or weight (kg) are used in the regressions.
All the regressions include controls for the rice yield in the birth year; fetal death rate in the birth year; school enrollment rate of the parental generation; school-age-specific fixed effects; and year fixed effects.
\end{minipage}
}
\end{center}
\end{table}
\begin{table}[]
\def\arraystretch{1.0}
\begin{center}
\captionsetup{justification=centering}
\caption{Effects of Fetal Earthquake Exposure on Height (cm) and Weight (kg): Testing the Potential Impacts on Surrounding Cohorts}
\label{tab:r_robustness2}
\footnotesize
\scalebox{1.0}[1]{
\begin{tabular}{lcccccccc}
\toprule
&\multicolumn{2}{c}{Boys}&\multicolumn{2}{c}{Girls}\\
\cmidrule(rr){2-3}\cmidrule(rr){4-5}
&\multicolumn{1}{c}{(1)}&\multicolumn{1}{c}{(2)}&\multicolumn{1}{c}{(3)}&\multicolumn{1}{c}{(4)}\\
&\multicolumn{1}{c}{Ages 6--8}&\multicolumn{1}{c}{Ages 9--11}&\multicolumn{1}{c}{Ages 6--8}&\multicolumn{1}{c}{Ages 9--11}\\\hline
Panel A: Effects on height							&&&&\\
\hspace{5pt}1922 birth cohort						&0.063[0.126]		&-0.038[0.534]		&0.042[0.704]		&0.023[0.805]\\
\hspace{5pt}1922 birth cohort $\times$ SIS7		&0.167[0.406]		&0.447[0.476]		&0.073[0.402]		&0.049[1.000]\\
\hspace{5pt}1923 birth cohort						&-0.143[0.040]**	&-0.173[0.070]*	&-0.218[0.042]**	&-0.222[0.003]***\\
\hspace{5pt}1923 birth cohort $\times$ SIS7		&0.067[0.526]		&0.170[0.480]		&-0.284[0.012]**	&-0.598[0.003]***\\
\hspace{5pt}1924 birth cohort						&-0.065[0.374]		&-0.091[0.354]		&0.134[0.242]		&-0.099[0.255]	\\
\hspace{5pt}1924 birth cohort $\times$ SIS7		&0.071[0.724]		&0.077[1.000]		&0.193[0.148]		&0.363[0.513]\\
\hspace{5pt}Observations							
&\multicolumn{1}{c}{14,139}
&\multicolumn{1}{c}{14,133}
&\multicolumn{1}{c}{14,145}
&\multicolumn{1}{c}{14,138}		\\
													&&&&\\
Panel B: Effects on weight							&&&&\\
\hspace{5pt}1922 birth cohort						&0.040[0.236]		&0.008[0.866]		&-0.006[0.848]		&0.020[0.799]\\
\hspace{5pt}1922 birth cohort $\times$ SIS7		&0.073[0.500]		&0.289[0.434]		&-0.062[0.408]		&-0.094[0.161]	\\
\hspace{5pt}1923 birth cohort						&-0.039[0.062]*	&-0.033[0.410]		&-0.104[0.006]***	&-0.124[0.013]***\\
\hspace{5pt}1923 birth cohort $\times$ SIS7		&0.012[0.790]		&0.136[0.480]		&-0.276[0.002]***	&-0.314[0.005]***\\
\hspace{5pt}1924 birth cohort						&0.011[0.810]		&-0.025[0.650]		&0.006[0.794]		&-0.046[0.131]\\
\hspace{5pt}1924 birth cohort $\times$ SIS7		&-0.020[1.000]		&0.029[1.000] 		&-0.116[0.218]		&0.024[0.967]\\
\hspace{5pt}Observations							
&\multicolumn{1}{c}{14,139}
&\multicolumn{1}{c}{14,133}
&\multicolumn{1}{c}{14,145}
&\multicolumn{1}{c}{14,138}		\\\bottomrule
\end{tabular}
}
{\scriptsize
\begin{minipage}{420pt}
\setstretch{0.85}
***, **, and * represent statistical significance at the 1\%, 5\%, and 10\% levels based on the $p$-values from the wild cluster bootstrap resampling method in brackets, respectively.
The data are clustered at the 13-county level in the bootstrap procedure.
The number of replications is fixed to 1,000 for all the specifications.\\
Notes:
The numbers of observations for each regression reported in columns (1)--(4) are $14,139$, $14,133$, $14,145$, and $14,138$, respectively.
The data on the school-year-age-level average values of height (cm) or weight (kg) are used in the regressions.
All the regressions include controls for the rice yield in the birth year; fetal death rate in the birth year; school enrollment rate of the parental generation; school-age-specific fixed effects; and year fixed effects.
\end{minipage}
}
\end{center}
\end{table}
\begin{table}[]
\def\arraystretch{1.0}
\begin{center}
\captionsetup{justification=centering}
\caption{Effects of Fetal Earthquake Exposure on Height (cm) and Weight (kg): Testing the Potential Impacts on Surrounding Cohorts Excluding the 1923 Birth Cohort}
\label{tab:r_placebo}
\footnotesize
\scalebox{1.0}[1]{
\begin{tabular}{lD{.}{.}{-2}D{.}{.}{-2}D{.}{.}{-2}D{.}{.}{-2}}
\toprule
&\multicolumn{2}{c}{Boys}&\multicolumn{2}{c}{Girls}\\
\cmidrule(rr){2-3}\cmidrule(rr){4-5}
&\multicolumn{1}{c}{(1)}&\multicolumn{1}{c}{(2)}&\multicolumn{1}{c}{(3)}&\multicolumn{1}{c}{(4)}\\
&\multicolumn{1}{c}{Ages 6--8}&\multicolumn{1}{c}{Ages 9--11}&\multicolumn{1}{c}{Ages 6--8}&\multicolumn{1}{c}{Ages 9--11}\\\hline

Panel A: Effects on height				&&&&\\
\hspace{5pt}1918 birth cohort			&-0.024		&0.066		&-0.028	&0.034		\\
										&[0.734]		&[0.328]	&[0.762]	&[0.709]	\\
\hspace{5pt}1919 birth cohort			&0.015			&-0.042	&0.057		&0.110		\\
										&[0.724]		&[0.528]	&[0.352]	&[0.213]	\\
\hspace{5pt}1920 birth cohort			&-0.098$*$		&0.031		&-0.080	&-0.037	\\
										&[0.080]		&[0.630]	&[0.212]	&[0.721]	\\
\hspace{5pt}1921 birth cohort			&0.024			&0.024		&0.011		&-0.035	\\
										&[0.692]		&[0.768]	&[0.884]	&[0.741]	\\
\hspace{5pt}1922 birth cohort			&0.078			&-0.017	&-0.017	&0.030		\\
										&[0.104]		&[0.654]	&[0.884]	&[0.737]	\\
\hspace{5pt}1924 birth cohort			&-0.079		&-0.084	&0.120		&-0.090	\\
										&[0.308]		&[0.396]	&[0.222]	&[0.327]	\\
\hspace{5pt}1925 birth cohort			&0.034			&0.056		&-0.005	&-0.000	\\
										&[0.538]		&[0.432]	&[0.916]	&[0.993]	\\
\hspace{5pt}Observations							
&\multicolumn{1}{c}{12,844}
&\multicolumn{1}{c}{12,849}
&\multicolumn{1}{c}{12,851}
&\multicolumn{1}{c}{12,860}		\\
Panel B: Effects on weight				&&&&\\
\hspace{5pt}1918 birth cohort			&-0.053$*$		&0.002		&-0.058	&0.006		\\
										&[0.092]		&[0.902]	&[0.196]	&[0.915]	\\
\hspace{5pt}1919 birth cohort			&-0.005		&0.007		&-0.008	&0.015		\\
										&[0.794]		&[0.674]	&[0.750]	&[0.745]	\\
\hspace{5pt}1920 birth cohort			&-0.015		&-0.007	&-0.006	&0.075		\\
										&[0.330]		&[0.810]	&[0.898]	&[0.277]	\\
\hspace{5pt}1921 birth cohort			&0.016			&0.006		&0.032		&-0.010	\\
										&[0.508]		&[0.832]	&[0.308]	&[0.891]	\\
\hspace{5pt}1922 birth cohort			&0.044			&0.015		&-0.014	&0.005		\\
										&[0.120]		&[0.668]	&[0.672]	&[0.931]	\\
\hspace{5pt}1924 birth cohort			&-0.012		&-0.023	&0.008		&-0.031	\\
										&[0.662]		&[0.650]	&[0.790]	&[0.403]	\\
\hspace{5pt}1925 birth cohort			&-0.008		&-0.004	&0.016		&-0.009	\\
										&[0.786]		&[0.888]	&[0.436]	&[0.801]	\\
\hspace{5pt}Observations							
&\multicolumn{1}{c}{12,844}
&\multicolumn{1}{c}{12,849}
&\multicolumn{1}{c}{12,851}
&\multicolumn{1}{c}{12,860}		\\\bottomrule
\end{tabular}
}
{\scriptsize
\begin{minipage}{350pt}
\setstretch{0.85}
* represents statistical significance at the 10\% level based on the $p$-values from the wild cluster bootstrap resampling method in brackets.
The data are clustered at the 13-county level in the bootstrap procedure.
The number of replications is fixed to 1,000 for all the specifications.\\
Notes:
Each estimate of the cohort effect is obtained from a regression for the sample excluding the 1923 birth cohort.
The numbers of observations for each regression reported in columns (1)--(4) are $12,844$, $12,849$, $12,851$, and $12,860$, respectively.
The data on the school-year-age-level average values of height (cm) or weight (kg) are used in the regressions.
All the regressions include controls for the rice yield in the birth year; fetal death rate in the birth year; school enrollment rate of the parental generation; school-age-specific fixed effects; and year fixed effects.
The birth cohorts before 1918 and after 1926 are not included in these analyses because they include individuals aged under four years, which is insufficient to calculate the mean birth cohort effects.
\end{minipage}
}
\end{center}
\end{table}

\end{spacing}
\clearpage
\renewcommand{\refname}{{\large References, Documents, Statistical Reports, and Database}}
\begin{spacing}{0.91}

\end{spacing}
\end{document}